\documentclass[12pt]{article}
\usepackage{setspace,graphicx,epstopdf,amsmath,amsfonts,amssymb,amsthm}
\usepackage{marginnote,datetime,enumitem,subfigure,rotating,fancyvrb}
\usepackage{hyperref,float}
\usepackage[longnamesfirst]{natbib}
\usepackage{mathtools}
\usepackage{caption}
\usepackage{float}
\usepackage{lscape}
\usepackage{graphicx}
\usepackage{flafter} 
\usepackage{tabularx} 
\usepackage{booktabs}
\usepackage{changepage}
\usepackage{setspace}
\usepackage{placeins}
\usepackage{threeparttable}
\usepackage{ragged2e}
\usepackage[export]{adjustbox}
\newcolumntype{Y}{>{\raggedleft\arraybackslash}X}
\usdate
\usepackage[a4paper,bindingoffset=0.2in,%
            left=1.1in,right=0.85in,top=1in,bottom=1in,%
            footskip=.25in]{geometry}

\usepackage{indentfirst} 
\usepackage{endnotes}    

\begin{document}

\setlist{noitemsep}  
\onehalfspacing      
\renewcommand{\footnote}{\endnote}  

\author{Mykola Pinchuk\thanks{\rm Simon Business School, University of Rochester. Email: Mykola.Pinchuk@ur.rochester.edu. \newline I would like to thank Yixin Chen, Bill Schwert, Giulio Trigilia, Shuaiyu Chen, David Swanson, Zhao Jin, Pingle Wang and Robert Mann for helpful comments.}}

\title{\Large \bf Customer Momentum}

\date{08 December 2018}             


\maketitle
\thispagestyle{empty}

\bigskip


\centerline{\bf ABSTRACT}

\small
\begin{onehalfspace}  
  \noindent This paper examines customer momentum, defined as a positive relationship between a firm's returns and past returns of its customers. I confirm previous evidence (Cohen and Frazzini 2008) that customer momentum is both statistically and economically significant. Long-short equally-weighted (value-weighted) decile portfolio generates a monthly return of 122 (106) basis points and a t-statistic above 4 (2.8) with respect to Fama-French factor models. The paper reports that customer momentum neither explains nor is explained by price momentum and earnings momentum. Customer momentum is partially driven by the lead-lag relationship between small and large stocks. I find that in the post-discovery sample, customer momentum has a smaller magnitude and loses statistical significance. The results are consistent with the hypothesis that after its discovery, customer momentum decreased due to exploitation by investors.
\end{onehalfspace}
\normalsize
\medskip

\clearpage
\setstretch{1.6}

\section{Introduction} \label{sec:Model}
The supply chain link is arguably the most important measure of the relationship between the firms. The distress of the main customer is bad news for the firm. In the world of costless information production and unlimited investor attention all value-relevant news, including information on the main customers, is immediately incorporated into the stock price. The costs of collecting this information in a world, where most analysts specialize in separate stocks or industries, suggest that the diffusion of such information need not be immediate.
\paragraph{}
According to the hypothesis of slow information diffusion, firms with high recent returns of their customers are more likely to have high abnormal returns. Cohen and Frazzini (2008) document a positive relationship between the contemporaneous return of the firm and past returns of its customers. They refer to this effect as customer momentum and interpret it by slow information diffusion. 
This paper has two main objectives. The first goal is to replicate the results in Cohen and Frazzini (2008) and verify them in the post-discovery sample. The second objective is to explore the interaction between different types of momentum effects in order to gain insight into whether they capture different phenomena. 
\paragraph{}
This paper examines the stock returns of the firms, linked by customer relationships. Consistent with the past evidence, it finds return predictability by past returns of the customers. Long-short equally-weighted (value-weighted) portfolio, formed on past returns of customers, has a monthly return of 122 (106) basis points, suggesting a large magnitude of customer momentum. These returns do not decrease significantly after controlling for traditional asset pricing factors. To further test whether customer momentum is captured by factor models, I perform Fama-MacBeth regressions and factor-spanning tests. In both cases factor models can not explain customer momentum, implying that customer momentum is another asset pricing anomaly. 
\paragraph{}
While presented results are consistent with limited attention, there may be liquidity or risk-based explanations of customer momentum. To gain insight into the possible interpretation of this effect as a risk premium, I examine the relationship of customer momentum with macroeconomic variables. Customer momentum has a very low correlation with consumption growth, GDP growth, and different proxies for labor income. While these relationships are not a formal test of risk-based explanation, they do not suggest that customer momentum captures risk under consumption-based asset pricing models. 
\paragraph{}
In the second part of the paper, I examine the relationship between customer momentum, price momentum, and earnings momentum. Returns of customers and suppliers exhibit a significantly positive contemporaneous correlation. Therefore, customer momentum can be driven by the combination of the price momentum of customers and the contemporaneous correlation between the returns of customers and suppliers. Previous research (Chordia and Shivakumar 2006, Novy-Marx 2015) suggests that earnings momentum captures price momentum. I confirm these findings and report that after controlling for earnings momentum, the effect of price momentum loses significance and reverses the sign. Customer momentum, defined by the returns of the customer portfolio within the most recent month, does not have explanatory power for price momentum or earnings momentum. Customer momentum is not explained by the other types of momentum, suggesting that it reflects a different effect. When customer momentum is measured as past returns of the customer portfolio for more than 1 recent month, it is captured by both price momentum and earnings momentum. These results suggest that the customer momentum should be measured using 1-1 lag since otherwise, it is a noisy proxy for price momentum. 
\paragraph{}
By using the data after 2004, this paper provides an out-of-sample analysis of the customer momentum effect, discovered by Cohen and Frazzini on 1981-2004 sample. Given the recent surge in the number of documented asset pricing anomalies, researchers expressed concerns that these anomalies are just statistical artifacts (Harvey, Liu and Zhu 2016). Alternatively, the anomalies could have been real at the time of publication but disappeared due to trading activity after the publication (Schwert 2003, Pontiff and MacLean 2015). Customer momentum has the same pattern and smaller magnitude in the recent sample. Returns of value-weighted long-short portfolios, formed on customer momentum, are not significantly different from zero. The customer momentum factor has a negative return during 2005-2018. These findings suggest that customer momentum probably is not a statistical artifact, but becomes less significant due to post-publication exploitation by investors.
\paragraph{}
Lead-lag relationship, defined as cross-correlations between returns of stocks at different lags (Lo and MacKinlay 1990), can be an alternative explanation of customer momentum. Since the sample consists of firms and customers, which account for more than 10\% of total sales, the mean customer is much larger than the mean supplier. Therefore, there should exist some correlation between past returns of large customers and current returns of small suppliers. After restricting the sample to the links with the smallest ratio of customer-to-supplier size, the magnitude of customer momentum decreases by a factor of 2-4. After adding an interaction term between past returns of customers and their relative size to Fama-MacBeth regressions, customer momentum reverses the sign. Overall, the results suggest that customer momentum is at least partially driven by the lead-lag relationship between the returns of large and small stocks.
\paragraph{}
Finally, this paper investigates the limited attention channel of customer momentum. To measure attention to the customer-supplier link, I calculate the normalized abnormal trading volume of the supplier during the day of the earnings announcement of its customer. No spike in the trading volume of the supplier during the customer's earnings announcement can imply low attention to the customer-supplier link. The results suggest that customer momentum across equally-weighted portfolios is significantly stronger for low-attention links, consistent with the limited attention hypothesis. However, no such pattern is present in value-weighted portfolios. Therefore, the findings provide mixed evidence on the limited attention explanation of customer momentum.
\paragraph{}
One potential concern is the low external validity of the results. Low quality of data on customer links leads to a small sample size. In the earlier years, the sample represents 3\%-10\% of the total market capitalization. Slight improvement in the quality of links leads to an increase in this fraction to 10\%-16\% in the late sample. Since the sample uses the data on returns of only American companies, the firms, whose main customers are incorporated outside the US, are excluded from the sample. Similarly, the sample excludes the firms, whose primary customers are private firms or governmental agencies. Finally, there are cross-industry differences in the number of customers and the distribution of the total sales between them. All these facts can result in bias.

\section{Literature review} \label{sec:Model}
There is growing finance literature, exploring limited investor attention and its consequences for asset pricing. Psychology (Kahneman 1973) suggests that individuals have limited cognitive capacity and can allocate attention only between small number of tasks. Merton (1987) was the first to suggest that investors pay little attention to the stocks, for which the cost of information collection overweights potential return. Hong and Stein (1999) propose the model of slow information diffusion, heterogeneous across assets. They assume that due to bounded rationality investors fail to extract the information, contained in stock prices. Peng and Xiong (2006) develop the model, in which investors with limited attention focus more on market-wide news than on firm-specific news. 
\paragraph{}
The literature on the limited investors' attention is mostly empirical. Huberman and Regev (2001) find that market had ignored important publicly available information on the firm for 5 months and reacted only after the information was published in news media. Hou and Moskowitz (2005) examine firms heterogeneity in price delay and find that small segment of neglected firms accounts for large proportion of the variation in returns. Barber and Odean (2007) find that individual investors buy stocks during attention-grabbing events, such as news, high abnormal trading volume or extreme one-day returns. Hong, Torous and Valkanov (2005) report that industry returns appear to predict market returns and conclude that this implies investor inattention. Menzly and Ozbas (2010) find that stocks in economically related customer-supplier industries cross-predict returns of each other.
\paragraph{}
Recent literature focuses on a search for the measures of investor inattention. Hirshleifer, Lim and Teoh (2009) find that during the days when a large fraction of firms report earnings, price reaction is smaller and post-earnings announcement drift (PEAD) is larger. Hong, Peng and Xiong (2009) report that stocks with large abnormal trading volume during the earnings announcements experience larger one-day returns and smaller PEAD. Ben-Rephael, Da and Israelsen (2017) propose new measure of attention of institutional investors and report similar findings when investor attention is high.  
\paragraph{}
While this article is the first paper to examine interplay between customer momentum, price momentum and earnings momentum, there are several studies on the relationship between price momentum and earnings momentum. Using the results from return spreads in independent three-by-three portfolio sorts on PEAD and price momentum, Chan, Jegadeesh and Lakonishok (1996) conclude that earnings momentum and price momentum are two separate phenomena. Chordia nad Shivakumar (2006) finds that while past returns and earnings surprises have separate explanatory power for the future returns, price momentum factor is subsumed by earnings momentum factor. Novy-Marx (2015) shows that earnings momentum, both as a characteristic and as a factor, completely captures price momentum.

\section{Data and sample selection} \label{sec:Model}
The Regulation SFAS 131 requires all public firms to report identities of the customers, accounting for more than 10\% of total sales. This data is publicly available from Compustat Segments. I use CRSP as a source of common stock returns and Compustat as a source of accounting data. The sample covers the period between January 1978 and June 2018. 
\paragraph{}
Compustat Segments contains identities of the main customers in the form of company names, originally reported by the firms. Unfortunately, firms do not always report proper company name of their customers. Prior to 1997, many firms reported names of their customers as abbreviations or initialisms. These factors complicate forming customer-supplier links. I was deliberately conservative in creating the customer links, which resulted in relatively small number of matches. To be consistent with asset pricing literature, I allow for 6-month lag between the fiscal year-ends of firms, reporting customers, and their stock returns. After matching the firms and the customers, I calculate equally-weighted returns of customers' portfolio. 
\paragraph{}
The sample consists of all firms with nonmissing customer links, present in merged CRSP/Compustat. I require firms to have market equity data on June of each year. Links data includes 255747 firm-month links and 4746 public firms across 1977-2018. The merged sample includes 145489 firm-month observations and 3266 firms. Table 1 reports sample coverage across years.
\begin{center}
[Insert Table 1] 
\end{center}
\paragraph{}
The Table 1 documents that the sample coverage varies between 1\% and 16\% of the total market capitalization. The years before 2000 have especially low sample coverage, lower than 5\%. There are multiple reasons for such low fraction of matched firms. Most firms, present in CRSP sample, do not report data on the principal customers. The most frequently reported customers are US government, different countries, states and industries. Large fraction of firms, reported as principal customers, are private companies or the firms, incorporated outside of the US. 
\paragraph{}
Quarterly Compustat is the source of data on earnings announcement. Standardized unexpected earnings  (SUE) are defined as the most recent year-to-year change in earnings per share (EPS), divided by the standard deviation of the changes in earnings during the recent eight announcements. I define SUE as missing if there are less than 6 quarterly announcements during 2-year window. I use Compustat item EPSPXQ for EPS and RDQ for earnings announcement dates. I calculate announcement cumulative abnormal returns (CAR3) as stock returns during 3-day window, centered around announcement date minus market return for the same period. The paper uses CRSP value-weighted return (vwretd) as a proxy for market return. If earnings are announced on non-trading day, I assume that they are announced during the following trading day.
\paragraph{}
When defining momentum returns, the paper follows asset pricing literature and denotes the return window by two numbers, corresponding to the first and the last months of the window before the current month. Therefore, 12-2 price momentum is the return of the firm during the last year, excluding the most recent month. 

\section{Main Results} \label{sec:Model}
This section documents and discusses the main results of the paper. First, I present the evidence on the magnitude and significance of the customer momentum in the full sample, i.e., 1979-2018. Then I report the findings on the customer momentum using only the sample after 2005, therefore performing out-of-sample test of customer momentum. The last subsection discusses an interaction between different types of momentum in order to determine whether all types of momentum represent separate phenomena.

\subsection{Customer momentum in full sample}
In each month, I divide firms into 10 decile portfolios, based on their customers' return in the previous month. Due to the low number of stocks in the early part of the sample, I form decile breakpoints using the whole sample. Table 2 presents the excess returns of decile portfolios, formed on customer momentum, calculated using different lags. 
\begin{center}
[Insert Table 2 here] 
\end{center}
\paragraph{}
I report the results for the different number of lags to facilitate comparison between customer momentum and price momentum. Price momentum is usually defined as 12-2 momentum, i.e., returns during the past year, excluding the most recent month. The reason for this calculation is to exclude short-term reversal, since the returns have weak negative autocorrelation at 1-month lag. Unlike price momentum, customer momentum seems to be significantly driven by the returns in the previous month. For most of lags, the magnitude of customer momentum increases if I calculate customer momentum over the window, which includes the most recent month. These results are consistent with economic motivation for customer momentum. Investor inattention, which is the most intuitive explanation of customer momentum, seems unlikely to last for several months. Therefore, for the remainder of this paper, unless otherwise indicated, customer momentum will mean positive relationship between returns of the target firm and the returns of the customers' portfolio during the most recent month. 
Panel A of Table 2 reports excess returns of equally-weighted decile portfolios, formed on customer momentum. The results suggest that long-short decile portfolio has monthly excess return of 1.22\%, which is both statistically and economically significant. Long-short value-weighted decile portfolios, formed on customer momentum, generate somewhat smaller monthly returns of 1.06\%. These results highlight large magnitude of customer momentum and suggest that it is stronger across small stocks. Returns of value-weighted decile portfolios appear to be rather noisy, which is probably caused by relatively small sample size.
\paragraph{}
In order to test whether return factors can explain returns of portfolios, formed on customer momentum, the paper calculates abnormal returns and reports them in Table 3. I use the following factor models:
\begin{itemize}
    \item {CAPM}
    \item {CAPM + Momentum}
    \item {Fama-French 3}
    \item {Fama-French 3 + Momentum}
    \item {Fama-French 5}
    \item {Fama-French 5 + Momentum}
\end{itemize}
\begin{center}
[Insert Table 3 here] 
\end{center}
\paragraph{}
Table 3 clearly shows that traditional asset pricing factors can not explain high returns of decile portfolios, formed on customer momentum. Abnormal returns of equally-weighted long-short customer momentum portfolio are not affected by Fama-French factors and are above 1.1\% for any specification. These results are highly statistically significant, featuring T-statistics above 4.5. Abnormal returns of value-weighted portfolios, formed on customer momentum, are significantly positive, varying between 0.91\% and 1.04\%.
\begin{center}
[Insert Table 4 and Table 5 here] 
\end{center}
\paragraph{}
Tables 4 and 5 report abnormal returns of decile portfolios, formed on customer momentum 6-1 and 12-1 respectively. The results are very different from Table 3. Now price momentum captures the effects of customer momentum and renders alphas statistically insignificant. These findings suggest that customer momentum has effect, independent from price momentum only if measured during the most recent month. Therefore, customer momentum at lags, higher than 1-1, seems to be noisy proxy for price momentum.
\begin{center}
[Insert Table 6 and Figure 1 here] 
\end{center}
\paragraph{}
Table 6 documents summary statistic of value-weighted long-short quintile and decile portfolios, based on customer momentum. Long-short decile portfolio has excess return 0.4\% larger than market excess return and Sharpe Ratio of 0.54. Unlike excess market return, long-short portfolios on customer momentum appear to be right-skewed. Figure 1 shows performance of \$1, invested in each portfolio. Both quintile and decile long-short portfolios are uncorrelated with market, implying zero market risk. 
\paragraph{}
The obvious question is how to interpret customer momentum. While inattention explanation seems more intuitive, there may be risk-based explanation of customer momentum. In order to find whether customer momentum is related to macroeconomic variables, I construct customer momentum factor and show its correlations with macroeconomic variables in Table 7.
\paragraph{}
To construct customer momentum factor, I use exactly the same procedure as French. I
 construct breakpoints for subsample of NYSE stocks and construct customer momentum factor (CMOM) using two-by-three sorts by customer momentum and size. 
\begin{center}
[Insert Table 7] 
\end{center}
\paragraph{}
CMOM has significant correlations only with other asset pricing factors. It is positively correlated with UMD and SMB and has negative correlation with HML and RMW. Given the nature and construction of CMOM its positive correlation with UMD is not surprising. The paper will discuss the relationship between CMOM and the other factors in more detail in subsection 4.3. Lack or correlation of CMOM with macroeconomic variables implies that there is no obvious risk-based explanation in line with consumption asset pricing models. 

\subsection{Customer momentum in 2005-2018}
Recent asset pricing studies showed that anomalies often vanish out-of-sample (Pontiff and MacLean 2015, Hou, Xue and Zhang 2018). Data mining (Harvey, Liu and Zhu, 2016) and arbitrage (Schwert 2003) were suggested as possible explanations. Since Cohen and Frazzini (2008) discovered customer momentum using the sample from 1981 to 2004, 2005-2018 sample can provide out-of-sample test of customer momentum anomaly.
\paragraph{}
Table 6 documents excess returns of decile portfolios, formed on customer momentum. Excess returns of equally-weighted long-short portfolio shrank from 1.5\% to 0.6\% and excess returns of value-weighted long-short portfolio decreased from 1.3\% to 0.6\%. The former return is marginally statistically significant, while the latter loses statistical significance. 
\begin{center}
[Insert Table 8, Table 9 and Table 10] 
\end{center}
\paragraph{}
Tables 8-10 suggest that traditional factors are still unable to capture the customer momentum. However, in post-discovery sample excess returns of long-short portfolios are already small, so even small effects of factors on abnormal returns cause them to lose statistical significance. None of the two long-short portfolios delivers statistically significant alpha at 5\% significance level.
\paragraph{}
Since customer momentum has the same pattern in post-discovery sample and is still not explained by traditional asset pricing factors, I hesitate to say that it is statistical artifact. Small sample size can cause low power of the tests above and low statistical significance. It is possible to interpret smaller magnitude of customer momentum in post-discovery sample as a result of trading activity of investors. If customer momentum is driven by mispricing, then it should be arbitraged away over time. Recent findings from mutual funds literature suggest that when evaluating mutual funds, investors care only about alpha with respect to CAPM rather than alpha with respect to factor models (Berk and Binsbergen 2015, Barber, Huang and Odean 2016). This evidence implies that investors behave as if they attribute all anomalies to mispricing. Thus even if customer momentum represents higher risk, it is more likely that investors will load on this risk after the discovery, leading to diminishing post-discovery customer momentum. Alternatively, if customer momentum is driven by risk, then investors, less exposed to such risk, can load on customer momentum to capture its risk premium. Therefore, we can explain a decrease in customer momentum out of sample by investor activity regardless of the interpretation of customer momentum.

\subsection{Price Momentum, Earnings Momentum and Customer Momentum}
This subsection discusses the interaction between customer momentum, price momentum and earnings momentum. Although price momentum measures autocorrelation of returns and customer momentum captures return predictability across stocks, these two concepts are closely related. Since customer and supplier are connected by supply chain link, their returns should exhibit strong comovement. Contemporaneous correlation between the returns of the firm and its customers' portfolio is 0.26. This number is significantly higher than the contemporaneous correlation between the returns of randomly chosen stocks in CRSP, equal to 0.12.\footnotemark
\footnotetext{In every month, this randomly-chosen sample has the same number of stocks as full customer momentum sample.}
High contemporaneous correlation between returns of customers and suppliers, combined with price momentum of suppliers, can mechanically explain customer momentum. Alternatively, customer momentum and contemporaneous correlation between returns of customers and suppliers can explain price momentum of the suppliers. Therefore, this subsection explores the relationship between price momentum and customer momentum.
\paragraph{}
Following Novy-Marx (2015), I measure earnings momentum using standardized unexpected earnings (SUE) and cumulative abnormal returns within 3-day window around earnings announcements (CAR3). 
\begin{center}
[Insert Table 11] 
\end{center}
\paragraph{}
Table 11 describes summary statistics of the measures of momentum. Notice that customer momentum is measured using 1-1 lag, while price momentum is calculated at 12-2 lag. Therefore, to facilitate comparison between price and customer momentum, I add short-term reversal, which is measured using the same lag as the customer momentum. Table 13 reports results of Fama-MacBeth regressions of returns on stock characteristics, including different types of momentum. 
\begin{center}
[Insert Table 13] 
\end{center}
\paragraph{}
Coefficient on customer momentum is robust to different regression specifications, implying that customer momentum is not explained by price momentum, earnings momentum, size, book-to-market or profitability. The magnitude of the coefficient suggests that 1 standard deviation increase in customer momentum is associated with 4\% larger annualized returns. Consistent with Novy-Marx (2015), measures of earnings momentum capture price momentum, which loses significance and reverses the sign. 
\paragraph{}
Since cross-sectional tests can suffer from measurement error and rely upon linear relationship between returns and explanatory variables, I perform further analysis using factor spanning tests. I construct CMOM factor, following the same methodology as UMD. 
\begin{center}
[Insert Table 14, Table15, Table16 and Table 17] 
\end{center}
\paragraph{}
Tables 14, 15, 16 and 17 present results of factor spanning tests. When controlling for the other factors, CMOM has alpha between 0.29\% and 0.47\%. Specifications 1 and 2 show that CMOM has small loadings on Fama-French 5 factors and marginally significant negative loading on RMW factor. Specification 4 documents that while UMD has small explanatory power for CMOM in terms of its coefficient and $R^2$, it significantly decreases alpha of CMOM. Overall the regressions, presented in Table 14 provide little evidence that UMD can capture CMOM. While controlling for earnings momentum, measured by CAR3, makes alphas of CMOM not statistically significant, this loss of significance is caused by increased standard error rather than smaller coefficient. Therefore, unlike price momentum, customer momentum is not explained by earnings momentum. 
\paragraph{}
CMOM has marginally significant explanatory power for UMD and slightly decreases alpha of UMD from 0.85\% to 0.79\%. Consistent with the results from Fama-MacBeth regressions, earnings momentum has dramatic effect on UMD, rendering its alpha insignificant. When measured as SUEF, earnings momentum has very large abnormal returns and is not explained by the other factors. Specifications (4)-(8) in the Table 17 show that UMD has high explanatory power for earnings momentum factor, measured by CAR3F. Controlling for UMD increases $R^2$ thrice and renders alpha of CAR3F statistically insignificant.
\begin{center}
[Insert Figure 2] 
\end{center}
\paragraph{}
Figure 2 shows the performance of \$1, invested in CMOM and UMD in 1979. Both factors have comparable performance and significantly positive correlation before 2004. After 2004 the correlations between the factors decreased to -0.03. Although CMOM did not suffer the crash in 2009, it performed even worse than UMD in after-discovery sample. Since this plot raises questions whether CMOM changed its pattern after 2004, I perform the same analysis of interaction between the factors separately for 1979-2004 and 2005-2018. I report the results in the following tables:
\begin{center}
[Insert Tables 22-31] 
\end{center}
\paragraph{}
Results of Fama-MacBeth regressions for the two subsamples are strikingly different. Coefficients on all explanatory variables shrink by at least 40\%. In post-discovery sample SUE is the only statistically significant determinant of stock returns. While small sample size can cause low power and low statistical significance, it can not explain dramatic decrease in the absolute value of coefficients on all the variables. Moreover, coefficients on book-to-market, profitability and price momentum change the sign. Coefficient on customer momentum decreases by 60\% and is not statistically significant. One standard deviation increase in customer momentum is correlated with 1.6\% higher annualized returns, suggesting small economic significance. These results are consistent with hypothesis that all the variables, predicting stock returns, regardless of whether they represent risk or mispricing, are losing predictive power over time due to exploitation by investors. Both variables, likely to correspond to risk factors (book-to-market, size) and mispricing (measures of momentum) experience large decrease in the explanatory power. Non-representative sample is the alternative explanation of the results.

\section{Further results}

\subsection{Customer momentum and lead-lag relationship between the returns}

Lo and MacKinlay (1990) find that weekly returns of large stocks often lead weekly returns of small stocks and partially explain this finding by the asynchronous trading. Small stocks are less liquid and can have very little trading volume for long periods, preventing their price from fully incorporating all available information. Since large and liquid stocks are likely to react to new information faster, we observe some cross-correlation between returns of large and small stocks. 
\paragraph{}
This paper uses the sample of monthly returns, so lead-lag effect should be weaker than in Lo and  MacKinlay (1990). To test this alternative explanation, I restrict the sample to the links with low ratio of customer size to supplier size. Then I calculate returns of long-short portfolios, formed on past returns of customers. 
\begin{center}
[Insert Table 32] 
\end{center}
After restricting the sample to the customers smaller than their suppliers, equally-weighted return of quintile long-short customer momentum portfolio is 1.37\%, consistent with the results in Table 5 of Cohen and Frazzini (2008). Value-weighted returns of this quintile long-short portfolio are 1.92\%. These results can suggest that customer momentum is present across suppliers and customers of comparable size. However, equally-weighted and value-weighted returns of decile long-short customer momentum portfolio are -0.16\% and 0.45\%. If we restrict sample to the links with the ratio of customer size to supplier size below 2, then returns of long-short portfolios, formed on past returns of customers, are between 0.45\% and 0.99\%. 
\paragraph{}
Since the previous results are very noisy due to small subsample size (less than 5\% of the full sample), I calculate returns of long-short customer momentum portfolios across 5 quintiles of relative customer size. 
\begin{center}
[Insert Table 33] 
\end{center}
Table 33 reports the returns of long-short quintile portfolios, formed on customer momentum across different quintiles of ratio of customer size to supplier size. For both equally-weighted and value-weighted portfolios, there is clear pattern of increasing returns of long-short portfolio with the increase in relative size of customer. This pattern is especially strong for equally-weighted returns, where returns of long-short customer momentum portfolios increase from 0.2\% for links with small relative customer size to 1.3\% for links with large relative customer size. Returns of long-short value-weighted portfolios increase from 0.19\% to 0.73\% as the relative customer size increases. These results appear to contradict the results in Table 32. Since the sample size for each quintile in Table 33 is 20\% of the full sample compared to 4\% of the full sample after restricting the sample to the links with the relative customer size below 1, the results from Table 33 should have somewhat higher credibility. 
\begin{center}
[Insert Table 34] 
\end{center}
The similar pattern emerges in the returns of long-short decile portfolios, formed on customer momentum across quintiles of relative customer size. Table 34 shows that equally-weighted returns of long-short customer momentum portfolios increase from 0.45\% for links with small relative customer size to 1.61\% for links with large relative customer size. Similarly, value-weighted returns of long-short customer momentum portfolios increase from 0.51\% for links with small relative customer size to 1.12\% for links with large relative customer size. Therefore, the results from Tables 33 and 34 are consistent with lead-lag relationship and  suggest that customer momentum is much stronger for small firms with large customers. Since the full sample was not large, these subsamples are very small. Thus the results are noisy and I will hesitate to interpret them as strong evidence for alternative explanation of customer momentum.
\paragraph{}
To further test alternative explanation of customer momentum by lead-lag relationship, I perform Fama-MacBeth regressions, controlling for the interaction term between past returns of customers and their relative size. If alternative explanation is true, then including the interaction term will render coefficient on past customer returns insignificant. Furthermore, under alternative hypothesis the coefficient on the interaction term should be significantly positive. 
\begin{center}
[Insert Table 35] 
\end{center}
Table 35 reports the results of Fama-MacBeth regression of returns on the past returns of customers' portfolio as well as the interaction term between past returns of customer portfolio and the relative size of average customer. Notice that without controlling for interaction term the coefficient on past customers' returns is robust and varies very little across specifications. Adding the interaction term flips the sign of the coefficient. Consistent with lead-lag hypothesis, coefficient on the interaction term is significantly positive. Together with the previous results, this evidence suggests that to the significant extent customer momentum is driven by lead-lag relationship between the returns of large and small stocks.

\subsection{Customer momentum, constructed using earnings surprises or post-earning abnormal returns of customers}

To provide additional robustness check, the paper explores other sorting variables, such as earnings surprises (SUE) and cumulative abnormal returns of customers around earnings announcements (CAR3). Tables 36 and 37 report the results:
\begin{center}
[Insert Table 36 and Table 37] 
\end{center}
Table 36 documents the findings on the returns of portfolios, formed on sorts on SUE of customers' portfolio. Excess returns of long-short portfolio from quintile sorts are 0.27\% and 0.36\% for equally-weighted and value-weighted portfolios respectively. For long-short decile portfolios the excess returns are slightly larger and are 0.54\% and 0.45\%. The similar analysis for sorts on CAR3 produces the returns of long-short portfolios below 0.06\%. All in all, these results suggest that customer momentum is present only when the sorting variable is past return of customers' portfolio. These results are not surprising, since SUE and CAR3, unlike returns, are available at effectively quarterly frequency. Since customer momentum is the strongest at 1-1 lag, SUE and CAR3 are too slow-moving variables to fully capture customer momentum. 

\subsection{Limited attention hypothesis of customer momentum}

Limited attention hypothesis explains customer momentum by low attention of market participants to the supply chain links between the firms. Cohen and Frazzini (2008) argue for this channel without testing it formally.
\paragraph{}
Short-term nature of customer momentum and investor attention implies that monthly frequency may be too low to capture this effect. Therefore, this section uses daily returns. In order to measure the attention to customer-supplier link, I use abnormal trading volume of suppliers around earnings announcements of the customers. To facilitate comparison between different customer-supplier links, I restrict the sample to the links with the unique customer. For each customer-supplier pair, I calculate normalized abnormal trading volume (NAV) of the supplier during the day of earnings announcement of its customer. I use supplier's NAV as a measure of the attention to supply chain link for the next 3 months or until the next earnings announcement.
\begin{center}
[Insert Table 38] 
\end{center}
Table 38 reports customer momentum across different window width of past returns of the customer and the contemporaneous returns of the supplier. The results suggest that strong customer momentum is present across all window widths.
\begin{center}
[Insert Table 39] 
\end{center}
Table 39 documents the difference of customer momentum between the supply chain links with low attention and high attention. First I form long-short decile portfolios on the customer momentum. The results, reported in Table 39, are the differences in the returns of these long-short portfolios within the low-attention supply chain links and high-attention supply chain links. Limited attention hypothesis predicts that returns of long-short portfolios, formed on customer momentum of low-attention suppliers, will be larger than the returns of long-short portfolios, formed on customer momentum of high-attention suppliers. Low attention to supply chain links is likely associated with slower information diffusion. Therefore, under limited attention hypothesis, customer momentum of low-attention suppliers will be the stronger at longer horizon.  
\paragraph{}
In the case of equally-weighted portfolios, customer momentum across different levels of attention is consistent with limited attention explanation. In particular, customer momentum is absent at short frequencies (1-10 days), but is economically and statistically significant at longer frequencies (20, 30 days). However, customer momentum across value-weighted portfolios does not exhibit similar pattern. When customer momentum is measured as the relationship between 30-days ahead returns of suppliers and returns of the customers within the recent 1, 5, 10 or 20 days, low-attention supply chain links exhibit customer momentum, lower or comparable to the customer momentum of high-attention supply chain links. Thus, the findings from value-weighted portfolios are inconsistent with limited attention hypothesis.

\newpage

\section{Conclusion} \label{sec:Model}
The paper reports that the firms with high returns of principal customers in the previous month are likely to have abnormally high returns in the current month. Equally-weighted (value-weighted) long-short decile portfolio of firms, sorted on customer momentum, has 1.22\% (1.06\%) monthly returns. These returns are not explained by Fama-French 5 factors and price momentum. Analysis of customer momentum in the post-discovery sample reveals that it has a smaller magnitude and is marginally statistically significant. The customer momentum factor has negative returns after 2005. These results suggest that after the discovery, customer momentum decreased due to exploitation by investors. 
\paragraph{}
The paper explores the relationship between customer momentum, price momentum, and earnings momentum and reports that earnings momentum is not explained by the other types of momentum. Therefore, evidence suggests that customer momentum captures a phenomenon, independent of the other types of momentum. Further results show that customer momentum can not explain price momentum or earnings momentum. 
\paragraph{}
Limited attention and the lead-lag relationship between large and small stocks due to non-synchronous trading (Lo and MacKinlay 1990) are the two likely explanations of customer momentum. The paper finds mixed evidence for the limited attention hypothesis. Customer momentum is 2-3 times larger among customer-supplier links with large relative customer size. This implies that customer momentum is driven by the fact that the average customer is 15 times larger than the average supplier in my sample.

\newpage
\section{References:}
\begin{enumerate}

    \item {Barber, Brad M., and Terrance Odean. "All that glitters: The effect of attention and news on the buying behavior of individual and institutional investors." The Review of Financial Studies 21, no. 2 (2007): 785-818.}
    \item {Barber, Brad M., Xing Huang, and Terrance Odean. "Which factors matter to investors? Evidence from mutual fund flows." The Review of Financial Studies 29, no. 10 (2016): 2600-2642.}
    \item {Ben-Rephael, Azi, Zhi Da, and Ryan D. Israelsen. "It depends on where you search: institutional investor attention and underreaction to news." The Review of Financial Studies 30, no. 9 (2017): 3009-3047.}
    \item {Berk, Jonathan B., and Jules H. Van Binsbergen. "Measuring skill in the mutual fund industry." Journal of Financial Economics 118, no. 1 (2015): 1-20.}
    \item {Chan, Louis KC, Narasimhan Jegadeesh, and Josef Lakonishok. "Momentum strategies." The Journal of Finance 51, no. 5 (1996): 1681-1713.}
    \item {Chordia, Tarun, and Lakshmanan Shivakumar. "Earnings and price momentum." Journal of financial economics 80, no. 3 (2006): 627-656.}
    \item {Cohen, Lauren, and Andrea Frazzini. "Economic links and predictable returns." The Journal of Finance 63, no. 4 (2008): 1977-2011.}
    \item {DellaVigna, Stefano, and Joshua M. Pollet. "Investor inattention and Friday earnings announcements." The Journal of Finance 64, no. 2 (2009): 709-749.}
    \item {Engelberg, Joseph, and Pengjie Gao. "In search of attention." The Journal of Finance 66, no. 5 (2011): 1461-1499.}
    \item {Fama, Eugene F., and James D. MacBeth. "Risk, return, and equilibrium: Empirical tests." Journal of political economy 81, no. 3 (1973): 607-636.}
    \item {Fama, E. F., \& French, K. R. (2015). A five-factor asset pricing model. Journal of Financial Economics, 116(1), 1–22. }
    \item {Harvey, Campbell R., Yan Liu, and Heqing Zhu. "… and the cross-section of expected returns." The Review of Financial Studies 29, no. 1 (2016): 5-68.}
    \item {Hirshleifer, David, Sonya Seongyeon Lim, and Siew Hong Teoh. "Driven to distraction: Extraneous events and underreaction to earnings news." The Journal of Finance 64, no. 5 (2009): 2289-2325.}
    \item {Hong, Harrison, and Jeremy C. Stein. "A unified theory of underreaction, momentum trading, and overreaction in asset markets." The Journal of finance 54, no. 6 (1999): 2143-2184.}
    \item {Hong, Harrison, Walter Torous, and Rossen Valkanov. "Do industries lead stock markets?." Journal of Financial Economics 83, no. 2 (2007): 367-396.}
    \item {Hou, Kewei, and Tobias J. Moskowitz. "Market frictions, price delay, and the cross-section of expected returns." The Review of Financial Studies 18, no. 3 (2005): 981-1020.}
    \item {Hou, Kewei, Wei Xiong, and Lin Peng. "A tale of two anomalies: The implications of investor attention for price and earnings momentum." (2009).}
    \item {Hou, Kewei, Chen Xue, and Lu Zhang. "Digesting anomalies: An investment approach." The Review of Financial Studies 28, no. 3 (2015): 650-705.}
    \item {Hou, Kewei, Chen Xue, and Lu Zhang. Replicating anomalies. No. w23394. National Bureau of Economic Research, 2018}
    \item {Huberman, Gur, and Tomer Regev. "Contagious speculation and a cure for cancer: A nonevent that made stock prices soar." The Journal of Finance 56, no. 1 (2001): 387-396.}
    \item {Jarrell, Gregg A., and Annette B. Poulsen. "Stock Trading before the Announcement of Tender Offers: Insider Trading or Market Anticipation?" Journal of Law, Economics, and Organization 5, no. 2 (1989).}
    \item {Jensen, Michael C., Fischer Black, and Myron S. Scholes. "The capital asset pricing model: Some empirical tests." (1972).}
    \item {Kahneman, Daniel. Attention and effort. Vol. 1063. Englewood Cliffs, NJ: Prentice-Hall, 1973.}
    \item {Loh, Roger K. "Investor inattention and the underreaction to stock recommendations." Financial Management 39, no. 3 (2010): 1223-1252.}
    \item {Lo, Andrew W., and A. Craig MacKinlay. "When are contrarian profits due to stock market overreaction?." The review of financial studies 3, no. 2 (1990): 175-205. }
    \item {McLean, R. David, and Jeffrey Pontiff. "Does academic research destroy stock return predictability?." The Journal of Finance 71, no. 1 (2016): 5-32.}
    \item {Menzly, Lior, and Oguzhan Ozbas. "Market segmentation and cross‐predictability of returns." The Journal of Finance 65, no. 4 (2010): 1555-1580.}
    \item {Merton, Robert C. "A simple model of capital market equilibrium with incomplete information." The journal of finance 42, no. 3 (1987): 483-510.}
    \item {Novy-Marx, Robert. "Is momentum really momentum?." Journal of Financial Economics 103, no. 3 (2012): 429-453.}
    \item {Novy-Marx, Robert. Fundamentally, momentum is fundamental momentum. No. w20984. National Bureau of Economic Research, 2015.}
    \item {Peng, Lin, and Wei Xiong. "Investor attention, overconfidence and category learning." Journal of Financial Economics 80, no. 3 (2006): 563-602.}
    \item {Schwert, G. William. "Anomalies and market efficiency." Handbook of the Economics of Finance 1 (2003): 939-974.}

\end{enumerate}

\newpage

\newgeometry{left=2.5cm, right=0.75cm, top=1.75cm, bottom=1.5cm}
\section*{Appendix A: Main Results}


\begin{table}[!htbp] \centering 
  \caption{\textbf{Sample coverage} }
  \label{} 
  \small
  \begin{threeparttable}
    {\medskip\footnotesize
    The table reports equally-weighted and value-weighted fraction of firms in the sample as a proportion of all firms in CRSP. Linked firms represents number of firms, for which there is customer link and the customer has nonmissing returns. Fourth column reports fraction of firms in the sample as a proportion of all firms in CRSP. Fifth column report the same proportion, weighted by market capitalization.}
    \medskip
\begin{tabular}{@{\extracolsep{5pt}} ccccc} 
\\[-1.8ex]\hline 
\hline \\[-1.8ex] 
Year & Linked firms & All firms & Fraction of firms & Fraction of ME \\ 
\hline \\[-1.8ex] 
$1978$ & $650$ & $41,492$ & $0.02$ & $0.01$ \\ 
$1979$ & $1,842$ & $41,086$ & $0.04$ & $0.02$ \\ 
$1980$ & $2,227$ & $41,127$ & $0.05$ & $0.02$ \\ 
$1981$ & $2,314$ & $42,128$ & $0.05$ & $0.03$ \\ 
$1982$ & $2,069$ & $40,247$ & $0.05$ & $0.03$ \\ 
$1983$ & $2,414$ & $45,018$ & $0.05$ & $0.03$ \\ 
$1984$ & $2,366$ & $44,428$ & $0.05$ & $0.03$ \\ 
$1985$ & $2,586$ & $44,742$ & $0.06$ & $0.03$ \\ 
$1986$ & $2,766$ & $45,345$ & $0.06$ & $0.04$ \\ 
$1987$ & $2,992$ & $46,486$ & $0.06$ & $0.04$ \\ 
$1988$ & $2,905$ & $44,777$ & $0.06$ & $0.04$ \\ 
$1989$ & $2,564$ & $43,784$ & $0.06$ & $0.03$ \\ 
$1990$ & $2,271$ & $38,616$ & $0.06$ & $0.03$ \\ 
$1991$ & $2,462$ & $39,513$ & $0.06$ & $0.03$ \\ 
$1992$ & $2,985$ & $43,745$ & $0.07$ & $0.03$ \\ 
$1993$ & $3,550$ & $49,552$ & $0.07$ & $0.03$ \\ 
$1994$ & $4,058$ & $53,529$ & $0.08$ & $0.04$ \\ 
$1995$ & $4,646$ & $57,190$ & $0.08$ & $0.04$ \\ 
$1996$ & $4,948$ & $61,169$ & $0.08$ & $0.04$ \\ 
$1997$ & $5,047$ & $63,063$ & $0.08$ & $0.03$ \\ 
$1998$ & $4,464$ & $60,455$ & $0.07$ & $0.03$ \\ 
$1999$ & $3,926$ & $56,776$ & $0.07$ & $0.03$ \\ 
$2000$ & $3,400$ & $53,294$ & $0.06$ & $0.05$ \\ 
$2001$ & $3,947$ & $46,025$ & $0.09$ & $0.08$ \\ 
$2002$ & $3,641$ & $41,961$ & $0.09$ & $0.05$ \\ 
$2003$ & $4,194$ & $42,272$ & $0.10$ & $0.07$ \\ 
$2004$ & $5,043$ & $45,086$ & $0.11$ & $0.10$ \\ 
$2005$ & $4,831$ & $44,242$ & $0.11$ & $0.12$ \\ 
$2006$ & $5,185$ & $44,853$ & $0.12$ & $0.14$ \\ 
$2007$ & $5,315$ & $44,033$ & $0.12$ & $0.16$ \\ 
$2008$ & $4,742$ & $37,656$ & $0.13$ & $0.14$ \\ 
$2009$ & $4,307$ & $32,263$ & $0.13$ & $0.14$ \\ 
$2010$ & $4,942$ & $34,418$ & $0.14$ & $0.15$ \\ 
$2011$ & $4,998$ & $34,060$ & $0.15$ & $0.15$ \\ 
$2012$ & $2,187$ & $33,231$ & $0.07$ & $0.07$ \\ 
$2013$ & $2,675$ & $34,276$ & $0.08$ & $0.08$ \\ 
$2014$ & $4,912$ & $35,297$ & $0.14$ & $0.15$ \\ 
$2015$ & $4,596$ & $35,148$ & $0.13$ & $0.14$ \\ 
$2016$ & $4,290$ & $34,039$ & $0.13$ & $0.15$ \\ 
$2017$ & $4,321$ & $33,800$ & $0.13$ & $0.14$ \\ 
$2018$ & $1,902$ & $16,616$ & $0.11$ & $0.10$ \\ 
\hline \\[-1.8ex] 
\end{tabular} 
\end{threeparttable}
\end{table}
\restoregeometry

\normalsize

\newgeometry{left=1.5cm, right=0.75cm, top=1.75cm, bottom=1.5cm}
\begin{table}[htbp] \centering 
  \caption{\textbf{Excess returns of decile portfolios, formed on customer momentum at different lags}} 
  \label{} 
  \small
    \begin{threeparttable}
    \begin{flushleft}
    {\medskip\small
    This table reports excess returns of decile portfolios, formed on customer momentum. In each month, firms are ranked by the past returns of the portfolio of their customers. The window over which the past returns are calculated, is defined by the lag. Lag is defined as (Beginning of return window as a number of months before the current month)-(End of return window as a number of months before the current month). L/S corresponds to long-short portfolio, defined as a difference in returns of tenth and first decile portfolios. Tenth portfolio (D10) is the portfolio of the firms with the largest customers' returns. Standard errors are calculated using Newey-West adjustment.}
    \medskip
    \end{flushleft}

\begin{tabularx}{\linewidth}{p{1cm}p{1.5cm}p{1cm}p{1cm}p{1cm}p{1cm}p{1cm}p{1cm}p{1cm}p{1cm}p{1cm}p{1cm}p{1cm}}
    \toprule
    \multicolumn{7}{l}{\textbf{Panel A: Equally-weighted decile portfolios}} \\
    \midrule  
\\[-1.8ex]\hline 
\hline \\[-1.8ex] 
lag & statistic & D1 & D2 & D3 & D4 & D5 & D6 & D7 & D8 & D9 & D10 & L/S \\ 
\hline \\[-1.8ex] 
1-1 & Mean & 0.59$^{*}$ & 0.93$^{***}$ & 1.03$^{***}$ & 1.06$^{***}$ & 1.08$^{***}$ & 0.93$^{***}$ & 1.40$^{***}$ & 1.36$^{***}$ & 1.50$^{***}$ & 1.81$^{***}$ & 1.22$^{***}$ \\ 
 & T-stat & [1.83] & [2.94] & [3.52] & [3.68] & [3.81] & [3.41] & [4.76] & [4.46] & [4.96] & [5.26] & [5.17] \\ 
2-1 & Mean & 0.51 & 0.84$^{***}$ & 1.08$^{***}$ & 1.10$^{***}$ & 0.89$^{***}$ & 1.03$^{***}$ & 0.90$^{***}$ & 1.38$^{***}$ & 1.47$^{***}$ & 1.74$^{***}$ & 1.23$^{***}$ \\ 
 & T-stat & [1.53] & [2.65] & [3.60] & [3.70] & [3.19] & [3.64] & [3.03] & [4.86] & [4.87] & [4.88] & [4.44] \\ 
3-1 & Mean & 0.53 & 0.69$^{**}$ & 0.97$^{***}$ & 1.21$^{***}$ & 0.92$^{***}$ & 0.94$^{***}$ & 1.03$^{***}$ & 1.31$^{***}$ & 1.25$^{***}$ & 1.60$^{***}$ & 1.07$^{***}$ \\ 
 & T-stat & [1.56] & [2.20] & [3.23] & [4.01] & [3.15] & [3.31] & [3.68] & [4.39] & [4.11] & [4.54] & [3.71] \\ 
7-1 & Mean & 0.75$^{**}$ & 0.86$^{***}$ & 0.72$^{**}$ & 0.70$^{**}$ & 0.72$^{**}$ & 1.01$^{***}$ & 1.11$^{***}$ & 0.96$^{***}$ & 1.27$^{***}$ & 1.63$^{***}$ & 0.88$^{***}$ \\ 
 & T-stat & [2.27] & [2.67] & [2.42] & [2.47] & [2.43] & [3.37] & [4.00] & [3.24] & [4.03] & [4.59] & [3.38] \\ 
12-1 & Mean & 0.58$^{*}$ & 0.33 & 0.83$^{***}$ & 0.90$^{***}$ & 0.47$^{*}$ & 0.90$^{***}$ & 0.94$^{***}$ & 0.81$^{***}$ & 1.30$^{***}$ & 1.68$^{***}$ & 1.10$^{***}$ \\ 
 & T-stat & [1.73] & [1.03] & [2.79] & [3.09] & [1.72] & [2.93] & [3.35] & [2.86] & [4.30] & [4.73] & [3.93] \\ 
2-2 & Mean & 1.07$^{***}$ & 1.01$^{***}$ & 1.36$^{***}$ & 1.28$^{***}$ & 1.20$^{***}$ & 1.12$^{***}$ & 1.24$^{***}$ & 1.21$^{***}$ & 1.12$^{***}$ & 1.43$^{***}$ & 0.36 \\ 
 & T-stat & [3.13] & [3.23] & [4.35] & [4.42] & [4.20] & [3.79] & [4.45] & [4.14] & [3.67] & [4.21] & [1.46] \\ 
7-2 & Mean & 0.84$^{**}$ & 0.74$^{**}$ & 0.80$^{***}$ & 1.21$^{***}$ & 1.16$^{***}$ & 0.85$^{***}$ & 0.94$^{***}$ & 1.28$^{***}$ & 1.08$^{***}$ & 1.43$^{***}$ & 0.59$^{**}$ \\ 
 & T-stat & [2.49] & [2.23] & [2.60] & [4.18] & [4.01] & [2.99] & [3.16] & [4.08] & [3.62] & [4.04] & [2.07] \\ 
12-2 & Mean & 0.55 & 0.60$^{*}$ & 1.10$^{***}$ & 0.83$^{***}$ & 0.61$^{**}$ & 0.97$^{***}$ & 0.84$^{***}$ & 0.85$^{***}$ & 1.33$^{***}$ & 1.71$^{***}$ & 1.16$^{***}$ \\ 
 & T-stat & [1.64] & [1.83] & [3.86] & [2.85] & [2.19] & [3.31] & [2.83] & [2.97] & [4.45] & [4.75] & [3.92] \\ 

\hline \\[-1.8ex] 
\end{tabularx} 

\begin{tabularx}{\linewidth}{p{1cm}p{1.5cm}p{1cm}p{1cm}p{1cm}p{1cm}p{1cm}p{1cm}p{1cm}p{1cm}p{1cm}p{1cm}p{1cm}}
    \toprule
    \multicolumn{7}{l}{\textbf{Panel B: Value-weighted decile portfolios}} \\
    \midrule  
\\[-1.8ex]\hline 
\hline \\[-1.8ex] 
lag & statistic & D1 & D2 & D3 & D4 & D5 & D6 & D7 & D8 & D9 & D10 & L/S \\ 
\hline \\[-1.8ex] 
1-1 & Mean & 0.20 & 0.38 & 0.54$^{*}$ & 0.95$^{***}$ & 0.52$^{*}$ & 0.78$^{***}$ & 0.69$^{**}$ & 0.57$^{*}$ & 0.78$^{**}$ & 1.26$^{***}$ & 1.06$^{***}$ \\ 
 & T-stat & [0.58] & [1.05] & [1.84] & [3.23] & [1.76] & [2.66] & [2.30] & [1.84] & [2.29] & [3.46] & [3.34] \\ 
2-1 & Mean & 0.31 & 0.50 & 0.69$^{**}$ & 0.48 & 0.85$^{***}$ & 0.64$^{**}$ & 0.37 & 0.73$^{**}$ & 1.20$^{***}$ & 1.36$^{***}$ & 1.05$^{***}$ \\ 
 & T-stat & [0.87] & [1.50] & [2.19] & [1.56] & [2.97] & [2.08] & [1.20] & [2.52] & [3.57] & [3.92] & [3.23] \\ 
3-1 & Mean & 0.29 & 0.40 & 0.73$^{**}$ & 0.68$^{**}$ & 0.45 & 0.59$^{**}$ & 0.56$^{*}$ & 1.02$^{***}$ & 0.71$^{**}$ & 1.47$^{***}$ & 1.18$^{***}$ \\ 
 & T-stat & [0.78] & [1.18] & [2.31] & [2.14] & [1.46] & [2.02] & [1.95] & [3.16] & [2.10] & [4.11] & [3.41] \\ 
7-1 & Mean & 0.57 & 0.64$^{*}$ & 0.64$^{**}$ & 0.52$^{*}$ & 0.56$^{*}$ & 0.65$^{**}$ & 0.98$^{***}$ & 0.39 & 1.03$^{***}$ & 1.24$^{***}$ & 0.67$^{*}$ \\ 
 & T-stat & [1.49] & [1.87] & [1.98] & [1.75] & [1.75] & [2.10] & [3.37] & [1.30] & [2.86] & [3.54] & [1.94] \\ 
12-1 & Mean & 0.67$^{*}$ & 0.09 & 0.82$^{***}$ & 0.85$^{***}$ & 0.46 & 0.75$^{**}$ & 0.79$^{***}$ & 0.40 & 1.07$^{***}$ & 1.37$^{***}$ & 0.71$^{*}$ \\ 
 & T-stat & [1.71] & [0.25] & [2.68] & [2.77] & [1.49] & [2.45] & [2.65] & [1.23] & [3.42] & [3.78] & [1.86] \\ 
2-2 & Mean & 0.65$^{*}$ & 0.47 & 1.01$^{***}$ & 0.48 & 0.74$^{**}$ & 0.53$^{*}$ & 0.64$^{**}$ & 0.89$^{***}$ & 0.69$^{**}$ & 1.03$^{***}$ & 0.39 \\ 
 & T-stat & [1.82] & [1.49] & [2.99] & [1.46] & [2.37] & [1.67] & [2.16] & [2.87] & [2.11] & [3.02] & [1.25] \\ 
7-2 & Mean & 0.71$^{*}$ & 0.32 & 0.59$^{*}$ & 0.96$^{***}$ & 0.93$^{***}$ & 0.63$^{**}$ & 0.58$^{**}$ & 0.97$^{***}$ & 0.78$^{**}$ & 1.19$^{***}$ & 0.48 \\ 
 & T-stat & [1.88] & [0.88] & [1.84] & [3.28] & [3.08] & [1.99] & [2.01] & [2.75] & [2.51] & [3.28] & [1.25] \\ 
12-2 & Mean & 0.70$^{*}$ & 0.18 & 0.87$^{***}$ & 0.69$^{**}$ & 0.73$^{**}$ & 0.81$^{**}$ & 0.62$^{*}$ & 0.48 & 1.05$^{***}$ & 1.26$^{***}$ & 0.56 \\ 
 & T-stat & [1.83] & [0.52] & [2.81] & [2.45] & [2.37] & [2.53] & [1.91] & [1.51] & [3.35] & [3.43] & [1.44] \\ 
\hline \\[-1.8ex] 
\end{tabularx} 
\end{threeparttable}
\end{table}
\restoregeometry

\newpage

\newgeometry{left=1.25cm, right=0.75cm, top=2cm, bottom=1.5cm}
\begin{table}[!htbp] 
  \centering 
  \caption{\textbf{Abnormal returns of decile portfolios, formed on customer momentum 1-1}}
  \label{} 
  \footnotesize
    \begin{threeparttable}
    \begin{flushleft}
    {\medskip\small
    This table reports abnormal returns of decile portfolios, formed on returns of customers' portfolio during the past month. The tables uses 3 different factor models to control for factor structure of returns, augmenting them with momentum factor UMD. Alpha is an intercept in Black-Jensen-Scholes regressions of returns on factors. D10 is the decile portfolio of the firms with the largest past customers' returns.
L/S corresponds to long-short portfolio, defined as a difference in returns of tenth and first decile portfolios.}
    \medskip
    \end{flushleft}

\begin{tabularx}{\linewidth}{l*{12}{Y}}
    \toprule
    \multicolumn{12}{l}{\textbf{Panel A: Equally-weighted decile portfolios}} \\
    \midrule  
\\[-1.8ex]\hline 
\hline \\[-1.8ex] 
Model & Statistic & D1 & D2 & D3 & D4 & D5 & D6 & D7 & D8 & D9 & D10 & L/S \\ 
\hline \\[-1.8ex] 
CAPM & alpha & -0.27 & 0.10 & 0.26 & 0.27 & 0.33$^{*}$ & 0.21 & 0.63$^{***}$ & 0.57$^{***}$ & 0.69$^{***}$ & 0.94$^{***}$ & 1.21$^{***}$ \\ 
 & T-stat & [-1.35] & [0.51] & [1.43] & [1.62] & [1.88] & [1.23] & [3.41] & [2.93] & [3.79] & [4.13] & [5.07] \\ 
CAPM + UMD & alpha & -0.20 & 0.09 & 0.24 & 0.26 & 0.38$^{**}$ & 0.21 & 0.59$^{***}$ & 0.58$^{***}$ & 0.67$^{***}$ & 0.90$^{***}$ & 1.09$^{***}$ \\ 
 & T-stat & [-0.98] & [0.43] & [1.32] & [1.55] & [2.16] & [1.23] & [3.19] & [2.92] & [3.67] & [3.90] & [4.57] \\ 
FF3 & alpha & -0.29$^{*}$ & 0.13 & 0.23$^{*}$ & 0.22$^{*}$ & 0.30$^{**}$ & 0.20 & 0.57$^{***}$ & 0.56$^{***}$ & 0.68$^{***}$ & 0.94$^{***}$ & 1.24$^{***}$ \\ 
 & T-stat & [-1.78] & [0.82] & [1.69] & [1.69] & [2.05] & [1.42] & [3.90] & [3.42] & [4.47] & [4.98] & [5.14] \\ 
FF3 + UMD & alpha & -0.14 & 0.19 & 0.27$^{*}$ & 0.25$^{*}$ & 0.41$^{***}$ & 0.25$^{*}$ & 0.57$^{***}$ & 0.62$^{***}$ & 0.72$^{***}$ & 0.96$^{***}$ & 1.11$^{***}$ \\ 
 & T-stat & [-0.88] & [1.21] & [1.93] & [1.89] & [2.81] & [1.78] & [3.82] & [3.75] & [4.63] & [4.98] & [4.54] \\ 
FF5 & alpha & -0.21 & 0.22 & 0.28$^{**}$ & 0.29$^{**}$ & 0.37$^{**}$ & 0.32$^{**}$ & 0.66$^{***}$ & 0.73$^{***}$ & 0.78$^{***}$ & 1.02$^{***}$ & 1.23$^{***}$ \\ 
 & T-stat & [-1.21] & [1.36] & [1.98] & [2.15] & [2.45] & [2.24] & [4.46] & [4.38] & [4.95] & [5.22] & [4.92] \\ 
FF5 + UMD & alpha & -0.11 & 0.26 & 0.30$^{**}$ & 0.30$^{**}$ & 0.44$^{***}$ & 0.34$^{**}$ & 0.65$^{***}$ & 0.76$^{***}$ & 0.80$^{***}$ & 1.03$^{***}$ & 1.14$^{***}$ \\ 
 & T-stat & [-0.65] & [1.57] & [2.11] & [2.23] & [2.95] & [2.40] & [4.35] & [4.51] & [5.00] & [5.20] & [4.55] \\ 
\hline \\[-1.8ex] 
\end{tabularx} 

\begin{tabularx}{\linewidth}{l*{12}{Y}}
    \toprule
    \multicolumn{12}{l}{\textbf{Panel B: Value-weighted decile portfolios}} \\
    \midrule
\\[-1.8ex]\hline 
\hline \\[-1.8ex] 
Model & Statistic & D1 & D2 & D3 & D4 & D5 & D6 & D7 & D8 & D9 & D10 & L/S \\ 
\hline \\[-1.8ex] 
CAPM & alpha & -0.67$^{***}$ & -0.43 & -0.19 & 0.21 & -0.21 & 0.05 & -0.05 & -0.20 & -0.08 & 0.36 & 1.03$^{***}$ \\ 
 & T-stat & [-2.82] & [-1.61] & [-0.94] & [1.09] & [-1.02] & [0.24] & [-0.25] & [-0.93] & [-0.35] & [1.46] & [3.23] \\ 
CAPM + UMD & alpha & -0.55$^{**}$ & -0.34 & -0.11 & 0.22 & -0.16 & 0.15 & 0.01 & -0.14 & -0.06 & 0.37 & 0.92$^{***}$ \\ 
 & T-stat & [-2.31] & [-1.24] & [-0.53] & [1.11] & [-0.77] & [0.78] & [0.07] & [-0.66] & [-0.25] & [1.47] & [2.85] \\ 
FF3 & alpha & -0.59$^{**}$ & -0.29 & -0.19 & 0.24 & -0.14 & 0.14 & -0.04 & -0.15 & 0.00 & 0.47$^{*}$ & 1.06$^{***}$ \\ 
 & T-stat & [-2.54] & [-1.13] & [-0.98] & [1.30] & [-0.68] & [0.72] & [-0.21] & [-0.71] & [0.00] & [1.92] & [3.28] \\ 
FF3 + UMD & alpha & -0.38$^{*}$ & -0.07 & -0.07 & 0.30 & -0.03 & 0.32$^{*}$ & 0.05 & -0.04 & 0.08 & 0.55$^{**}$ & 0.93$^{***}$ \\ 
 & T-stat & [-1.65] & [-0.26] & [-0.33] & [1.60] & [-0.14] & [1.65] & [0.25] & [-0.19] & [0.35] & [2.21] & [2.83] \\ 
FF5 & alpha & -0.43$^{*}$ & -0.06 & -0.09 & 0.47$^{**}$ & -0.05 & 0.36$^{*}$ & 0.08 & 0.02 & 0.25 & 0.56$^{**}$ & 0.99$^{***}$ \\ 
 & T-stat & [-1.77] & [-0.25] & [-0.42] & [2.46] & [-0.23] & [1.79] & [0.37] & [0.08] & [1.13] & [2.24] & [2.95] \\ 
FF5 + UMD & alpha & -0.30 & 0.07 & -0.01 & 0.48$^{**}$ & 0.02 & 0.46$^{**}$ & 0.13 & 0.08 & 0.28 & 0.61$^{**}$ & 0.91$^{***}$ \\ 
 & T-stat & [-1.23] & [0.27] & [-0.03] & [2.53] & [0.11] & [2.34] & [0.63] & [0.37] & [1.24] & [2.41] & [2.69] \\ 
\hline \\[-1.8ex] 
\end{tabularx} 
\end{threeparttable}
\end{table}
\restoregeometry

\newgeometry{left=1cm, right=0.5cm, top=2cm, bottom=1.5cm}
\begin{table}[!htbp] \centering 
  \caption{\textbf{Abnormal returns of decile portfolios, formed on customer momentum 6-1}}
  \label{} 
  \footnotesize
      \begin{flushleft}
    {\medskip\small
    This table reports abnormal returns of decile portfolios, formed on returns of customers' portfolio during the past 6 months. The tables uses 3 different factor models to control for factor structure of returns, augmenting them with momentum factor UMD. Alpha is an intercept in Black-Jensen-Scholes regressions of returns on factors.
    D10 is the decile portfolio of the firms with the largest past customers' returns.
    L/S corresponds to long-short portfolio, defined as a difference in returns of tenth and first decile portfolios.}
    \medskip
    \end{flushleft}

\begin{tabularx}{\linewidth}{l*{12}{Y}}
    \toprule
    \multicolumn{12}{l}{\textbf{Panel A: Equally-weighted decile portfolios}} \\
    \midrule
\\[-1.8ex]\hline 
\hline \\[-1.8ex] 
Model & Statistic & D1 & D2 & D3 & D4 & D5 & D6 & D7 & D8 & D9 & D10 & L/S \\ 
\hline \\[-1.8ex] 
CAPM & alpha & -0.09 & 0.06 & -0.00 & -0.02 & 0.01 & 0.28 & 0.41$^{**}$ & 0.22 & 0.53$^{**}$ & 0.83$^{***}$ & 0.92$^{***}$ \\ 
 & T-stat & [-0.43] & [0.28] & [-0.02] & [-0.11] & [0.06] & [1.44] & [2.40] & [1.20] & [2.47] & [3.29] & [3.51] \\ 
CAPM + UMD & alpha & 0.04 & 0.24 & 0.13 & 0.06 & 0.08 & 0.25 & 0.39$^{**}$ & 0.14 & 0.41$^{*}$ & 0.62$^{**}$ & 0.58$^{**}$ \\ 
 & T-stat & [0.20] & [1.24] & [0.68] & [0.33] & [0.41] & [1.26] & [2.25] & [0.74] & [1.92] & [2.49] & [2.36] \\ 
FF3 & alpha & -0.04 & 0.08 & -0.01 & -0.04 & -0.06 & 0.24 & 0.34$^{**}$ & 0.19 & 0.55$^{***}$ & 0.82$^{***}$ & 0.86$^{***}$ \\ 
 & T-stat & [-0.26] & [0.45] & [-0.06] & [-0.31] & [-0.32] & [1.43] & [2.35] & [1.13] & [3.05] & [4.36] & [3.33] \\ 
FF3 + UMD & alpha & 0.20 & 0.37$^{**}$ & 0.21 & 0.10 & 0.05 & 0.23 & 0.33$^{**}$ & 0.10 & 0.46$^{**}$ & 0.62$^{***}$ & 0.42$^{*}$ \\ 
 & T-stat & [1.28] & [2.16] & [1.30] & [0.70] & [0.31] & [1.35] & [2.24] & [0.62] & [2.53] & [3.34] & [1.76] \\ 
FF5 & alpha & 0.10 & 0.29 & 0.13 & 0.03 & 0.02 & 0.34$^{**}$ & 0.34$^{**}$ & 0.05 & 0.49$^{***}$ & 0.87$^{***}$ & 0.78$^{***}$ \\ 
 & T-stat & [0.59] & [1.55] & [0.75] & [0.20] & [0.13] & [1.98] & [2.23] & [0.29] & [2.64] & [4.47] & [2.89] \\ 
FF5 + UMD & alpha & 0.25 & 0.47$^{***}$ & 0.27 & 0.12 & 0.09 & 0.32$^{*}$ & 0.33$^{**}$ & 0.00 & 0.43$^{**}$ & 0.72$^{***}$ & 0.47$^{*}$ \\ 
 & T-stat & [1.61] & [2.69] & [1.64] & [0.86] & [0.51] & [1.86] & [2.16] & [0.03] & [2.33] & [3.81] & [1.90] \\ 
\hline \\[-1.8ex] 
\end{tabularx} 

\begin{tabularx}{\linewidth}{l*{12}{Y}}
    \toprule
    \multicolumn{12}{l}{\textbf{Panel B: Value-weighted decile portfolios}} \\
    \midrule
\\[-1.8ex]\hline 
\hline \\[-1.8ex] 
Model & Statistic & D1 & D2 & D3 & D4 & D5 & D6 & D7 & D8 & D9 & D10 & L/S \\ 
\hline \\[-1.8ex] 
CAPM & alpha & -0.32 & -0.13 & -0.09 & -0.20 & -0.15 & -0.05 & 0.28 & -0.29 & 0.25 & 0.45$^{*}$ & 0.78$^{**}$ \\ 
 & T-stat & [-1.24] & [-0.56] & [-0.39] & [-1.05] & [-0.64] & [-0.22] & [1.47] & [-1.38] & [0.93] & [1.82] & [2.23] \\ 
CAPM + UMD & alpha & -0.07 & 0.12 & 0.05 & -0.08 & -0.06 & -0.07 & 0.27 & -0.29 & 0.11 & 0.30 & 0.37 \\ 
 & T-stat & [-0.29] & [0.52] & [0.23] & [-0.43] & [-0.28] & [-0.32] & [1.42] & [-1.34] & [0.41] & [1.21] & [1.13] \\ 
FF3 & alpha & -0.24 & -0.05 & -0.04 & -0.19 & -0.18 & -0.04 & 0.26 & -0.30 & 0.38 & 0.47$^{**}$ & 0.71$^{**}$ \\ 
 & T-stat & [-0.96] & [-0.22] & [-0.17] & [-1.01] & [-0.76] & [-0.18] & [1.39] & [-1.40] & [1.49] & [2.06] & [2.05] \\ 
FF3 + UMD & alpha & 0.14 & 0.32 & 0.18 & -0.01 & -0.06 & -0.05 & 0.27 & -0.29 & 0.28 & 0.33 & 0.19 \\ 
 & T-stat & [0.60] & [1.39] & [0.79] & [-0.08] & [-0.27] & [-0.23] & [1.37] & [-1.34] & [1.09] & [1.43] & [0.57] \\ 
FF5 & alpha & 0.00 & 0.25 & 0.07 & -0.08 & -0.13 & 0.04 & 0.30 & -0.44$^{**}$ & 0.41 & 0.48$^{**}$ & 0.48 \\ 
 & T-stat & [-0.01] & [1.03] & [0.30] & [-0.39] & [-0.53] & [0.19] & [1.53] & [-2.02] & [1.59] & [2.00] & [1.34] \\ 
FF5 + UMD & alpha & 0.25 & 0.48$^{**}$ & 0.22 & 0.04 & -0.05 & 0.02 & 0.30 & -0.42$^{*}$ & 0.34 & 0.38 & 0.13 \\ 
 & T-stat & [1.01] & [2.08] & [0.92] & [0.19] & [-0.21] & [0.11] & [1.50] & [-1.91] & [1.31] & [1.57] & [0.38] \\
\hline \\[-1.8ex] 
\end{tabularx} 
\end{table}
\restoregeometry

\newgeometry{left=1.25cm, right=0.75cm, top=2cm, bottom=1.5cm}
\begin{table}[!htbp] \centering 
  \caption{\textbf{Abnormal returns of decile portfolios, formed on customer momentum 12-1}}
  \label{} 
  \footnotesize
      \begin{flushleft}
    {\medskip\small
    This table reports abnormal returns of decile portfolios, formed on returns of customers' portfolio during the past 12 months. The tables uses 3 different factor models to control for factor structure of returns, augmenting them with momentum factor UMD. Alpha is an intercept in Black-Jensen-Scholes regressions of returns on factors.
    D10 is the decile portfolio of the firms with the largest past customers' returns.
    L/S corresponds to long-short portfolio, defined as a difference in returns of tenth and first decile portfolios.}
    \medskip
    \end{flushleft}

\begin{tabularx}{\linewidth}{l*{12}{Y}}
    \toprule
    \multicolumn{12}{l}{\textbf{Panel A: Equally-weighted decile portfolios}} \\
    \midrule
\\[-1.8ex]\hline 
\hline \\[-1.8ex] 
Model & Statistic & D1 & D2 & D3 & D4 & D5 & D6 & D7 & D8 & D9 & D10 & L/S \\ 
\hline \\[-1.8ex] 
CAPM & alpha & -0.20 & -0.40$^{*}$ & 0.13 & 0.22 & -0.16 & 0.22 & 0.25 & 0.14 & 0.61$^{***}$ & 0.94$^{***}$ & 1.14$^{***}$ \\ 
 & T-stat & [-0.89] & [-1.88] & [0.66] & [1.16] & [-0.87] & [1.02] & [1.49] & [0.78] & [3.02] & [3.58] & [4.04] \\ 
CAPM + UMD & alpha & 0.03 & -0.17 & 0.22 & 0.32$^{*}$ & 0.01 & 0.27 & 0.23 & 0.04 & 0.44$^{**}$ & 0.67$^{***}$ & 0.64$^{***}$ \\ 
 & T-stat & [0.12] & [-0.83] & [1.17] & [1.68] & [0.07] & [1.24] & [1.37] & [0.24] & [2.25] & [2.64] & [2.64] \\ 
FF3 & alpha & -0.19 & -0.39$^{**}$ & 0.09 & 0.16 & -0.23 & 0.17 & 0.18 & 0.14 & 0.58$^{***}$ & 0.95$^{***}$ & 1.14$^{***}$ \\ 
 & T-stat & [-1.00] & [-2.01] & [0.55] & [0.96] & [-1.38] & [0.97] & [1.21] & [0.90] & [3.39] & [4.55] & [4.05] \\ 
FF3 + UMD & alpha & 0.15 & -0.05 & 0.24 & 0.30$^{*}$ & -0.02 & 0.27 & 0.16 & 0.05 & 0.41$^{**}$ & 0.67$^{***}$ & 0.51$^{**}$ \\ 
 & T-stat & [0.89] & [-0.26] & [1.44] & [1.86] & [-0.10] & [1.48] & [1.10] & [0.33] & [2.39] & [3.31] & [2.12] \\ 
FF5 & alpha & 0.11 & -0.27 & 0.07 & 0.26 & -0.22 & 0.20 & 0.13 & 0.05 & 0.54$^{***}$ & 0.97$^{***}$ & 0.86$^{***}$ \\ 
 & T-stat & [0.56] & [-1.37] & [0.40] & [1.57] & [-1.27] & [1.06] & [0.89] & [0.29] & [3.00] & [4.46] & [2.98] \\ 
FF5 + UMD & alpha & 0.33$^{*}$ & -0.03 & 0.18 & 0.36$^{**}$ & -0.06 & 0.26 & 0.13 & -0.01 & 0.41$^{**}$ & 0.76$^{***}$ & 0.43$^{*}$ \\ 
 & T-stat & [1.89] & [-0.19] & [1.06] & [2.17] & [-0.38] & [1.41] & [0.83] & [-0.05] & [2.34] & [3.67] & [1.72] \\ 
\hline \\[-1.8ex] 
\end{tabularx} 

\begin{tabularx}{\linewidth}{l*{12}{Y}}
    \toprule
    \multicolumn{12}{l}{\textbf{Panel B: Value-weighted decile portfolios}} \\
    \midrule
\\[-1.8ex]\hline 
\hline \\[-1.8ex] 
Model & Statistic & D1 & D2 & D3 & D4 & D5 & D6 & D7 & D8 & D9 & D10 & L/S \\ 
\hline \\[-1.8ex] 
CAPM & alpha & -0.17 & -0.64$^{**}$ & 0.13 & 0.17 & -0.19 & 0.09 & 0.12 & -0.27 & 0.41$^{*}$ & 0.63$^{**}$ & 0.79$^{**}$ \\ 
 & T-stat & [-0.59] & [-2.35] & [0.63] & [0.82] & [-0.85] & [0.42] & [0.60] & [-1.11] & [1.81] & [2.31] & [2.08] \\ 
CAPM + UMD & alpha & 0.15 & -0.31 & 0.21 & 0.25 & 0.01 & 0.11 & 0.11 & -0.38 & 0.28 & 0.40 & 0.25 \\ 
 & T-stat & [0.55] & [-1.20] & [0.98] & [1.16] & [0.03] & [0.48] & [0.54] & [-1.54] & [1.24] & [1.50] & [0.71] \\ 
FF3 & alpha & -0.09 & -0.53$^{**}$ & 0.12 & 0.18 & -0.17 & 0.12 & 0.11 & -0.20 & 0.41$^{*}$ & 0.74$^{***}$ & 0.83$^{**}$ \\ 
 & T-stat & [-0.34] & [-1.99] & [0.59] & [0.83] & [-0.76] & [0.56] & [0.55] & [-0.80] & [1.85] & [2.93] & [2.17] \\ 
FF3 + UMD & alpha & 0.37 & -0.05 & 0.23 & 0.29 & 0.09 & 0.17 & 0.11 & -0.29 & 0.27 & 0.52$^{**}$ & 0.16 \\ 
 & T-stat & [1.40] & [-0.19] & [1.07] & [1.36] & [0.40] & [0.79] & [0.54] & [-1.18] & [1.19] & [2.09] & [0.44] \\ 
FF5 & alpha & 0.22 & -0.32 & 0.06 & 0.28 & -0.06 & 0.12 & 0.16 & -0.32 & 0.35 & 0.89$^{***}$ & 0.67$^{*}$ \\ 
 & T-stat & [0.77] & [-1.17] & [0.28] & [1.29] & [-0.27] & [0.55] & [0.74] & [-1.30] & [1.53] & [3.42] & [1.68] \\ 
FF5 + UMD & alpha & 0.53$^{**}$ & 0.01 & 0.14 & 0.36 & 0.12 & 0.16 & 0.15 & -0.38 & 0.26 & 0.71$^{***}$ & 0.19 \\ 
 & T-stat & [1.98] & [0.06] & [0.66] & [1.62] & [0.51] & [0.72] & [0.71] & [-1.50] & [1.10] & [2.80] & [0.52] \\ 
\hline \\[-1.8ex] 
\end{tabularx} 
\end{table}
\restoregeometry

\begin{table}[!htbp] \centering 
  \caption{\textbf{Summary statistic of long-short portfolios, formed on Customer momentum}} 
  \label{} 
        \begin{flushleft}
    {\medskip\small
 The table reports summary statistic, including percentiles of returns of long-short portfolios and a market portfolio. L/S5 corresponds to long-short portfolio, defined as a difference in returns of extreme quintile portfolios. L/S10 corresponds to long-short portfolio, defined as a difference in returns of extreme decile portfolios. Mkt is the value-weighted return of all CRSP stocks(vwretd). Sharpe ratio is computed at annual frequency.}
    \medskip
    \end{flushleft}
\begin{tabular}{@{\extracolsep{5pt}} cccc} 
\\[-1.8ex]\hline 
\hline \\[-1.8ex] 
 & LS5 & LS10 & Mkt \\ 
\hline \\[-1.8ex] 
Mean & $0.64$ & $1.06$ & $0.64$ \\ 
SD & $5.62$ & $6.95$ & $4.40$ \\ 
min & $-18$ & $-20.72$ & $-23.14$ \\ 
p05 & $-7.41$ & $-8.98$ & $-7.14$ \\ 
p25 & $-2.43$ & $-3.05$ & $-1.95$ \\ 
p50 & $0.45$ & $0.60$ & $1.03$ \\ 
p75 & $3.28$ & $4.70$ & $3.51$ \\ 
p95 & $9.08$ & $11.80$ & $7.02$ \\ 
max & $35.08$ & $41.87$ & $12.43$ \\ 
Sharpe ratio & $0.45$ & $0.54$ & $0.50$ \\ 
\hline \\[-1.8ex] 
\end{tabular} 
\end{table}

\begin{figure*}
\textbf{Performance of \$1 (log scale)}
\vskip 12 pt
\begin{flushleft}
{The graph describes growth of the investment of \$1 in one of the three portfolios. L/S5 (L/S10) corresponds to long-short portfolio, defined as a difference in returns of extreme quintile (decile) portfolios. Mkt is the value-weighted return of all CRSP stocks(vwretd).}
\end{flushleft}
\centering
\includegraphics[width=1\textwidth]{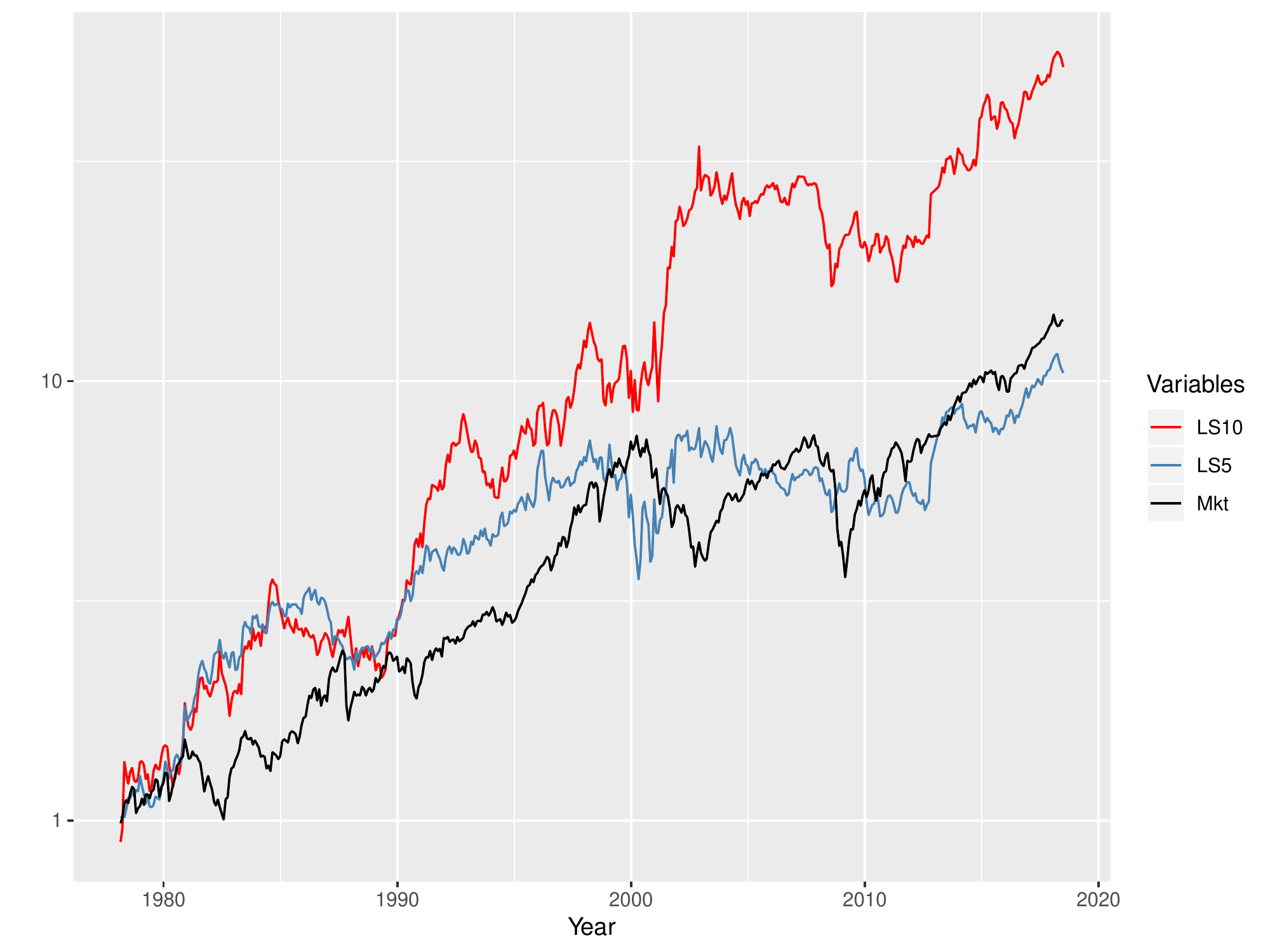}
{\textbf{Fig. 1.} }
\end{figure*}

\begin{landscape}
\begin{table}[ht]
  \caption{\textbf{Correlation between factors and macroeconomic variables, 1979-2018}} 
  \label{}
      \begin{flushleft}
    {\medskip\small
    The table reports the correlations between factors and macroeconomic variables. 
    CMOM is customer momentum factor, constructed from 2x3 sorts on size and returns of customers' porfolio in the last month. GDP\_gr, PCEC\_gr, PAYEMS\_gr and UNRATE\_gr are growth rates of GDP, consumption, labor income and unemployment rate. ***, ** and * represent statistically significant results at 0.1\%, 1\% and 10\% significance level.}
    \medskip
    \end{flushleft}
\centering
\begin{tabular}{rlllllllllll}
  \hline
 & CMOM & Mkt.RF & SMB & HML & RMW & CMA & RF & UMD & GDP\_gr & PCEC\_gr & PAYEMS\_gr \\ 
  \hline
CMOM &  &  &  &  &  &  &  &  &  &  &  \\ 
  Mkt.RF &  0.08  &  &  &  &  &  &  &  &  &  &  \\ 
  SMB &  0.25*** &  0.16*  &  &  &  &  &  &  &  &  &  \\ 
  HML & -0.17**  & -0.29*** & -0.26*** &  &  &  &  &  &  &  &  \\ 
  RMW & -0.23*** & -0.28*** & -0.45*** &  0.31*** &  &  &  &  &  &  &  \\ 
  CMA &  0.00  & -0.40*** & -0.10  &  0.69*** &  0.09  &  &  &  &  &  &  \\ 
  RF &  0.05  & -0.11  & -0.17**  &  0.07  & -0.02  &  0.13  &  &  &  &  &  \\ 
  UMD &  0.29*** & -0.04  &  0.24*** & -0.35*** &  0.00  & -0.12  & -0.01  &  &  &  &  \\ 
  GDP\_gr & -0.05  &  0.18**  & -0.05  & -0.04  &  0.02  & -0.05  &  0.42*** & -0.10  &  &  &  \\ 
  PCEC\_gr & -0.02  &  0.31*** &  0.04  & -0.18**  & -0.09  & -0.08  &  0.45*** &  0.15*  &  0.76*** &  &  \\ 
  PAYEMS\_gr & -0.01  &  0.17**  & -0.09  & -0.08  & -0.02  & -0.03  &  0.03  &  0.03  &  0.53*** &  0.38*** &  \\ 
  UNRATE\_gr &  0.10  & -0.19**  &  0.03  &  0.03  & -0.02  &  0.03  &  0.16*  &  0.03  & -0.32*** & -0.24*** & -0.44*** \\ 
   \hline
\end{tabular}
\end{table}
\end{landscape}

\clearpage

\newgeometry{left=1.25cm, right=0.75cm, top=2cm, bottom=1.5cm}
\begin{table}[!htbp] \centering 
  \caption{\textbf{Mean excess returns of decile portfolios, 1978-2018} }
  \label{} 
  \small
    \begin{flushleft}
    {\medskip\small
    This table reports excess returns of decile portfolios, formed on customer momentum. In each month, firms are ranked by the past returns of the portfolio of their customers. The window over which the past returns are calculated, is defined by the lag. Lag is defined as (Beginning of return window as a number of months before the current month)-(End of return window as a number of months before the current month). L/S corresponds to long-short portfolio, defined as a difference in returns of tenth and first decile portfolios. Tenth portfolio (D10) is the portfolio of the firms with the largest customers' returns. Standard errors are calculated using Newey-West adjustment.}
    \medskip
    \end{flushleft}

\begin{tabularx}{\linewidth}{l*{12}{Y}}
    \toprule
    \multicolumn{12}{l}{\textbf{Panel A1: Equally-weighted portfolios, 1978-2004}} \\
    \midrule
\\[-1.8ex]\hline 
\hline \\[-1.8ex] 
lag & statistic & D1 & D2 & D3 & D4 & D5 & D6 & D7 & D8 & D9 & D10 & L/S \\ 
\hline \\[-1.8ex] 
1-1 & Mean & 0.58 & 0.96$^{**}$ & 1.00$^{***}$ & 1.01$^{***}$ & 1.14$^{***}$ & 0.93$^{***}$ & 1.60$^{***}$ & 1.40$^{***}$ & 1.55$^{***}$ & 2.10$^{***}$ & 1.52$^{***}$ \\ 
 & T-stat & [1.39] & [2.30] & [2.62] & [2.72] & [3.08] & [2.67] & [4.20] & [3.57] & [3.88] & [4.59] & [4.74] \\ 
\hline \\[-1.8ex] 
\end{tabularx} 

\begin{tabularx}{\linewidth}{l*{12}{Y}}
    \toprule
    \multicolumn{12}{l}{\textbf{Panel A2: Equally-weighted portfolios, 2005-2018}} \\
    \midrule
\\[-1.8ex]\hline 
\hline \\[-1.8ex] 
lag & statistic & D1 & D2 & D3 & D4 & D5 & D6 & D7 & D8 & D9 & D10 & L/S \\ 
\hline \\[-1.8ex] 
1-1 & Mean & 0.64 & 0.92$^{**}$ & 1.11$^{**}$ & 1.19$^{***}$ & 0.97$^{**}$ & 0.97$^{**}$ & 1.05$^{**}$ & 1.32$^{***}$ & 1.46$^{***}$ & 1.26$^{***}$ & 0.62$^{**}$ \\ 
 & T-stat & [1.30] & [2.00] & [2.57] & [2.70] & [2.37] & [2.23] & [2.31] & [2.86] & [3.35] & [2.67] & [2.12] \\ 
\hline \\[-1.8ex] 
\end{tabularx} 

\begin{tabularx}{\linewidth}{l*{12}{Y}}
    \toprule
    \multicolumn{12}{l}{\textbf{Panel B1: Value-weighted portfolios, 1978-2004}} \\
    \midrule
\\[-1.8ex]\hline 
\hline \\[-1.8ex] 
lag & statistic & D1 & D2 & D3 & D4 & D5 & D6 & D7 & D8 & D9 & D10 & L/S \\ 
\hline \\[-1.8ex] 
1-1 & Mean & -0.08 & 0.43 & 0.38 & 0.82$^{**}$ & 0.45 & 0.63 & 0.71$^{*}$ & 0.60 & 0.87$^{*}$ & 1.22$^{**}$ & 1.30$^{***}$ \\ 
 & T-stat & [-0.17] & [0.88] & [0.99] & [2.06] & [1.14] & [1.57] & [1.79] & [1.48] & [1.86] & [2.47] & [3.03] \\ 
\hline \\[-1.8ex] 
\end{tabularx} 

\begin{tabularx}{\linewidth}{l*{12}{Y}}
    \toprule
    \multicolumn{12}{l}{\textbf{Panel B2: Value-weighted portfolios, 2005-2018}} \\
    \midrule
\\[-1.8ex]\hline 
\hline \\[-1.8ex] 
lag & statistic & D1 & D2 & D3 & D4 & D5 & D6 & D7 & D8 & D9 & D10 & L/S \\ 
\hline \\[-1.8ex] 
1-1 & Mean & 0.79 & 0.28 & 0.89$^{**}$ & 1.22$^{***}$ & 0.69$^{*}$ & 1.14$^{***}$ & 0.68 & 0.54 & 0.65 & 1.41$^{***}$ & 0.62 \\ 
 & T-stat & [1.60] & [0.65] & [2.04] & [3.12] & [1.70] & [3.08] & [1.64] & [1.23] & [1.56] & [2.97] & [1.54] \\ 
\hline \\[-1.8ex] 
\end{tabularx} 
\end{table}
\restoregeometry

\newpage

\newgeometry{left=1.25cm, right=0.75cm, top=1.5cm, bottom=1.5cm}
\begin{table}[!htbp] \centering 
  \caption{\textbf{Abnormal returns of equally-weighted decile portfolios, formed on customer momentum} }
  \label{} 
      \begin{flushleft}
    {\medskip\small
    This table reports abnormal returns of decile portfolios, formed on returns of customers' portfolio during the past 1 month. The tables uses 3 different factor models to control for factor structure of returns, augmenting them with momentum factor UMD. Alpha is an intercept in Black-Jensen-Scholes regressions of returns on factors.
    D10 is the decile portfolio of the firms with the largest past customers' returns.
    L/S corresponds to long-short portfolio, defined as a difference in returns of tenth and first decile portfolios.}
    \medskip
    \end{flushleft}
  \small

\begin{tabularx}{\linewidth}{l*{12}{Y}}
    \toprule
    \multicolumn{12}{l}{\textbf{Panel A: 1978-2004}} \\
    \midrule
\\[-1.8ex]\hline 
\hline \\[-1.8ex] 
Model & Statistic & D1 & D2 & D3 & D4 & D5 & D6 & D7 & D8 & D9 & D10 & L/S \\ 
\hline \\[-1.8ex] 
CAPM & alpha & -0.28 & 0.11 & 0.23 & 0.22 & 0.38 & 0.20 & 0.83$^{***}$ & 0.62$^{**}$ & 0.69$^{***}$ & 1.21$^{***}$ & 1.49$^{***}$ \\ 
 & T-stat & [-1.05] & [0.40] & [0.91] & [0.96] & [1.59] & [0.93] & [3.30] & [2.35] & [2.90] & [3.84] & [4.60] \\ 
CAPM + UMD & alpha & -0.21 & 0.12 & 0.17 & 0.19 & 0.49$^{**}$ & 0.18 & 0.73$^{***}$ & 0.59$^{**}$ & 0.67$^{***}$ & 1.09$^{***}$ & 1.30$^{***}$ \\ 
 & T-stat & [-0.76] & [0.43] & [0.67] & [0.85] & [2.03] & [0.80] & [2.87] & [2.19] & [2.74] & [3.41] & [3.98] \\ 
FF3 & alpha & -0.30 & 0.18 & 0.11 & 0.11 & 0.26 & 0.21 & 0.74$^{***}$ & 0.60$^{***}$ & 0.68$^{***}$ & 1.24$^{***}$ & 1.54$^{***}$ \\ 
 & T-stat & [-1.33] & [0.81] & [0.59] & [0.61] & [1.31] & [1.11] & [3.63] & [2.73] & [3.31] & [4.65] & [4.63] \\ 
FF3 + UMD & alpha & -0.11 & 0.31 & 0.15 & 0.17 & 0.47$^{**}$ & 0.26 & 0.72$^{***}$ & 0.67$^{***}$ & 0.74$^{***}$ & 1.23$^{***}$ & 1.34$^{***}$ \\ 
 & T-stat & [-0.47] & [1.43] & [0.79] & [0.92] & [2.38] & [1.38] & [3.44] & [2.97] & [3.53] & [4.49] & [3.96] \\ 
FF5 & alpha & -0.22 & 0.29 & 0.18 & 0.18 & 0.40$^{*}$ & 0.36$^{*}$ & 0.83$^{***}$ & 0.77$^{***}$ & 0.82$^{***}$ & 1.30$^{***}$ & 1.52$^{***}$ \\ 
 & T-stat & [-0.93] & [1.30] & [0.92] & [1.00] & [1.93] & [1.89] & [4.02] & [3.43] & [3.91] & [4.72] & [4.41] \\ 
FF5 + UMD & alpha & -0.09 & 0.38$^{*}$ & 0.19 & 0.22 & 0.53$^{***}$ & 0.38$^{*}$ & 0.80$^{***}$ & 0.79$^{***}$ & 0.85$^{***}$ & 1.28$^{***}$ & 1.37$^{***}$ \\ 
 & T-stat & [-0.38] & [1.67] & [1.01] & [1.16] & [2.63] & [1.95] & [3.82] & [3.48] & [3.97] & [4.59] & [3.96] \\ 
\hline \\[-1.8ex] 
\end{tabularx} 

\begin{tabularx}{\linewidth}{l*{12}{Y}}
    \toprule
    \multicolumn{12}{l}{\textbf{Panel B: 2005-2018}} \\
    \midrule
\\[-1.8ex]\hline 
\hline \\[-1.8ex] 
Model & Statistic & D1 & D2 & D3 & D4 & D5 & D6 & D7 & D8 & D9 & D10 & L/S \\ 
\hline \\[-1.8ex] 
CAPM & alpha & -0.24 & 0.11 & 0.33 & 0.39$^{*}$ & 0.23 & 0.22 & 0.22 & 0.51$^{**}$ & 0.70$^{***}$ & 0.42$^{*}$ & 0.67$^{**}$ \\ 
 & T-stat & [-0.93] & [0.42] & [1.47] & [1.74] & [1.08] & [0.87] & [0.97] & [1.98] & [2.82] & [1.66] & [2.25] \\ 
CAPM + UMD & alpha & -0.19 & 0.08 & 0.36 & 0.39$^{*}$ & 0.23 & 0.24 & 0.25 & 0.55$^{**}$ & 0.72$^{***}$ & 0.47$^{*}$ & 0.66$^{**}$ \\ 
 & T-stat & [-0.73] & [0.31] & [1.56] & [1.74] & [1.05] & [0.97] & [1.11] & [2.12] & [2.86] & [1.87] & [2.22] \\ 
FF3 & alpha & -0.23 & 0.08 & 0.30 & 0.40$^{**}$ & 0.20 & 0.20 & 0.23 & 0.51$^{**}$ & 0.71$^{***}$ & 0.43$^{**}$ & 0.67$^{**}$ \\ 
 & T-stat & [-1.10] & [0.41] & [1.57] & [2.23] & [1.13] & [1.00] & [1.32] & [2.19] & [3.30] & [2.06] & [2.23] \\ 
FF3 + UMD & alpha & -0.20 & 0.07 & 0.32$^{*}$ & 0.39$^{**}$ & 0.20 & 0.22 & 0.25 & 0.54$^{**}$ & 0.72$^{***}$ & 0.47$^{**}$ & 0.66$^{**}$ \\ 
 & T-stat & [-0.94] & [0.33] & [1.71] & [2.20] & [1.14] & [1.10] & [1.42] & [2.31] & [3.32] & [2.25] & [2.20] \\ 
FF5 & alpha & -0.12 & 0.11 & 0.39$^{**}$ & 0.45$^{**}$ & 0.16 & 0.25 & 0.26 & 0.67$^{***}$ & 0.73$^{***}$ & 0.47$^{**}$ & 0.59$^{*}$ \\ 
 & T-stat & [-0.55] & [0.52] & [1.99] & [2.42] & [0.89] & [1.18] & [1.39] & [2.81] & [3.23] & [2.16] & [1.91] \\ 
FF5 + UMD & alpha & -0.09 & 0.10 & 0.41$^{**}$ & 0.44$^{**}$ & 0.17 & 0.27 & 0.27 & 0.69$^{***}$ & 0.73$^{***}$ & 0.50$^{**}$ & 0.59$^{*}$ \\ 
 & T-stat & [-0.42] & [0.45] & [2.10] & [2.39] & [0.90] & [1.26] & [1.46] & [2.91] & [3.24] & [2.32] & [1.89] \\ 
\hline \\[-1.8ex] 
\end{tabularx}
\end{table}
\restoregeometry

\newpage

\newgeometry{left=1.25cm, right=0.75cm, top=1.5cm, bottom=1.5cm}
\begin{table}[!htbp] \centering 
  \caption{\textbf{Abnormal returns of value-weighted decile portfolios, formed on customer momentum} }
  \label{}
      \begin{flushleft}
    {\medskip\small
    This table reports abnormal returns of decile portfolios, formed on returns of customers' portfolio during the past 1 month. The tables uses 3 different factor models to control for factor structure of returns, augmenting them with momentum factor UMD. Alpha is an intercept in Black-Jensen-Scholes regressions of returns on factors.
    D10 is the decile portfolio of the firms with the largest past customers' returns.
    L/S corresponds to long-short portfolio, defined as a difference in returns of tenth and first decile portfolios.}
    \medskip
    \end{flushleft}
  \small

\begin{tabularx}{\linewidth}{l*{12}{Y}}
    \toprule
    \multicolumn{12}{l}{\textbf{Panel A: 1978-2004}} \\
    \midrule
\\[-1.8ex]\hline 
\hline \\[-1.8ex] 
Model & Statistic & D1 & D2 & D3 & D4 & D5 & D6 & D7 & D8 & D9 & D10 & L/S \\ 
\hline \\[-1.8ex] 
CAPM & alpha & -1.00$^{***}$ & -0.43 & -0.35 & 0.04 & -0.32 & -0.18 & -0.07 & -0.19 & -0.08 & 0.27 & 1.27$^{***}$ \\ 
 & T-stat & [-3.19] & [-1.14] & [-1.31] & [0.16] & [-1.17] & [-0.70] & [-0.24] & [-0.67] & [-0.25] & [0.78] & [2.94] \\ 
CAPM + UMD & alpha & -0.84$^{***}$ & -0.20 & -0.24 & 0.09 & -0.20 & 0.04 & 0.08 & -0.12 & -0.05 & 0.27 & 1.11$^{**}$ \\ 
 & T-stat & [-2.64] & [-0.52] & [-0.90] & [0.33] & [-0.72] & [0.16] & [0.30] & [-0.40] & [-0.17] & [0.78] & [2.52] \\ 
FF3 & alpha & -0.85$^{***}$ & -0.12 & -0.39 & 0.05 & -0.26 & -0.05 & -0.12 & -0.10 & 0.06 & 0.43 & 1.28$^{***}$ \\ 
 & T-stat & [-2.68] & [-0.32] & [-1.47] & [0.20] & [-0.94] & [-0.19] & [-0.43] & [-0.36] & [0.19] & [1.27] & [2.87] \\ 
FF3 + UMD & alpha & -0.59$^{*}$ & 0.27 & -0.22 & 0.19 & -0.08 & 0.24 & 0.06 & 0.04 & 0.15 & 0.51 & 1.11$^{**}$ \\ 
 & T-stat & [-1.86] & [0.76] & [-0.83] & [0.72] & [-0.27] & [0.92] & [0.22] & [0.14] & [0.46] & [1.47] & [2.42] \\ 
FF5 & alpha & -0.66$^{**}$ & 0.12 & -0.28 & 0.33 & -0.16 & 0.23 & 0.04 & 0.07 & 0.41 & 0.58$^{*}$ & 1.24$^{***}$ \\ 
 & T-stat & [-2.02] & [0.31] & [-1.01] & [1.26] & [-0.57] & [0.85] & [0.13] & [0.24] & [1.33] & [1.65] & [2.68] \\ 
FF5 + UMD & alpha & -0.49 & 0.37 & -0.17 & 0.38 & -0.04 & 0.41 & 0.15 & 0.15 & 0.42 & 0.62$^{*}$ & 1.11$^{**}$ \\ 
 & T-stat & [-1.52] & [1.01] & [-0.62] & [1.47] & [-0.14] & [1.52] & [0.51] & [0.51] & [1.34] & [1.73] & [2.38] \\
\hline \\[-1.8ex] 
\end{tabularx} 

\begin{tabularx}{\linewidth}{l*{12}{Y}}
    \toprule
    \multicolumn{12}{l}{\textbf{Panel B: 2005-2018}} \\
    \midrule
\\[-1.8ex]\hline 
\hline \\[-1.8ex] 
Model & Statistic & D1 & D2 & D3 & D4 & D5 & D6 & D7 & D8 & D9 & D10 & L/S \\ 
\hline \\[-1.8ex] 
CAPM & alpha & 0.00 & -0.42 & 0.15 & 0.55$^{**}$ & 0.04 & 0.58$^{**}$ & 0.01 & -0.18 & -0.02 & 0.62$^{**}$ & 0.61 \\ 
 & T-stat & [0.01] & [-1.50] & [0.57] & [2.41] & [0.15] & [2.23] & [0.05] & [-0.64] & [-0.07] & [2.11] & [1.49] \\ 
CAPM + UMD & alpha & 0.07 & -0.44 & 0.18 & 0.53$^{**}$ & 0.03 & 0.58$^{**}$ & 0.01 & -0.14 & 0.04 & 0.64$^{**}$ & 0.57 \\ 
 & T-stat & [0.23] & [-1.57] & [0.71] & [2.32] & [0.12] & [2.23] & [0.03] & [-0.51] & [0.13] & [2.20] & [1.38] \\ 
FF3 & alpha & -0.01 & -0.46$^{*}$ & 0.14 & 0.55$^{**}$ & -0.03 & 0.54$^{**}$ & -0.01 & -0.18 & -0.01 & 0.56$^{**}$ & 0.57 \\ 
 & T-stat & [-0.02] & [-1.69] & [0.53] & [2.38] & [-0.12] & [2.14] & [-0.04] & [-0.63] & [-0.03] & [1.97] & [1.38] \\ 
FF3 + UMD & alpha & 0.05 & -0.47$^{*}$ & 0.17 & 0.53$^{**}$ & -0.02 & 0.55$^{**}$ & -0.01 & -0.15 & 0.03 & 0.60$^{**}$ & 0.55 \\ 
 & T-stat & [0.15] & [-1.70] & [0.65] & [2.31] & [-0.07] & [2.19] & [-0.04] & [-0.54] & [0.11] & [2.11] & [1.33] \\ 
FF5 & alpha & 0.08 & -0.38 & 0.19 & 0.56$^{**}$ & 0.08 & 0.61$^{**}$ & 0.07 & -0.15 & 0.04 & 0.58$^{*}$ & 0.50 \\ 
 & T-stat & [0.24] & [-1.32] & [0.68] & [2.36] & [0.30] & [2.35] & [0.26] & [-0.52] & [0.13] & [1.95] & [1.16] \\ 
FF5 + UMD & alpha & 0.12 & -0.38 & 0.21 & 0.55$^{**}$ & 0.09 & 0.62$^{**}$ & 0.07 & -0.13 & 0.07 & 0.61$^{**}$ & 0.48 \\ 
 & T-stat & [0.37] & [-1.33] & [0.78] & [2.31] & [0.33] & [2.39] & [0.26] & [-0.45] & [0.24] & [2.07] & [1.12] \\
\hline \\[-1.8ex] 
\end{tabularx} 
\end{table}
\restoregeometry

\newpage

\newgeometry{left=2cm, right=1cm, top=1cm, bottom=1.5cm}
\begin{table}[!htbp] \centering 
  \caption{\textbf{Summary statistics of characteristics, 1979-2018} }
  \label{} 
          \begin{flushleft}
    {\medskip\small
    This table reports summary statistic of characteristics, including their percentiles.
    OP is operating profitability. CAR3 is abnormal return during 3-day window, centered around earnings announcement. Standardized unexpected earnings  (SUE) are defined as the most recent year-to-year change in earnings per share (EPS), divided by the standard deviation of the changes in earnings during the recent eight announcements. I define SUE as missing if there are less than 6 quarterly announcement during 2-year window. Cmom1-1 is the return of the customers' portfolio during the last month. Mom12-2 and Mom1-1 are past returns of the firm at a given lag.}
    \medskip
    \end{flushleft}
\begin{tabular}{@{\extracolsep{5pt}} ccccccccc} 
\\[-1.8ex]\hline 
\hline \\[-1.8ex] 
 & Cmom1-1 & Mom12-2 & Mom1-1 & log(ME) & log(B/M) & Op & CAR3 & SUE \\ 
\hline \\[-1.8ex] 
Mean & $0.91$ & $18.09$ & $1.37$ & $6.34$ & $-0.72$ & $0.25$ & $0$ & $0.07$ \\ 
SD & $7.84$ & $49.08$ & $11.22$ & $1.96$ & $0.73$ & $0.25$ & $0.07$ & $1.09$ \\ 
min & $-36.71$ & $-86.71$ & $-46.17$ & $1.15$ & $-4.07$ & $-1.31$ & $-0.32$ & $-3.04$ \\ 
p05 & $-11.76$ & $-40.77$ & $-16.27$ & $3.27$ & $-2$ & $-0.05$ & $-0.11$ & $-1.86$ \\ 
p25 & $-3.62$ & $-11.31$ & $-5.25$ & $4.94$ & $-1.19$ & $0.14$ & $-0.04$ & $-0.54$ \\ 
p50 & $0.98$ & $10.35$ & $0.87$ & $6.29$ & $-0.68$ & $0.24$ & $0$ & $0.07$ \\ 
p75 & $5.46$ & $37.08$ & $7.41$ & $7.59$ & $-0.22$ & $0.35$ & $0.04$ & $0.67$ \\ 
p95 & $13.50$ & $101.50$ & $20.53$ & $9.72$ & $0.40$ & $0.58$ & $0.12$ & $1.93$ \\ 
max & $38.21$ & $831.90$ & $76.52$ & $13.01$ & $2.13$ & $4.29$ & $0.41$ & $4.81$ \\
\hline \\[-1.8ex] 
\end{tabular} 
\end{table}

\begin{table}[!htbp] \centering 
  \caption{\textbf{Summary statistics of factors, 1979-2018} }
  \label{} 
    \begin{flushleft}
    {\medskip\small
    This table reports summary statistics of factors, including their percentiles. 
    CMOM, SUEF and CAR3F are factors, constructed from 2x3 sorts on ME and respective variable (returns of customers' portfolio during the recent month, SUE and CAR3). The factors use NYSE breakpoints and are constructed similarly to UMD. Standard errors are calculated using Newey-West correction.}
    \medskip
    \end{flushleft}
\begin{tabular}{@{\extracolsep{5pt}} cccccccccc} 
\\[-1.8ex]\hline 
\hline \\[-1.8ex] 
 & Mkt.RF & SMB & HML & RMW & CMA & UMD & CMOM & SUEF & CAR3F \\ 
\hline \\[-1.8ex] 
Mean & $0.68^{***}$ & $0.17$ & $0.26$ & $0.34^{***}$ & $0.27^{**}$ & $0.61^{***}$ & $0.40^{***}$ & $1.27^{***}$ & $0.18^{**}$ \\ 
SE & $0.21$ & $0.13$ & $0.16$ & $0.13$ & $0.10$ & $0.22$ & $0.14$ & $0.07$ & $0.07$ \\ 
T-stat & $[3.22]$ & $[1.31]$ & $[1.57]$ & $[2.67]$ & $[2.55]$ & $[2.78]$ & $[2.88]$ & $[17.31]$ & $[2.46]$ \\ 
SD & $4.38$ & $2.89$ & $2.92$ & $2.41$ & $1.99$ & $4.45$ & $3.50$ & $1.94$ & $1.57$ \\ 
min & $-23.24$ & $-15.33$ & $-11.10$ & $-19.06$ & $-6.88$ & $-34.39$ & $-10.55$ & $-8.34$ & $-10.98$ \\ 
p05 & $-7.09$ & $-4.07$ & $-3.95$ & $-2.82$ & $-2.66$ & $-6.46$ & $-4.85$ & $-1.48$ & $-2.17$ \\ 
p25 & $-1.94$ & $-1.51$ & $-1.38$ & $-0.81$ & $-0.99$ & $-1.01$ & $-1.54$ & $0.26$ & $-0.59$ \\ 
p50 & $1.09$ & $0.06$ & $0.04$ & $0.34$ & $0.16$ & $0.70$ & $0.06$ & $1.22$ & $0.31$ \\ 
p75 & $3.50$ & $1.93$ & $1.72$ & $1.38$ & $1.49$ & $2.88$ & $2.32$ & $2.18$ & $1.01$ \\ 
p95 & $7.07$ & $4.62$ & $5.30$ & $3.71$ & $3.29$ & $6.63$ & $5.93$ & $4.47$ & $2.46$ \\ 
max & $12.47$ & $18.75$ & $12.90$ & $13.51$ & $9.56$ & $18.33$ & $19.36$ & $10.12$ & $5.25$ \\
\hline \\[-1.8ex] 
\end{tabular} 
\end{table}
\restoregeometry

\newpage

\textbf{Performance of \$1 (log scale)}
\begin{figure*}[ht!]
\centering
\includegraphics[width=1\textwidth]{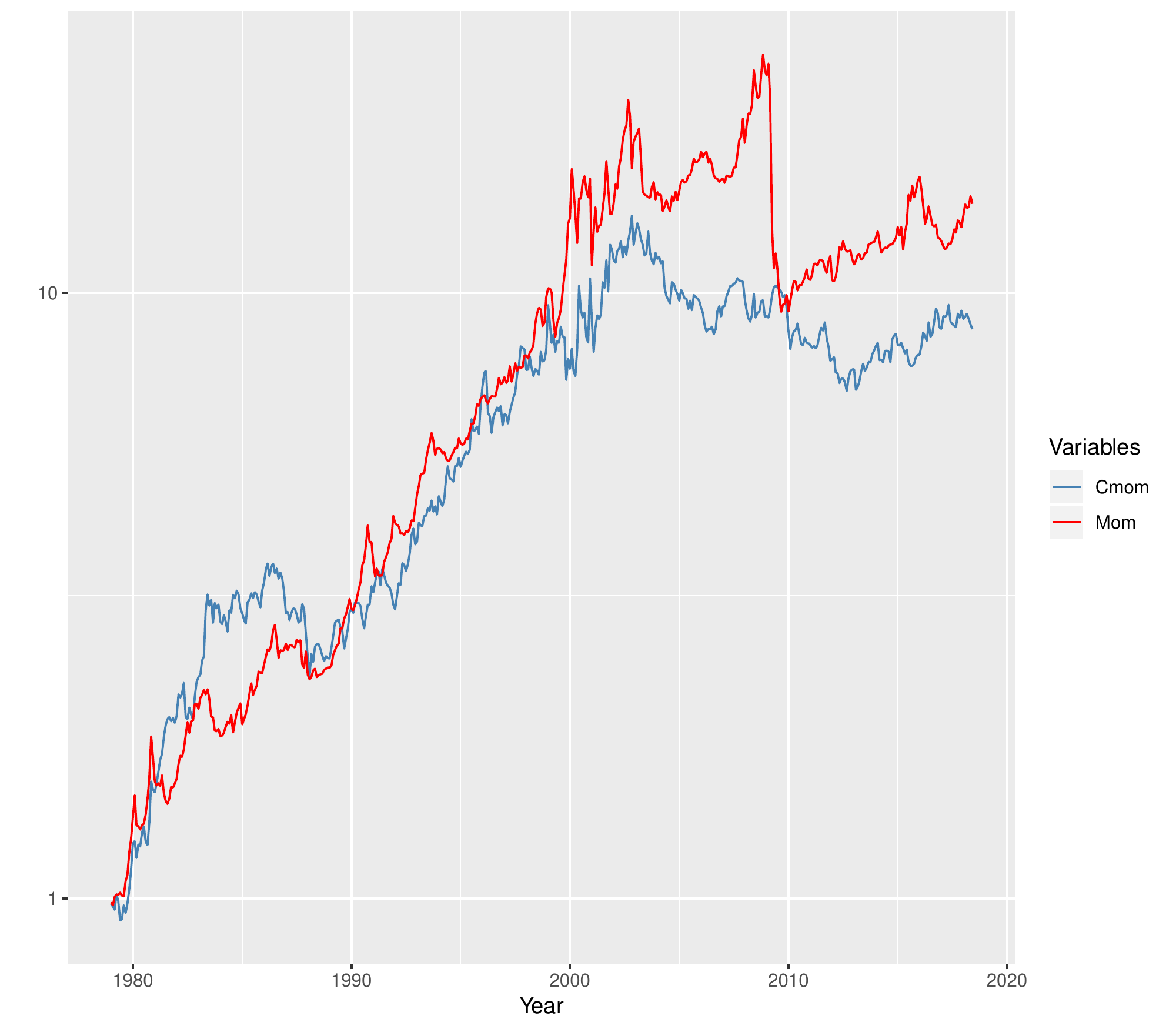}
\end{figure*}
\\
{\textbf{Fig. 2.} \\
This graph compares performance of momentum factors. "Cmom" denotes customer momentum factor CMOM. "Mom" stands for UMD. To facilitate comparison, customer momentum factor is scaled to have the same sample standard deviation as price momentum.}

\newgeometry{left=1.25cm, right=0.75cm, top=1cm, bottom=1.5cm}
\begin{landscape}
\renewcommand{\arraystretch}{0.9}
\begin{table}[!htbp] \centering 
  \caption{\textbf{Fama-MacBeth regressions, 1979-2018} }
  \label{} 
     \begin{flushleft}
    {\small
    This table reports the results of Fama-MacBeth regression of returns on the characteristics. Cmom1-1 is returns of customers' portfolio during the last month. Mom1-1 and Mom12-2 are past returns of stock at lags 1-1 and 12-2. OP is operating profitability. CAR3 is abnormal return during 3-day window, centered around earnings announcement. Standardized unexpected earnings  (SUE) are defined as the most recent year-to-year change in earnings per share (EPS), divided by the standard deviation of the changes in earnings during the recent eight announcements.}
    \end{flushleft}
\begin{tabular}{@{\extracolsep{5pt}}lccccccccc} 
\\[-1.8ex]\hline 
\hline \\[-1.8ex] 
 & \multicolumn{9}{c}{\textit{Dependent variable: Return}} \\ 
\cline{2-10} 
\\[-1.8ex] & (1) & (2) & (3) & (4) & (5) & (6) & (7) & (8) & (9)\\ 
\hline \\[-1.8ex] 
 Cmom1-1 & 0.04$^{***}$ & 0.04$^{***}$ &  &  &  & 0.04$^{***}$ &  & 0.04$^{***}$ & 0.04$^{***}$ \\ 
  & [3.20] & [3.07] &  &  &  & [3.21] &  & [3.34] & [3.52] \\ 
  & & & & & & & & & \\ 
 log(ME) &  & -0.12$^{**}$ &  & -0.13$^{***}$ & -0.10$^{**}$ & -0.11$^{**}$ & -0.12$^{**}$ & -0.14$^{***}$ & -0.13$^{**}$ \\ 
  &  & [-2.18] &  & [-2.58] & [-2.03] & [-2.12] & [-2.38] & [-2.73] & [-2.50] \\ 
  & & & & & & & & & \\ 
 log(B/M) &  & 0.13 &  & 0.16 & 0.16 & 0.13 & 0.21 & 0.17 & 0.20 \\ 
  &  & [0.92] &  & [1.08] & [1.07] & [0.89] & [1.37] & [1.16] & [1.27] \\ 
  & & & & & & & & & \\ 
 OP &  &  &  & -0.04 & 0.32 & 0.29 & 0.44 & 0.01 & 0.46 \\ 
  &  &  &  & [-0.09] & [0.71] & [0.62] & [0.96] & [0.03] & [0.98] \\ 
  & & & & & & & & & \\ 
 Mom12-2 &  &  & 0.01$^{**}$ & 0.004 &  &  & -0.003 & 0.003 & -0.004 \\ 
  &  &  & [2.06] & [1.34] &  &  & [-1.10] & [0.96] & [-1.54] \\ 
  & & & & & & & & & \\ 
 Mom1-1 &  &  & -0.02$^{***}$ & -0.03$^{***}$ &  &  & -0.05$^{***}$ & -0.04$^{***}$ & -0.05$^{***}$ \\ 
  &  &  & [-2.70] & [-4.24] &  &  & [-5.87] & [-4.55] & [-6.12] \\ 
  & & & & & & & & & \\ 
 CAR3 &  &  &  &  & 2.55$^{**}$ & 2.83$^{**}$ & 3.16$^{***}$ &  & 3.63$^{***}$ \\ 
  &  &  &  &  & [2.21] & [2.47] & [2.59] &  & [2.93] \\ 
  & & & & & & & & & \\ 
 SUE &  &  &  &  & 0.87$^{***}$ & 0.87$^{***}$ & 0.92$^{***}$ &  & 0.92$^{***}$ \\ 
  &  &  &  &  & [12.92] & [12.76] & [13.13] &  & [12.98] \\ 
  & & & & & & & & & \\ 
 Constant & 1.35$^{***}$ & 2.04$^{***}$ & 1.27$^{***}$ & 2.00$^{***}$ & 1.83$^{***}$ & 1.81$^{***}$ & 1.94$^{***}$ & 2.07$^{***}$ & 1.99$^{***}$ \\ 
  & [5.12] & [4.91] & [5.11] & [5.02] & [4.46] & [4.40] & [4.89] & [5.15] & [4.99] \\ 
  & & & & & & & & & \\ 
\hline \\[-1.8ex] 
Observations & 25,421 & 25,421 & 25,421 & 25,421 & 25,421 & 25,421 & 25,421 & 25,421 & 25,421 \\ 
Adjusted R$^{2}$ & 0.007 & 0.028 & 0.031 & 0.060 & 0.051 & 0.057 & 0.077 & 0.066 & 0.083 \\ 
\hline 
\hline \\[-1.8ex] 

\end{tabular} 
\end{table} 
\end{landscape}
\restoregeometry

\newgeometry{left=2cm, right=1cm, top=1cm, bottom=1.5cm}
\begin{table}[!htbp] \centering 
  \caption{\textbf{Factor spanning test of Customer momentum, 1979-2018}} 
  \label{} 
              \begin{flushleft}
    {\medskip\small
    This table reports the results of time-series regression of customer momentum factor CMOM on the other factors.
    CMOM, SUEF and CAR3F are factors, constructed from 2x3 sorts on ME and respective variable (past returns of customers' portfolio during the past month, SUE and CAR3). The factors use NYSE breakpoints and are constructed similarly to UMD. }
    \medskip
    \end{flushleft}
\begin{tabular}{@{\extracolsep{0pt}}lccccccc} 
\\[-1.8ex]\hline 
\hline \\[-1.8ex] 
 & \multicolumn{7}{c}{\textit{Dependent variable:}} \\ 
\cline{2-8} 
\\[-1.8ex] & \multicolumn{7}{c}{CMOM} \\ 
\\[-1.8ex] & (1) & (2) & (3) & (4) & (5) & (6) & (7)\\ 
\hline \\[-1.8ex] 
 Mkt.RF & 0.03 & 0.01 & 0.05 & 0.03 & 0.05 & 0.03 & 0.03 \\ 
  & [0.70] & [0.21] & [1.11] & [0.50] & [1.06] & [0.51] & [0.60] \\ 
  & & & & & & & \\ 
 SMB & 0.04 & -0.01 & 0.03 & -0.02 & 0.05 & 0.01 & -0.01 \\ 
  & [0.45] & [-0.08] & [0.43] & [-0.31] & [0.62] & [0.12] & [-0.21] \\ 
  & & & & & & & \\ 
 HML & -0.09 & -0.06 & -0.05 & 0.03 & -0.08 & -0.01 & 0.03 \\ 
  & [-1.04] & [-0.60] & [-0.56] & [0.28] & [-0.90] & [-0.10] & [0.28] \\ 
  & & & & & & & \\ 
 CMA &  & -0.04 &  & -0.11 &  & -0.11 & -0.12 \\ 
  &  & [-0.24] &  & [-0.71] &  & [-0.72] & [-0.80] \\ 
  & & & & & & & \\ 
 RMW &  & -0.15$^{*}$ &  & -0.19$^{**}$ &  & -0.17$^{**}$ & -0.19$^{**}$ \\ 
  &  & [-1.77] &  & [-2.49] &  & [-2.04] & [-2.39] \\ 
  & & & & & & & \\ 
 UMD &  &  & 0.12$^{*}$ & 0.13$^{**}$ &  &  & 0.10 \\ 
  &  &  & [1.84] & [2.07] &  &  & [1.40] \\ 
  & & & & & & & \\ 
 SUEF &  &  &  &  & -0.04 & 0.01 & -0.06 \\ 
  &  &  &  &  & [-0.36] & [0.12] & [-0.53] \\ 
  & & & & & & & \\ 
 CAR3f &  &  &  &  & 0.35$^{**}$ & 0.36$^{**}$ & 0.26$^{*}$ \\ 
  &  &  &  &  & [2.18] & [2.25] & [1.71] \\ 
  & & & & & & & \\ 
 Constant & 0.39$^{***}$ & 0.47$^{***}$ & 0.29$^{**}$ & 0.38$^{**}$ & 0.35$^{*}$ & 0.39$^{*}$ & 0.43$^{**}$ \\ 
  & [2.84] & [3.13] & [2.10] & [2.55] & [1.69] & [1.82] & [2.00] \\ 
  & & & & & & & \\ 
\hline \\[-1.8ex] 
Observations & 470 & 470 & 470 & 470 & 470 & 470 & 470 \\ 
R$^{2}$ & 0.011 & 0.019 & 0.031 & 0.045 & 0.033 & 0.044 & 0.055 \\ 
Adjusted R$^{2}$ & 0.005 & 0.008 & 0.023 & 0.032 & 0.023 & 0.029 & 0.038 \\ 
Residual Std. Error & 3.495 & 3.489 & 3.463 & 3.447 & 3.464 & 3.452 & 3.436 \\ 
F Statistic & 1.771 & 1.796 & 3.765 & 3.611 & 3.171 & 3.022 & 3.340 \\ 
\hline 
\hline \\[-1.8ex] 
\end{tabular} 
\end{table}
\restoregeometry

\begin{table}[!htbp] \centering 
  \caption{\textbf{Factor spanning test of Momentum, 1979-2018}} 
  \label{} 
              \begin{flushleft}
    {\medskip\small
    This table reports the results of time-series regression of momentum factor UMD on the other factors.
    CMOM, SUEF and CAR3F are factors, constructed from 2x3 sorts on ME and respective variable (past returns of customers' portfolio during the past month, SUE and CAR3). The factors use NYSE breakpoints and are constructed similarly to UMD. }
    \medskip
    \end{flushleft}
\begin{tabular}{@{\extracolsep{5pt}}lccccc} 
\\[-1.8ex]\hline 
\hline \\[-1.8ex] 
 & \multicolumn{5}{c}{\textit{Dependent variable:}} \\ 
\cline{2-6} 
\\[-1.8ex] & \multicolumn{5}{c}{UMD} \\ 
\\[-1.8ex] & (1) & (2) & (3) & (4) & (5)\\ 
\hline \\[-1.8ex] 
 Mkt.RF & -0.20$^{**}$ & -0.11 & -0.07 & -0.21$^{**}$ & -0.07 \\ 
  & [-2.24] & [-1.30] & [-1.20] & [-2.23] & [-1.29] \\ 
  & & & & & \\ 
 SMB & 0.04 & 0.12 & 0.18 & 0.04 & 0.17 \\ 
  & [0.34] & [0.95] & [1.50] & [0.32] & [1.51] \\ 
  & & & & & \\ 
 HML & -0.41$^{**}$ & -0.66$^{***}$ & -0.32$^{**}$ & -0.39$^{**}$ & -0.31$^{**}$ \\ 
  & [-2.22] & [-3.49] & [-2.00] & [-2.20] & [-2.04] \\ 
  & & & & & \\ 
 CMA &  & 0.50 &  &  &  \\ 
  &  & [1.50] &  &  &  \\ 
  & & & & & \\ 
 RMW &  & 0.33 &  &  &  \\ 
  &  & [1.41] &  &  &  \\ 
  & & & & & \\ 
 SUEF &  &  & 0.73$^{***}$ &  & 0.73$^{***}$ \\ 
  &  &  & [4.84] &  & [4.92] \\ 
  & & & & & \\ 
 CAR3f &  &  & 1.01$^{***}$ &  & 0.98$^{***}$ \\ 
  &  &  & [3.81] &  & [3.59] \\ 
  & & & & & \\ 
 CMOM &  &  &  & 0.17$^{**}$ & 0.10 \\ 
  &  &  &  & [2.05] & [1.34] \\ 
  & & & & & \\ 
 Constant & 0.85$^{***}$ & 0.60$^{**}$ & -0.39 & 0.79$^{***}$ & -0.42$^{*}$ \\ 
  & [4.61] & [2.31] & [-1.55] & [4.37] & [-1.74] \\ 
  & & & & & \\ 
\hline \\[-1.8ex] 
Observations & 474 & 474 & 474 & 470 & 470 \\ 
R$^{2}$ & 0.083 & 0.126 & 0.362 & 0.101 & 0.368 \\ 
Adjusted R$^{2}$ & 0.078 & 0.117 & 0.355 & 0.093 & 0.360 \\ 
Residual Std. Error & 4.270 & 4.179 & 3.571 & 4.248 & 3.570 \\ 
F Statistic & 14.270 & 13.500 & 53.050 & 13.060 & 44.890 \\ 
\hline 
\hline \\[-1.8ex] 

\end{tabular} 
\end{table}

\newgeometry{left=2cm, right=1cm, top=1cm, bottom=1.5cm}
\begin{table}[!htbp] \centering 
  \caption{\textbf{Factor spanning test of Earnings momentum factor (SUEF), 1979-2018}} 
  \label{} 
              \begin{flushleft}
    {\medskip\small
    This table reports the results of time-series regression of earnings momentum factor SUEF on the other factors.
    CMOM, SUEF and CAR3F are factors, constructed from 2x3 sorts on ME and respective variable (past returns of customers' portfolio during the past month, SUE and CAR3). The factors use NYSE breakpoints and are constructed similarly to UMD. }
    \medskip
    \end{flushleft}
\begin{tabular}{@{\extracolsep{4pt}}lp{1.2cm}p{1.2cm}p{1.2cm}p{1.2cm}p{1.2cm}p{1.2cm}p{1.2cm}p{1.2cm}} 
\\[-1.8ex]\hline 
\hline \\[-1.8ex] 
 & \multicolumn{8}{c}{\textit{Dependent variable:}} \\ 
\cline{2-9} 
\\[-1.8ex] & \multicolumn{8}{c}{SUEF} \\ 
\\[-1.8ex] & (1) & (2) & (3) & (4) & (5) & (6) & (7) & (8)\\ 
\hline \\[-1.8ex] 
 CAR3f & 0.41$^{***}$ &  &  &  &  &  &  &  \\ 
  & [4.45] &  &  &  &  &  &  &  \\ 
  & & & & & & & & \\ 
 Mkt.RF &  & -0.08$^{**}$ & -0.04 & -0.05 & -0.02 & -0.08$^{**}$ & -0.04 & -0.02 \\ 
  &  & [-2.36] & [-1.16] & [-1.47] & [-0.56] & [-2.33] & [-1.15] & [-0.56] \\ 
  & & & & & & & & \\ 
 SMB &  & -0.12$^{***}$ & -0.08$^{**}$ & -0.13$^{***}$ & -0.10$^{***}$ & -0.12$^{***}$ & -0.08$^{**}$ & -0.10$^{***}$ \\ 
  &  & [-3.42] & [-2.28] & [-3.36] & [-3.07] & [-3.39] & [-2.26] & [-3.04] \\ 
  & & & & & & & & \\ 
 HML &  & -0.06 & -0.19$^{**}$ & 0.02 & -0.08 & -0.06 & -0.19$^{**}$ & -0.08 \\ 
  &  & [-0.76] & [-2.39] & [0.26] & [-1.01] & [-0.72] & [-2.34] & [-0.99] \\ 
  & & & & & & & & \\ 
 CMA &  &  & 0.27$^{***}$ &  & 0.19$^{**}$ &  & 0.27$^{***}$ & 0.18$^{**}$ \\ 
  &  &  & [2.79] &  & [2.45] &  & [2.82] & [2.38] \\ 
  & & & & & & & & \\ 
 RMW &  &  & 0.19$^{***}$ &  & 0.13$^{**}$ &  & 0.19$^{***}$ & 0.13$^{**}$ \\ 
  &  &  & [3.22] &  & [2.18] &  & [3.22] & [2.16] \\ 
  & & & & & & & & \\ 
 UMD &  &  &  & 0.19$^{***}$ & 0.17$^{***}$ &  &  & 0.17$^{***}$ \\ 
  &  &  &  & [5.87] & [5.63] &  &  & [5.73] \\ 
  & & & & & & & & \\ 
 CMOM &  &  &  &  &  & 0.01 & 0.02 & -0.01 \\ 
  &  &  &  &  &  & [0.48] & [0.88] & [-0.33] \\ 
  & & & & & & & & \\ 
 Constant & 1.19$^{***}$ & 1.36$^{***}$ & 1.22$^{***}$ & 1.20$^{***}$ & 1.12$^{***}$ & 1.36$^{***}$ & 1.21$^{***}$ & 1.12$^{***}$ \\ 
  & [17.83] & [17.05] & [14.04] & [13.63] & [13.26] & [16.65] & [13.42] & [12.95] \\ 
  & & & & & & & & \\ 
\hline \\[-1.8ex] 
Observations & 474 & 474 & 474 & 474 & 474 & 470 & 470 & 470 \\ 
R$^{2}$ & 0.110 & 0.078 & 0.147 & 0.246 & 0.278 & 0.078 & 0.148 & 0.277 \\ 
Adjusted R$^{2}$ & 0.108 & 0.072 & 0.138 & 0.240 & 0.268 & 0.070 & 0.137 & 0.266 \\ 
Residual Std. Error & 1.833 & 1.870 & 1.803 & 1.693 & 1.660 & 1.877 & 1.809 & 1.668 \\ 
F Statistic & 58.360 & 13.320 & 16.110 & 38.300 & 29.930 & 9.855 & 13.360 & 25.270 \\ 
\hline 
\hline \\[-1.8ex] 

\end{tabular} 
\end{table}
\restoregeometry

\newgeometry{left=2cm, right=1cm, top=1cm, bottom=1.5cm}
\begin{table}[!htbp] \centering 
  \caption{\textbf{Factor spanning test of Earnings momentum (CAR3), 1979-2018}} 
  \label{} 
              \begin{flushleft}
    {\medskip\small
    This table reports the results of time-series regression of earnings momentum factor CAR3 on the other factors.
    CMOM, SUEF and CAR3F are factors, constructed from 2x3 sorts on ME and respective variable (past returns of customers' portfolio during the past month, SUE and CAR3). The factors use NYSE breakpoints and are constructed similarly to UMD. }
    \medskip
    \end{flushleft}
\begin{tabular}{@{\extracolsep{0pt}}lp{1.3cm}p{1.3cm}p{1.3cm}p{1.3cm}p{1.3cm}p{1.3cm}p{1.3cm}p{1.3cm}} 
\\[-1.8ex]\hline 
\hline \\[-1.8ex] 
 & \multicolumn{8}{c}{\textit{Dependent variable:}} \\ 
\cline{2-9} 
\\[-1.8ex] & \multicolumn{8}{c}{CAR3F} \\ 
\\[-1.8ex] & (1) & (2) & (3) & (4) & (5) & (6) & (7) & (8)\\ 
\hline \\[-1.8ex] 
 SUEF & 0.27$^{***}$ &  &  &  &  &  &  &  \\ 
  & [4.16] &  &  &  &  &  &  &  \\ 
  & & & & & & & & \\ 
 Mkt.RF &  & -0.07$^{**}$ & -0.04 & -0.04 & -0.02 & -0.07$^{**}$ & -0.04 & -0.03 \\ 
  &  & [-2.17] & [-1.49] & [-1.35] & [-0.96] & [-2.20] & [-1.51] & [-1.02] \\ 
  & & & & & & & & \\ 
 SMB &  & -0.04 & -0.03 & -0.05$^{*}$ & -0.05$^{**}$ & -0.04 & -0.03 & -0.05$^{**}$ \\ 
  &  & [-1.21] & [-1.08] & [-1.87] & [-2.05] & [-1.38] & [-1.15] & [-2.11] \\ 
  & & & & & & & & \\ 
 HML &  & -0.04 & -0.13$^{**}$ & 0.03 & -0.03 & -0.04 & -0.13$^{**}$ & -0.03 \\ 
  &  & [-0.92] & [-2.48] & [0.74] & [-0.72] & [-0.77] & [-2.31] & [-0.72] \\ 
  & & & & & & & & \\ 
 CMA &  &  & 0.19$^{**}$ &  & 0.11$^{**}$ &  & 0.19$^{***}$ & 0.11$^{***}$ \\ 
  &  &  & [2.44] &  & [2.42] &  & [2.58] & [2.58] \\ 
  & & & & & & & & \\ 
 RMW &  &  & 0.06 &  & 0.01 &  & 0.07 & 0.02 \\ 
  &  &  & [1.34] &  & [0.29] &  & [1.49] & [0.43] \\ 
  & & & & & & & & \\ 
 UMD &  &  &  & 0.17$^{***}$ & 0.16$^{***}$ &  &  & 0.16$^{***}$ \\ 
  &  &  &  & [5.58] & [5.64] &  &  & [5.31] \\ 
  & & & & & & & & \\ 
 CMOM &  &  &  &  &  & 0.07$^{***}$ & 0.07$^{***}$ & 0.04$^{*}$ \\ 
  &  &  &  &  &  & [2.76] & [2.97] & [1.85] \\ 
  & & & & & & & & \\ 
 Constant & -0.16 & 0.25$^{***}$ & 0.18$^{**}$ & 0.11 & 0.08 & 0.22$^{***}$ & 0.15$^{*}$ & 0.07 \\ 
  & [-1.26] & [3.50] & [2.24] & [1.47] & [1.13] & [3.23] & [1.88] & [0.94] \\ 
  & & & & & & & & \\ 
\hline \\[-1.8ex] 
Observations & 474 & 474 & 474 & 474 & 474 & 470 & 470 & 470 \\ 
R$^{2}$ & 0.110 & 0.046 & 0.077 & 0.246 & 0.255 & 0.067 & 0.101 & 0.265 \\ 
Adjusted R$^{2}$ & 0.108 & 0.040 & 0.067 & 0.240 & 0.246 & 0.059 & 0.090 & 0.254 \\ 
Residual Std. Error & 1.486 & 1.542 & 1.519 & 1.372 & 1.366 & 1.530 & 1.505 & 1.362 \\ 
F Statistic & 58.360 & 7.492 & 7.831 & 38.320 & 26.670 & 8.381 & 8.712 & 23.840 \\ 
\hline 
\hline \\[-1.8ex] 

\end{tabular} 
\end{table}
\restoregeometry

\newpage

\section*{Appendix B: Further Results}

\begin{table}[!htbp] \centering 
  \caption{\textbf{Mean excess returns of equally-weighted quintile portfolios, formed on customer momentum 1-1} }
  \label{} 
      \begin{flushleft}
    {\medskip\small
   This table reports excess returns of quintile portfolios, formed on customer momentum. In each month, firms are ranked by the past returns of the portfolio of their customers. The window over which the past returns are calculated, is defined by the lag. Lag is defined as (Beginning of return window as a number of months before the current month)-(End of return window as a number of months before the current month). L/S corresponds to long-short portfolio, defined as a difference in returns of fifth and first quintile portfolios. Fifth portfolio (Q5) is the portfolio of the firms with the largest customers' returns. Standard errors are calculated using Newey-West adjustment.}
    \medskip
    \end{flushleft}
\begin{tabular}{@{\extracolsep{5pt}} cccccccc} 
\\[-1.8ex]\hline 
\hline \\[-1.8ex] 
Lag & Statistic & Q1 & Q2 & Q3 & Q4 & Q5 & L/S \\ 
\hline \\[-1.8ex] 
1-1 & Mean & 0.72$^{**}$ & 1.03$^{***}$ & 1.00$^{***}$ & 1.38$^{***}$ & 1.64$^{***}$ & 0.93$^{***}$ \\ 
 & T-stat & [2.36] & [3.72] & [3.80] & [4.93] & [5.42] & [5.51] \\ 
2-1 & Mean & 0.66$^{**}$ & 1.11$^{***}$ & 0.98$^{***}$ & 1.14$^{***}$ & 1.60$^{***}$ & 0.94$^{***}$ \\ 
 & T-stat & [2.12] & [3.94] & [3.69] & [4.19] & [5.20] & [4.95] \\ 
3-1 & Mean & 0.64$^{**}$ & 1.09$^{***}$ & 0.94$^{***}$ & 1.15$^{***}$ & 1.46$^{***}$ & 0.82$^{***}$ \\ 
 & T-stat & [2.06] & [3.86] & [3.47] & [4.24] & [4.68] & [4.05] \\ 
4-1 & Mean & 0.64$^{**}$ & 0.86$^{***}$ & 1.14$^{***}$ & 0.97$^{***}$ & 1.46$^{***}$ & 0.82$^{***}$ \\ 
 & T-stat & [2.00] & [3.21] & [4.24] & [3.58] & [4.69] & [3.86] \\ 
7-1 & Mean & 0.77$^{**}$ & 0.71$^{***}$ & 0.89$^{***}$ & 1.04$^{***}$ & 1.46$^{***}$ & 0.69$^{***}$ \\ 
 & T-stat & [2.49] & [2.60] & [3.20] & [3.88] & [4.68] & [3.30] \\ 
12-1 & Mean & 0.50 & 0.85$^{***}$ & 0.66$^{**}$ & 0.87$^{***}$ & 1.58$^{***}$ & 1.08$^{***}$ \\ 
 & T-stat & [1.61] & [3.12] & [2.47] & [3.29] & [4.95] & [4.65] \\ 
2-2 & Mean & 1.03$^{***}$ & 1.32$^{***}$ & 1.16$^{***}$ & 1.23$^{***}$ & 1.31$^{***}$ & 0.29 \\ 
 & T-stat & [3.31] & [4.66] & [4.26] & [4.58] & [4.36] & [1.49] \\ 
4-2 & Mean & 0.88$^{***}$ & 1.00$^{***}$ & 1.29$^{***}$ & 1.02$^{***}$ & 1.27$^{***}$ & 0.39$^{*}$ \\ 
 & T-stat & [2.75] & [3.55] & [4.72] & [3.90] & [4.09] & [1.93] \\ 
7-2 & Mean & 0.78$^{**}$ & 1.03$^{***}$ & 1.02$^{***}$ & 1.09$^{***}$ & 1.31$^{***}$ & 0.53$^{**}$ \\ 
 & T-stat & [2.45] & [3.67] & [3.80] & [3.93] & [4.29] & [2.47] \\ 
12-2 & Mean & 0.61$^{*}$ & 0.95$^{***}$ & 0.82$^{***}$ & 0.81$^{***}$ & 1.60$^{***}$ & 0.99$^{***}$ \\ 
 & T-stat & [1.95] & [3.52] & [3.03] & [3.02] & [5.06] & [4.46] \\ 
3-3 & Mean & 1.08$^{***}$ & 1.05$^{***}$ & 1.28$^{***}$ & 1.41$^{***}$ & 1.33$^{***}$ & 0.25 \\ 
 & T-stat & [3.57] & [3.80] & [4.55] & [4.97] & [4.35] & [1.34] \\ 
7-3 & Mean & 0.97$^{***}$ & 0.91$^{***}$ & 1.14$^{***}$ & 1.22$^{***}$ & 1.27$^{***}$ & 0.30 \\ 
 & T-stat & [3.06] & [3.31] & [4.09] & [4.48] & [4.00] & [1.34] \\ 
12-3 & Mean & 0.70$^{**}$ & 1.00$^{***}$ & 0.73$^{***}$ & 0.97$^{***}$ & 1.49$^{***}$ & 0.79$^{***}$ \\ 
 & T-stat & [2.31] & [3.50] & [2.67] & [3.59] & [4.85] & [3.66] \\ 
12-7 & Mean & 0.92$^{***}$ & 0.99$^{***}$ & 0.97$^{***}$ & 1.08$^{***}$ & 1.52$^{***}$ & 0.60$^{***}$ \\ 
 & T-stat & [3.01] & [3.60] & [3.47] & [3.74] & [4.95] & [3.03] \\  
\hline \\[-1.8ex] 
\end{tabular} 
\end{table}

\begin{table}[!htbp] \centering 
  \caption{\textbf{Mean excess returns of value-weighted quintile portfolios, formed on customer momentum 1-1}} 
  \label{} 
        \begin{flushleft}
    {\medskip\small
This table reports excess returns of quintile portfolios, formed on customer momentum. In each month, firms are ranked by the past returns of the portfolio of their customers. The window over which the past returns are calculated, is defined by the lag. Lag is defined as (Beginning of return window as a number of months before the current month)-(End of return window as a number of months before the current month). L/S corresponds to long-short portfolio, defined as a difference in returns of fifth and first quintile portfolios. Fifth portfolio (Q5) is the portfolio of the firms with the largest customers' returns. Standard errors are calculated using Newey-West adjustment.}
    \medskip
    \end{flushleft}
\begin{tabular}{@{\extracolsep{5pt}} cccccccc} 
\\[-1.8ex]\hline 
\hline \\[-1.8ex] 
Lag & Statistic & Q1 & Q2 & Q3 & Q4 & Q5 & L/S \\ 
\hline \\[-1.8ex] 
1-1 & Mean & 0.29 & 0.76$^{***}$ & 0.58$^{**}$ & 0.49$^{*}$ & 0.93$^{***}$ & 0.64$^{**}$ \\ 
 & T-stat & [0.92] & [2.90] & [2.15] & [1.80] & [2.94] & [2.50] \\ 
2-1 & Mean & 0.33 & 0.40 & 0.82$^{***}$ & 0.50$^{*}$ & 1.18$^{***}$ & 0.84$^{***}$ \\ 
 & T-stat & [1.05] & [1.40] & [2.94] & [1.84] & [3.79] & [3.37] \\ 
3-1 & Mean & 0.42 & 0.66$^{**}$ & 0.57$^{**}$ & 0.65$^{**}$ & 1.00$^{***}$ & 0.59$^{**}$ \\ 
 & T-stat & [1.29] & [2.34] & [2.13] & [2.43] & [3.27] & [2.27] \\ 
4-1 & Mean & 0.49 & 0.46$^{*}$ & 0.77$^{***}$ & 0.63$^{**}$ & 0.97$^{***}$ & 0.48$^{*}$ \\ 
 & T-stat & [1.46] & [1.70] & [2.93] & [2.18] & [3.22] & [1.67] \\ 
7-1 & Mean & 0.57$^{*}$ & 0.61$^{**}$ & 0.55$^{*}$ & 0.67$^{**}$ & 1.05$^{***}$ & 0.48$^{*}$ \\ 
 & T-stat & [1.75] & [2.12] & [1.93] & [2.55] & [3.31] & [1.66] \\ 
12-1 & Mean & 0.37 & 0.76$^{***}$ & 0.49$^{*}$ & 0.64$^{**}$ & 1.23$^{***}$ & 0.86$^{***}$ \\ 
 & T-stat & [1.07] & [2.84] & [1.75] & [2.21] & [3.95] & [2.62] \\ 
2-2 & Mean & 0.55$^{*}$ & 0.60$^{**}$ & 0.67$^{**}$ & 0.73$^{***}$ & 0.85$^{***}$ & 0.29 \\ 
 & T-stat & [1.81] & [2.06] & [2.28] & [2.70] & [2.88] & [1.22] \\ 
4-2 & Mean & 0.65$^{**}$ & 0.59$^{**}$ & 0.90$^{***}$ & 0.46$^{*}$ & 0.81$^{**}$ & 0.16 \\ 
 & T-stat & [1.98] & [2.05] & [3.30] & [1.77] & [2.57] & [0.62] \\ 
7-2 & Mean & 0.47 & 0.75$^{***}$ & 0.69$^{**}$ & 0.67$^{**}$ & 0.94$^{***}$ & 0.47 \\ 
 & T-stat & [1.39] & [2.70] & [2.47] & [2.45] & [3.13] & [1.60] \\ 
12-2 & Mean & 0.42 & 0.76$^{***}$ & 0.70$^{**}$ & 0.50$^{*}$ & 1.17$^{***}$ & 0.74$^{**}$ \\ 
 & T-stat & [1.26] & [2.86] & [2.42] & [1.77] & [3.82] & [2.42] \\ 
3-3 & Mean & 0.63$^{**}$ & 0.40 & 0.78$^{***}$ & 1.05$^{***}$ & 0.92$^{***}$ & 0.29 \\ 
 & T-stat & [2.03] & [1.42] & [2.70] & [3.52] & [3.06] & [1.10] \\ 
7-3 & Mean & 0.54 & 0.76$^{***}$ & 1.07$^{***}$ & 0.65$^{**}$ & 0.76$^{**}$ & 0.22 \\ 
 & T-stat & [1.63] & [2.62] & [3.90] & [2.31] & [2.47] & [0.75] \\ 
12-3 & Mean & 0.54 & 0.71$^{**}$ & 0.64$^{**}$ & 0.64$^{**}$ & 1.15$^{***}$ & 0.61$^{**}$ \\ 
 & T-stat & [1.63] & [2.47] & [2.17] & [2.19] & [3.78] & [2.10] \\ 
12-7 & Mean & 0.65$^{**}$ & 0.51$^{*}$ & 0.68$^{**}$ & 0.91$^{***}$ & 0.86$^{***}$ & 0.21 \\ 
 & T-stat & [2.11] & [1.72] & [2.35] & [3.22] & [2.68] & [0.70] \\
\hline \\[-1.8ex] 
\end{tabular} 
\end{table}

\newgeometry{left=1.5cm, right=0.75cm, top=3cm, bottom=1.5cm}
\begin{table}[!htbp] \centering 
  \caption{\textbf{Summary statistics of characteristics} }
  \label{} 
            \begin{flushleft}
    {\medskip\small
    This table reports summary statistics of characteristics, including their percentiles for different time subsamples.
    OP is operating profitability. CAR3 is abnormal return during 3-day window, centered around earnings announcement. Standardized unexpected earnings  (SUE) are defined as the most recent year-to-year change in earnings per share (EPS), divided by the standard deviation of the changes in earnings during the recent eight announcements. I define SUE as missing if there are less than 6 quarterly announcement during 2-year window. Cmom1-1 is the return of the customers' portfolio during the last month. Mom12-2 and Mom1-1 are past returns of the firm at a given lag.}
    \medskip
    \end{flushleft}
    \begin{tabularx}{\linewidth}{l*{12}{Y}}
    \toprule
    \multicolumn{7}{l}{\textbf{Panel A: 1979-2004}} \\
    \midrule
\\[-1.8ex]\hline 
\hline \\[-1.8ex] 
 & Cmom1-1 & Mom12-2 & Mom1-1 & log(ME) & log(B/M) & Op & CAR3 & SUE \\ 
\hline \\[-1.8ex] 
Mean & $1.05$ & $21.75$ & $1.54$ & $5.49$ & $-0.54$ & $0.25$ & $0.003$ & $0.04$ \\ 
SD & $8.41$ & $56.62$ & $12.05$ & $1.73$ & $0.73$ & $0.21$ & $0.06$ & $1.10$ \\ 
min & $-36.36$ & $-86.71$ & $-46.17$ & $1.15$ & $-3.40$ & $-1.25$ & $-0.32$ & $-3.04$ \\ 
p05 & $-12.56$ & $-41.63$ & $-17.03$ & $2.80$ & $-1.82$ & $-0.06$ & $-0.10$ & $-1.89$ \\ 
p25 & $-3.79$ & $-11.51$ & $-5.62$ & $4.26$ & $-1.00$ & $0.15$ & $-0.03$ & $-0.59$ \\ 
p50 & $1.04$ & $11.43$ & $0.83$ & $5.32$ & $-0.48$ & $0.26$ & $0.001$ & $0.06$ \\ 
p75 & $5.88$ & $41.24$ & $7.96$ & $6.62$ & $-0.04$ & $0.35$ & $0.03$ & $0.64$ \\ 
p95 & $14.93$ & $118.60$ & $22.30$ & $8.52$ & $0.58$ & $0.56$ & $0.11$ & $1.89$ \\ 
max & $38.21$ & $831.90$ & $76.52$ & $13.01$ & $2.13$ & $2.90$ & $0.41$ & $4.81$ \\
\hline \\[-1.8ex] 
\end{tabularx} 

  \begin{tabularx}{\linewidth}{l*{12}{Y}}
    \toprule
    \multicolumn{7}{l}{\textbf{Panel B: 2005-2018}} \\
    \midrule
\\[-1.8ex]\hline 
\hline \\[-1.8ex] 
 & Cmom1-1 & Mom12-2 & Mom1-1 & log(ME) & log(B/M) & Op & CAR3 & SUE \\ 
\hline \\[-1.8ex] 
Mean & $0.75$ & $13.99$ & $1.17$ & $7.30$ & $-0.92$ & $0.25$ & $0.003$ & $0.10$ \\ 
SD & $7.14$ & $38.52$ & $10.19$ & $1.74$ & $0.68$ & $0.28$ & $0.08$ & $1.08$ \\ 
min & $-36.71$ & $-80.19$ & $-39.70$ & $2.37$ & $-4.07$ & $-1.31$ & $-0.29$ & $-3.03$ \\ 
p05 & $-11.10$ & $-39.79$ & $-15.44$ & $4.74$ & $-2.13$ & $-0.05$ & $-0.12$ & $-1.82$ \\ 
p25 & $-3.40$ & $-11.09$ & $-4.87$ & $6.13$ & $-1.32$ & $0.14$ & $-0.04$ & $-0.48$ \\ 
p50 & $0.89$ & $9.25$ & $0.89$ & $7.13$ & $-0.85$ & $0.22$ & $0.001$ & $0.08$ \\ 
p75 & $5.04$ & $32.59$ & $6.85$ & $8.27$ & $-0.46$ & $0.34$ & $0.05$ & $0.71$ \\ 
p95 & $12.22$ & $84.00$ & $18.38$ & $10.80$ & $0.09$ & $0.62$ & $0.13$ & $1.97$ \\ 
max & $37.86$ & $272.40$ & $61.13$ & $12.27$ & $1.11$ & $4.29$ & $0.35$ & $3.41$ \\
\hline \\[-1.8ex] 
\end{tabularx} 
\end{table}
\restoregeometry

\newgeometry{left=2cm, right=1cm, top=1.5cm, bottom=1cm}
\begin{table}[!htbp] \centering 
  \caption{\textbf{Summary statistics of factors}} 
  \label{} 
                \begin{flushleft}
    {\medskip\small
    This table reports summary statistics of factors, including their percentiles. 
    CMOM, SUEF and CAR3F are factors, constructed from 2x3 sorts on ME and respective variable (returns of customers' portfolio during the recent month, SUE and CAR3). The factors use NYSE breakpoints and are constructed similarly to UMD. Standard errors are calculated using Newey-West correction.}
    \medskip
    \end{flushleft}
  \begin{tabularx}{\linewidth}{l*{12}{Y}}
    \toprule
    \multicolumn{7}{l}{\textbf{Panel A: 1979-2018}} \\
    \midrule
\\[-1.8ex]\hline 
\hline \\[-1.8ex] 
 & Mkt.RF & SMB & HML & RMW & CMA & UMD & CMOM & SUEF & CAR3F \\ 
\hline \\[-1.8ex] 
Mean & $0.68^{***}$ & $0.17$ & $0.26$ & $0.34^{***}$ & $0.27^{**}$ & $0.61^{***}$ & $0.40^{***}$ & $1.27^{***}$ & $0.18^{**}$ \\ 
SE & $0.21$ & $0.13$ & $0.16$ & $0.13$ & $0.10$ & $0.22$ & $0.14$ & $0.07$ & $0.07$ \\ 
T-stat & $[3.22]$ & $[1.31]$ & $[1.57]$ & $[2.67]$ & $[2.55]$ & $[2.78]$ & $[2.88]$ & $[17.31]$ & $[2.46]$ \\ 
SD & $4.38$ & $2.89$ & $2.92$ & $2.41$ & $1.99$ & $4.45$ & $3.50$ & $1.94$ & $1.57$ \\ 
min & $-23.24$ & $-15.33$ & $-11.10$ & $-19.06$ & $-6.88$ & $-34.39$ & $-10.55$ & $-8.34$ & $-10.98$ \\ 
p05 & $-7.09$ & $-4.07$ & $-3.95$ & $-2.82$ & $-2.66$ & $-6.46$ & $-4.85$ & $-1.48$ & $-2.17$ \\ 
p25 & $-1.94$ & $-1.51$ & $-1.38$ & $-0.81$ & $-0.99$ & $-1.01$ & $-1.54$ & $0.26$ & $-0.59$ \\ 
p50 & $1.09$ & $0.06$ & $0.04$ & $0.34$ & $0.16$ & $0.70$ & $0.06$ & $1.22$ & $0.31$ \\ 
p75 & $3.50$ & $1.93$ & $1.72$ & $1.38$ & $1.49$ & $2.88$ & $2.32$ & $2.18$ & $1.01$ \\ 
p95 & $7.07$ & $4.62$ & $5.30$ & $3.71$ & $3.29$ & $6.63$ & $5.93$ & $4.47$ & $2.46$ \\ 
max & $12.47$ & $18.75$ & $12.90$ & $13.51$ & $9.56$ & $18.33$ & $19.36$ & $10.12$ & $5.25$ \\
\hline \\[-1.8ex] 
\end{tabularx} 
  \begin{tabularx}{\linewidth}{l*{12}{Y}}
    \toprule
    \multicolumn{7}{l}{\textbf{Panel B: 1979-2004}} \\
    \midrule
\\[-1.8ex]\hline 
\hline \\[-1.8ex] 
 & Mkt.RF & SMB & HML & RMW & CMA & UMD & CMOM & SUEF & CAR3F \\ 
\hline \\[-1.8ex] 
Mean & $0.67^{***}$ & $0.19$ & $0.42^{*}$ & $0.38^{**}$ & $0.40^{***}$ & $0.87^{***}$ & $0.63^{***}$ & $1.27^{***}$ & $0.30^{***}$ \\ 
SE & $0.25$ & $0.18$ & $0.22$ & $0.18$ & $0.14$ & $0.23$ & $0.18$ & $0.09$ & $0.08$ \\ 
T-stat & $[2.65]$ & $[1.05]$ & $[1.93]$ & $[2.15]$ & $[2.88]$ & $[3.75]$ & $[3.52]$ & $[13.95]$ & $[3.86]$ \\ 
SD & $4.55$ & $3.12$ & $3.09$ & $2.75$ & $2.24$ & $4.41$ & $3.96$ & $1.86$ & $1.48$ \\ 
min & $-23.24$ & $-15.33$ & $-10.57$ & $-19.06$ & $-6.88$ & $-25.06$ & $-10.55$ & $-8.34$ & $-7.68$ \\ 
p05 & $-7.10$ & $-4.32$ & $-4.11$ & $-3.29$ & $-2.89$ & $-6.61$ & $-5.39$ & $-1.34$ & $-2.10$ \\ 
p25 & $-2.12$ & $-1.53$ & $-1.28$ & $-0.85$ & $-0.98$ & $-0.68$ & $-1.60$ & $0.27$ & $-0.46$ \\ 
p50 & $1.13$ & $0.05$ & $0.50$ & $0.38$ & $0.36$ & $1.04$ & $0.26$ & $1.11$ & $0.37$ \\ 
p75 & $3.73$ & $2.11$ & $1.95$ & $1.47$ & $1.68$ & $2.95$ & $2.95$ & $2.31$ & $1.12$ \\ 
p95 & $7.16$ & $4.63$ & $5.57$ & $3.96$ & $4.07$ & $7.31$ & $6.19$ & $4.31$ & $2.55$ \\ 
max & $12.47$ & $18.75$ & $12.90$ & $13.51$ & $9.56$ & $18.33$ & $19.36$ & $8.48$ & $5.25$ \\
\hline \\[-1.8ex] 
\end{tabularx} 
  \begin{tabularx}{\linewidth}{l*{12}{Y}}
    \toprule
    \multicolumn{7}{l}{\textbf{Panel C: 2005-2018}} \\
    \midrule
\\[-1.8ex]\hline 
\hline \\[-1.8ex] 
 & Mkt.RF & SMB & HML & RMW & CMA & UMD & CMOM & SUEF & CAR3F \\ 
\hline \\[-1.8ex] 
Mean & $0.69^{*}$ & $0.13$ & $-0.06$ & $0.25^{*}$ & $0.002$ & $0.12$ & $-0.06$ & $1.27^{***}$ & $-0.04$ \\ 
SE & $0.38$ & $0.15$ & $0.22$ & $0.14$ & $0.12$ & $0.45$ & $0.17$ & $0.12$ & $0.15$ \\ 
T-stat & $[1.82]$ & $[0.87]$ & $[-0.25]$ & $[1.77]$ & $[0.01]$ & $[0.27]$ & $[-0.34]$ & $[10.17]$ & $[-0.27]$ \\ 
SD & $4.05$ & $2.39$ & $2.54$ & $1.56$ & $1.38$ & $4.48$ & $2.29$ & $2.10$ & $1.72$ \\ 
min & $-17.23$ & $-4.76$ & $-11.10$ & $-3.71$ & $-3.32$ & $-34.39$ & $-6.09$ & $-6.74$ & $-10.98$ \\ 
p05 & $-6.35$ & $-3.72$ & $-3.33$ & $-1.97$ & $-2.16$ & $-5.40$ & $-4.08$ & $-1.82$ & $-2.31$ \\ 
p25 & $-1.46$ & $-1.44$ & $-1.45$ & $-0.67$ & $-1.00$ & $-1.33$ & $-1.37$ & $0.26$ & $-0.90$ \\ 
p50 & $0.99$ & $0.08$ & $-0.23$ & $0.30$ & $-0.06$ & $0.38$ & $-0.05$ & $1.33$ & $0.19$ \\ 
p75 & $3.12$ & $1.60$ & $1.14$ & $1.17$ & $0.90$ & $2.52$ & $1.30$ & $2.17$ & $0.82$ \\ 
p95 & $6.79$ & $3.73$ & $3.81$ & $3.04$ & $2.26$ & $5.31$ & $3.54$ & $4.83$ & $2.10$ \\ 
max & $11.35$ & $6.81$ & $8.27$ & $4.82$ & $3.67$ & $12.54$ & $6.37$ & $10.12$ & $4.93$ \\
\hline \\[-1.8ex] 
\end{tabularx} 
\end{table}
\restoregeometry

\newgeometry{left=2cm, right=1cm, top=3cm, bottom=1.5cm}
\begin{table}[!htbp] \centering 
  \caption{\textbf{Fama-MacBeth regressions, 1979-2004} }
  \label{} 
            \begin{flushleft}
    {\medskip\small
This table reports the results of Fama-MacBeth regression of returns on the characteristics. Cmom1-1 is returns of customers' portfolio during the last month. Mom1-1 and Mom12-2 are past returns of stock at lags 1-1 and 12-2. OP is operating profitability. CAR3 is abnormal return during 3-day window, centered around earnings announcement. Standardized unexpected earnings  (SUE) are defined as the most recent year-to-year change in earnings per share (EPS), divided by the standard deviation of the changes in earnings during the recent eight announcements.}
    \medskip
    \end{flushleft}
  \small
\begin{tabular}{@{\extracolsep{0pt}}lp{1.3cm}p{1.3cm}p{1.3cm}p{1.3cm}p{1.3cm}p{1.3cm}p{1.3cm}p{1.3cm}p{1.3cm}} 
\\[-1.8ex]\hline 
\hline \\[-1.8ex] 
 & \multicolumn{9}{c}{\textit{Dependent variable:}} \\ 
\cline{2-10} 
\\[-1.8ex] & (1) & (2) & (3) & (4) & (5) & (6) & (7) & (8) & (9)\\ 
\hline \\[-1.8ex] 
 Cmom1-1 & 0.05$^{***}$ & 0.06$^{***}$ &  &  &  & 0.05$^{***}$ &  & 0.05$^{***}$ & 0.06$^{***}$ \\ 
  & [2.93] & [2.82] &  &  &  & [3.11] &  & [3.15] & [3.33] \\ 
  & & & & & & & & & \\ 
 log(ME) &  & -0.14$^{*}$ &  & -0.16$^{**}$ & -0.12$^{*}$ & -0.13$^{*}$ & -0.14$^{**}$ & -0.17$^{**}$ & -0.15$^{**}$ \\ 
  &  & [-1.90] &  & [-2.39] & [-1.74] & [-1.83] & [-2.06] & [-2.53] & [-2.18] \\ 
  & & & & & & & & & \\ 
 log(B/M) &  & 0.27 &  & 0.36$^{*}$ & 0.30 & 0.26 & 0.40$^{*}$ & 0.38$^{*}$ & 0.38$^{*}$ \\ 
  &  & [1.46] &  & [1.74] & [1.44] & [1.26] & [1.92] & [1.84] & [1.81] \\ 
  & & & & & & & & & \\ 
 OP &  &  &  & 0.08 & 0.48 & 0.45 & 0.74 & 0.18 & 0.78 \\ 
  &  &  &  & [0.13] & [0.77] & [0.70] & [1.14] & [0.28] & [1.18] \\ 
  & & & & & & & & & \\ 
 Mom12-2 &  &  & 0.01$^{***}$ & 0.01$^{**}$ &  &  & 0.001 & 0.01$^{**}$ & -0.001 \\ 
  &  &  & [3.44] & [2.47] &  &  & [0.16] & [2.05] & [-0.30] \\ 
  & & & & & & & & & \\ 
 Mom1-1 &  &  & -0.03$^{***}$ & -0.05$^{***}$ &  &  & -0.06$^{***}$ & -0.05$^{***}$ & -0.07$^{***}$ \\ 
  &  &  & [-2.83] & [-4.63] &  &  & [-5.91] & [-4.89] & [-6.08] \\ 
  & & & & & & & & & \\ 
 CAR3 &  &  &  &  & 3.58$^{**}$ & 4.02$^{**}$ & 4.14$^{**}$ &  & 4.83$^{***}$ \\ 
  &  &  &  &  & [2.22] & [2.51] & [2.44] &  & [2.80] \\ 
  & & & & & & & & & \\ 
 SUE &  &  &  &  & 1.01$^{***}$ & 1.00$^{***}$ & 1.04$^{***}$ &  & 1.04$^{***}$ \\ 
  &  &  &  &  & [11.34] & [11.05] & [10.97] &  & [10.73] \\ 
  & & & & & & & & & \\ 
 Constant & 1.50$^{***}$ & 2.38$^{***}$ & 1.47$^{***}$ & 2.45$^{***}$ & 2.10$^{***}$ & 2.06$^{***}$ & 2.29$^{***}$ & 2.50$^{***}$ & 2.34$^{***}$ \\ 
  & [4.49] & [4.72] & [4.64] & [4.99] & [4.26] & [4.12] & [4.70] & [5.10] & [4.77] \\ 
  & & & & & & & & & \\ 
\hline \\[-1.8ex] 
Observations & 25,421 & 25,421 & 25,421 & 25,421 & 25,421 & 25,421 & 25,421 & 25,421 & 25,421 \\ 
Adjusted R$^{2}$ & 0.008 & 0.031 & 0.033 & 0.067 & 0.059 & 0.066 & 0.086 & 0.073 & 0.093 \\ 

\hline 
\hline \\[-1.8ex] 

\end{tabular} 
\end{table}
\restoregeometry

\newgeometry{left=2cm, right=1cm, top=3cm, bottom=1.5cm}
\begin{table}[!htbp] \centering 
  \caption{\textbf{Fama-MacBeth regressions, 2005-2018}} 
  \label{} 
    \begin{flushleft}
    {\medskip\small
This table reports the results of Fama-MacBeth regression of returns on the characteristics. Cmom1-1 is returns of customers' portfolio during the last month. Mom1-1 and Mom12-2 are past returns of stock at lags 1-1 and 12-2. OP is operating profitability. CAR3 is abnormal return during 3-day window, centered around earnings announcement. Standardized unexpected earnings  (SUE) are defined as the most recent year-to-year change in earnings per share (EPS), divided by the standard deviation of the changes in earnings during the recent eight announcements.}
    \medskip
    \end{flushleft}
\begin{tabular}{@{\extracolsep{0pt}}lp{1.3cm}p{1.3cm}p{1.3cm}p{1.3cm}p{1.3cm}p{1.3cm}p{1.3cm}p{1.3cm}p{1.3cm}} 
\\[-1.8ex]\hline 
\hline \\[-1.8ex] 
 & \multicolumn{9}{c}{\textit{Dependent variable:}} \\ 
\cline{2-10} 
\\[-1.8ex] & (1) & (2) & (3) & (4) & (5) & (6) & (7) & (8) & (9)\\ 
\hline \\[-1.8ex] 
 Cmom1-1 & 0.02 & 0.02 &  &  &  & 0.02 &  & 0.02 & 0.02 \\ 
  & [1.29] & [1.22] &  &  &  & [0.95] &  & [1.15] & [1.17] \\ 
  & & & & & & & & & \\ 
 log(ME) &  & -0.07 &  & -0.06 & -0.06 & -0.07 & -0.07 & -0.06 & -0.07 \\ 
  &  & [-1.16] &  & [-0.98] & [-1.12] & [-1.13] & [-1.26] & [-1.02] & [-1.30] \\ 
  & & & & & & & & & \\ 
 log(B/M) &  & -0.18 &  & -0.25 & -0.13 & -0.13 & -0.20 & -0.26 & -0.20 \\ 
  &  & [-1.18] &  & [-1.48] & [-0.75] & [-0.75] & [-1.18] & [-1.51] & [-1.19] \\ 
  & & & & & & & & & \\ 
 OP &  &  &  & -0.31 & -0.03 & -0.07 & -0.19 & -0.35 & -0.23 \\ 
  &  &  &  & [-0.76] & [-0.08] & [-0.16] & [-0.47] & [-0.86] & [-0.56] \\ 
  & & & & & & & & & \\ 
 Mom12-2 &  &  & -0.01 & -0.01 &  &  & -0.01$^{**}$ & -0.01 & -0.01$^{**}$ \\ 
  &  &  & [-1.22] & [-1.15] &  &  & [-2.17] & [-1.24] & [-2.28] \\ 
  & & & & & & & & & \\ 
 Mom1-1 &  &  & -0.003 & -0.004 &  &  & -0.02 & -0.01 & -0.02 \\ 
  &  &  & [-0.30] & [-0.36] &  &  & [-1.49] & [-0.48] & [-1.63] \\ 
  & & & & & & & & & \\ 
 CAR3 &  &  &  &  & 0.36 & 0.28 & 1.06 &  & 1.06 \\ 
  &  &  &  &  & [0.33] & [0.26] & [0.87] &  & [0.89] \\ 
  & & & & & & & & & \\ 
 SUE &  &  &  &  & 0.58$^{***}$ & 0.59$^{***}$ & 0.65$^{***}$ &  & 0.67$^{***}$ \\ 
  &  &  &  &  & [6.52] & [6.75] & [8.17] &  & [8.42] \\ 
  & & & & & & & & & \\ 
 Constant & 1.04$^{**}$ & 1.34$^{*}$ & 0.84$^{**}$ & 1.06 & 1.25$^{*}$ & 1.29$^{*}$ & 1.20$^{*}$ & 1.14$^{*}$ & 1.26$^{*}$ \\ 
  & [2.46] & [1.81] & [2.17] & [1.56] & [1.70] & [1.77] & [1.76] & [1.65] & [1.83] \\ 
  & & & & & & & & & \\ 
\hline \\[-1.8ex] 
Observations & 22,666 & 22,666 & 22,666 & 22,666 & 22,666 & 22,666 & 22,666 & 22,666 & 22,666 \\ 
Adjusted R$^{2}$ & 0.005 & 0.021 & 0.026 & 0.046 & 0.034 & 0.039 & 0.056 & 0.050 & 0.060 \\ 

\hline 
\hline \\[-1.8ex] 

\end{tabular} 
\end{table}

\newpage

\begin{table}[!htbp] \centering 
  \caption{\textbf{Factor spanning test for Momentum, 1979-2004} }
  \label{} 
              \begin{flushleft}
    {\medskip\small
    This table reports the results of time-series regression of momentum factor UMD on the other factors.
    CMOM, SUEF and CAR3F are factors, constructed from 2x3 sorts on ME and respective variable (past returns of customers' portfolio during the past month, SUE and CAR3). The factors use NYSE breakpoints and are constructed similarly to UMD. }
    \medskip
    \end{flushleft}
\begin{tabular}{@{\extracolsep{5pt}}lccccc} 
\\[-1.8ex]\hline 
\hline \\[-1.8ex] 
 & \multicolumn{5}{c}{\textit{Dependent variable:}} \\ 
\cline{2-6} 
\\[-1.8ex] & \multicolumn{5}{c}{UMD} \\ 
\\[-1.8ex] & (1) & (2) & (3) & (4) & (5)\\ 
\hline \\[-1.8ex] 
 Mkt.RF & -0.12 & -0.03 & -0.05 & -0.13 & -0.06 \\ 
  & [-1.21] & [-0.32] & [-0.72] & [-1.28] & [-0.85] \\ 
  & & & & & \\ 
 SMB & 0.11 & 0.18 & 0.20 & 0.10 & 0.19 \\ 
  & [0.68] & [1.26] & [1.43] & [0.65] & [1.42] \\ 
  & & & & & \\ 
 HML & -0.26 & -0.58$^{***}$ & -0.34$^{*}$ & -0.25 & -0.33$^{*}$ \\ 
  & [-1.13] & [-2.60] & [-1.85] & [-1.17] & [-1.93] \\ 
  & & & & & \\ 
 CMA &  & 0.55 &  &  &  \\ 
  &  & [1.32] &  &  &  \\ 
  & & & & & \\ 
 RMW &  & 0.36 &  &  &  \\ 
  &  & [1.36] &  &  &  \\ 
  & & & & & \\ 
 SUEF &  &  & 0.78$^{***}$ &  & 0.79$^{***}$ \\ 
  &  &  & [4.56] &  & [4.73] \\ 
  & & & & & \\ 
 CAR3f &  &  & 0.88$^{***}$ &  & 0.81$^{***}$ \\ 
  &  &  & [2.82] &  & [2.74] \\ 
  & & & & & \\ 
 CMOM &  &  &  & 0.21$^{**}$ & 0.15$^{*}$ \\ 
  &  &  &  & [2.44] & [1.85] \\ 
  & & & & & \\ 
 Constant & 1.04$^{***}$ & 0.75$^{**}$ & -0.23 & 0.91$^{***}$ & -0.32 \\ 
  & [4.13] & [2.17] & [-0.68] & [3.60] & [-0.96] \\ 
  & & & & & \\
\hline \\[-1.8ex] 
Observations & 312 & 312 & 312 & 312 & 312 \\ 
R$^{2}$ & 0.038 & 0.095 & 0.286 & 0.072 & 0.303 \\ 
Adjusted R$^{2}$ & 0.028 & 0.080 & 0.275 & 0.060 & 0.289 \\ 
Residual Std. Error & 4.351 & 4.234 & 3.759 & 4.278 & 3.722 \\ 
F Statistic & 4.006 & 6.398 & 24.540 & 5.996 & 22.070 \\ 
\hline 
\hline \\[-1.8ex] 

\end{tabular} 
\end{table}

\begin{table}[!htbp] \centering 
  \caption{\textbf{Factor spanning test for Momentum, 2005-2018} }
  \label{} 
              \begin{flushleft}
    {\medskip\small
    This table reports the results of time-series regression of momentum factor UMD on the other factors.
    CMOM, SUEF and CAR3F are factors, constructed from 2x3 sorts on ME and respective variable (past returns of customers' portfolio during the past month, SUE and CAR3). The factors use NYSE breakpoints and are constructed similarly to UMD. }
    \medskip
    \end{flushleft}
\begin{tabular}{@{\extracolsep{5pt}}lccccc} 
\\[-1.8ex]\hline 
\hline \\[-1.8ex] 
 & \multicolumn{5}{c}{\textit{Dependent variable:}} \\ 
\cline{2-6} 
\\[-1.8ex] & \multicolumn{5}{c}{UMD} \\ 
\\[-1.8ex] & (1) & (2) & (3) & (4) & (5)\\ 
\hline \\[-1.8ex] 
 Mkt.RF & -0.27$^{**}$ & -0.24$^{*}$ & -0.13$^{*}$ & -0.26$^{**}$ & -0.12$^{*}$ \\ 
  & [-2.01] & [-1.74] & [-1.87] & [-2.04] & [-1.77] \\ 
  & & & & & \\ 
 SMB & 0.01 & 0.02 & 0.07 & 0.01 & 0.08 \\ 
  & [0.06] & [0.12] & [0.61] & [0.09] & [0.67] \\ 
  & & & & & \\ 
 HML & -0.61$^{***}$ & -0.66$^{***}$ & -0.11 & -0.64$^{***}$ & -0.13 \\ 
  & [-3.28] & [-2.89] & [-0.47] & [-3.16] & [-0.57] \\ 
  & & & & & \\ 
 CMA &  & 0.19 &  &  &  \\ 
  &  & [0.50] &  &  &  \\ 
  & & & & & \\ 
 RMW &  & 0.12 &  &  &  \\ 
  &  & [0.45] &  &  &  \\ 
  & & & & & \\ 
 SUEF &  &  & 0.82$^{***}$ &  & 0.81$^{***}$ \\ 
  &  &  & [5.31] &  & [5.26] \\ 
  & & & & & \\ 
 CAR3f &  &  & 1.13$^{***}$ &  & 1.16$^{***}$ \\ 
  &  &  & [3.12] &  & [3.32] \\ 
  & & & & & \\ 
 CMOM &  &  &  & -0.14 & -0.16 \\ 
  &  &  &  & [-0.62] & [-1.14] \\ 
  & & & & & \\ 
 Constant & 0.27 & 0.22 & -0.80$^{**}$ & 0.25 & -0.78$^{**}$ \\ 
  & [0.74] & [0.62] & [-2.56] & [0.64] & [-2.51] \\ 
  & & & & & \\ 
\hline \\[-1.8ex] 
Observations & 162 & 162 & 162 & 158 & 158 \\ 
R$^{2}$ & 0.224 & 0.228 & 0.524 & 0.227 & 0.531 \\ 
Adjusted R$^{2}$ & 0.210 & 0.204 & 0.509 & 0.207 & 0.513 \\ 
Residual Std. Error & 3.984 & 3.999 & 3.139 & 4.032 & 3.161 \\ 
F Statistic & 15.230 & 9.234 & 34.390 & 11.250 & 28.510 \\ 
\hline 
\hline \\[-1.8ex] 

\end{tabular} 
\end{table}

\begin{table}[!htbp] \centering 
  \caption{\textbf{Factor spanning test for Customer momentum, 1979-2004} }
  \label{} 
              \begin{flushleft}
    {\medskip\small
    This table reports the results of time-series regression of customer momentum factor CMOM on the other factors.
    CMOM, SUEF and CAR3F are factors, constructed from 2x3 sorts on ME and respective variable (past returns of customers' portfolio during the past month, SUE and CAR3). The factors use NYSE breakpoints and are constructed similarly to UMD. }
    \medskip
    \end{flushleft}
\begin{tabular}{@{\extracolsep{0pt}}lccccccc} 
\\[-1.8ex]\hline 
\hline \\[-1.8ex] 
 & \multicolumn{7}{c}{\textit{Dependent variable:}} \\ 
\cline{2-8} 
\\[-1.8ex] & \multicolumn{7}{c}{CMOM} \\ 
\\[-1.8ex] & (1) & (2) & (3) & (4) & (5) & (6) & (7)\\ 
\hline \\[-1.8ex] 
 Mkt.RF & 0.04 & 0.02 & 0.06 & 0.03 & 0.06 & 0.02 & 0.03 \\ 
  & [0.66] & [0.26] & [0.90] & [0.37] & [0.76] & [0.31] & [0.37] \\ 
  & & & & & & & \\ 
 SMB & 0.05 & -0.01 & 0.03 & -0.05 & 0.05 & -0.005 & -0.05 \\ 
  & [0.55] & [-0.14] & [0.42] & [-0.60] & [0.61] & [-0.06] & [-0.58] \\ 
  & & & & & & & \\ 
 HML & -0.07 & -0.001 & -0.02 & 0.12 & -0.08 & 0.04 & 0.12 \\ 
  & [-0.53] & [-0.005] & [-0.21] & [0.75] & [-0.64] & [0.22] & [0.72] \\ 
  & & & & & & & \\ 
 CMA &  & -0.07 &  & -0.19 &  & -0.17 & -0.20 \\ 
  &  & [-0.34] &  & [-0.93] &  & [-0.85] & [-0.98] \\ 
  & & & & & & & \\ 
 RMW &  & -0.22$^{**}$ &  & -0.29$^{***}$ &  & -0.25$^{***}$ & -0.28$^{***}$ \\ 
  &  & [-2.39] &  & [-3.47] &  & [-2.63] & [-3.30] \\ 
  & & & & & & & \\ 
 UMD &  &  & 0.17$^{**}$ & 0.21$^{***}$ &  &  & 0.18$^{**}$ \\ 
  &  &  & [2.51] & [2.95] &  &  & [2.10] \\ 
  & & & & & & & \\ 
 SUEF &  &  &  &  & -0.07 & 0.02 & -0.10 \\ 
  &  &  &  &  & [-0.48] & [0.16] & [-0.67] \\ 
  & & & & & & & \\ 
 CAR3f &  &  &  &  & 0.49$^{**}$ & 0.50$^{**}$ & 0.35$^{*}$ \\ 
  &  &  &  &  & [2.36] & [2.40] & [1.93] \\ 
  & & & & & & & \\ 
 Constant & 0.62$^{***}$ & 0.73$^{***}$ & 0.44$^{**}$ & 0.57$^{***}$ & 0.55$^{*}$ & 0.58$^{**}$ & 0.63$^{**}$ \\ 
  & [3.18] & [3.50] & [2.24] & [2.76] & [1.93] & [1.97] & [2.18] \\ 
  & & & & & & & \\ 
\hline \\[-1.8ex] 
Observations & 312 & 312 & 312 & 312 & 312 & 312 & 312 \\ 
R$^{2}$ & 0.012 & 0.028 & 0.048 & 0.077 & 0.041 & 0.062 & 0.090 \\ 
Adjusted R$^{2}$ & 0.002 & 0.012 & 0.035 & 0.058 & 0.025 & 0.041 & 0.066 \\ 
Residual Std. Error & 3.958 & 3.938 & 3.892 & 3.844 & 3.911 & 3.880 & 3.829 \\ 
F Statistic & 1.211 & 1.749 & 3.828 & 4.217 & 2.615 & 2.894 & 3.745 \\ 
\hline 
\hline \\[-1.8ex] 

\end{tabular} 
\end{table}

\begin{table}[!htbp] \centering 
  \caption{\textbf{Factor spanning test for Customer momentum, 2005-2018}} 
  \label{} 
              \begin{flushleft}
    {\medskip\small
    This table reports the results of time-series regression of customer momentum factor CMOM on the other factors.
    CMOM, SUEF and CAR3F are factors, constructed from 2x3 sorts on ME and respective variable (past returns of customers' portfolio during the past month, SUE and CAR3). The factors use NYSE breakpoints and are constructed similarly to UMD. }
    \medskip
    \end{flushleft}
\begin{tabular}{@{\extracolsep{0pt}}lccccccc} 
\\[-1.8ex]\hline 
\hline \\[-1.8ex] 
 & \multicolumn{7}{c}{\textit{Dependent variable:}} \\ 
\cline{2-8} 
\\[-1.8ex] & \multicolumn{7}{c}{CMOM} \\ 
\\[-1.8ex] & (1) & (2) & (3) & (4) & (5) & (6) & (7)\\ 
\hline \\[-1.8ex] 
 Mkt.RF & 0.03 & 0.04 & 0.02 & 0.03 & 0.03 & 0.05 & 0.04 \\ 
  & [0.50] & [0.82] & [0.31] & [0.63] & [0.65] & [0.99] & [0.83] \\ 
  & & & & & & & \\ 
 SMB & 0.01 & 0.05 & 0.02 & 0.05 & 0.02 & 0.06 & 0.06 \\ 
  & [0.17] & [0.63] & [0.18] & [0.63] & [0.20] & [0.65] & [0.72] \\ 
  & & & & & & & \\ 
 HML & -0.16$^{*}$ & -0.10 & -0.18$^{**}$ & -0.13$^{*}$ & -0.17$^{***}$ & -0.11 & -0.12 \\ 
  & [-1.91] & [-1.06] & [-2.49] & [-1.74] & [-2.75] & [-1.49] & [-1.63] \\ 
  & & & & & & & \\ 
 CMA &  & -0.18 &  & -0.17 &  & -0.19 & -0.19 \\ 
  &  & [-1.20] &  & [-1.22] &  & [-1.27] & [-1.31] \\ 
  & & & & & & & \\ 
 RMW &  & 0.21 &  & 0.22 &  & 0.21 & 0.22$^{*}$ \\ 
  &  & [1.59] &  & [1.62] &  & [1.63] & [1.71] \\ 
  & & & & & & & \\ 
 UMD &  &  & -0.05 & -0.05 &  &  & -0.08$^{*}$ \\ 
  &  &  & [-0.81] & [-0.96] &  &  & [-1.65] \\ 
  & & & & & & & \\ 
 SUEF &  &  &  &  & -0.05 & -0.06 & 0.01 \\ 
  &  &  &  &  & [-0.45] & [-0.56] & [0.11] \\ 
  & & & & & & & \\ 
 CAR3f &  &  &  &  & 0.07 & 0.08 & 0.18 \\ 
  &  &  &  &  & [0.38] & [0.47] & [1.19] \\ 
  & & & & & & & \\ 
 Constant & -0.10 & -0.17 & -0.08 & -0.16 & -0.03 & -0.10 & -0.17 \\ 
  & [-0.56] & [-0.93] & [-0.49] & [-0.88] & [-0.15] & [-0.42] & [-0.68] \\ 
  & & & & & & & \\
\hline \\[-1.8ex] 
Observations & 158 & 158 & 158 & 158 & 158 & 158 & 158 \\ 
R$^{2}$ & 0.027 & 0.048 & 0.033 & 0.055 & 0.030 & 0.053 & 0.066 \\ 
Adjusted R$^{2}$ & 0.008 & 0.017 & 0.008 & 0.017 & -0.002 & 0.009 & 0.016 \\ 
Residual Std. Error & 2.286 & 2.275 & 2.286 & 2.275 & 2.297 & 2.285 & 2.277 \\ 
F Statistic & 1.410 & 1.544 & 1.312 & 1.462 & 0.942 & 1.193 & 1.313 \\ 
\hline 
\hline \\[-1.8ex] 

\end{tabular} 
\end{table}

\newgeometry{left=1.5cm, right=1cm, top=3cm, bottom=1.5cm}
\begin{table}[!htbp] \centering 
  \caption{\textbf{Factor spanning test for Earnings momentum (SUE), 1979-2004}} 
  \label{} 
              \begin{flushleft}
    {\medskip\small
    This table reports the results of time-series regression of earning momentum factor SUEF on the other factors.
    CMOM, SUEF and CAR3F are factors, constructed from 2x3 sorts on ME and respective variable (past returns of customers' portfolio during the past month, SUE and CAR3). The factors use NYSE breakpoints and are constructed similarly to UMD. }
    \medskip
    \end{flushleft}
  \footnotesize
\begin{tabular}{@{\extracolsep{0pt}}lcccccccc} 
\\[-1.8ex]\hline 
\hline \\[-1.8ex] 
 & \multicolumn{8}{c}{\textit{Dependent variable:}} \\ 
\cline{2-9} 
\\[-1.8ex] & \multicolumn{8}{c}{SUEF} \\ 
\\[-1.8ex] & (1) & (2) & (3) & (4) & (5) & (6) & (7) & (8)\\ 
\hline \\[-1.8ex] 
 CAR3f & 0.48$^{***}$ &  &  &  &  &  &  &  \\ 
  & [4.14] &  &  &  &  &  &  &  \\ 
  & & & & & & & & \\ 
 Mkt.RF &  & -0.05 & -0.01 & -0.03 & -0.0002 & -0.05 & -0.01 & -0.0000 \\ 
  &  & [-1.58] & [-0.20] & [-1.19] & [-0.01] & [-1.59] & [-0.22] & [-0.001] \\ 
  & & & & & & & & \\ 
 SMB &  & -0.10$^{***}$ & -0.07$^{*}$ & -0.12$^{***}$ & -0.09$^{**}$ & -0.10$^{***}$ & -0.07$^{*}$ & -0.09$^{**}$ \\ 
  &  & [-2.58] & [-1.73] & [-3.01] & [-2.41] & [-2.62] & [-1.76] & [-2.41] \\ 
  & & & & & & & & \\ 
 HML &  & 0.08 & -0.09$^{*}$ & 0.12$^{***}$ & -0.01 & 0.08 & -0.09$^{*}$ & -0.01 \\ 
  &  & [1.39] & [-1.74] & [2.60] & [-0.12] & [1.40] & [-1.71] & [-0.11] \\ 
  & & & & & & & & \\ 
 CMA &  &  & 0.29$^{***}$ &  & 0.21$^{***}$ &  & 0.29$^{***}$ & 0.21$^{***}$ \\ 
  &  &  & [3.12] &  & [3.28] &  & [3.27] & [3.25] \\ 
  & & & & & & & & \\ 
 RMW &  &  & 0.19$^{***}$ &  & 0.13$^{**}$ &  & 0.19$^{***}$ & 0.13$^{**}$ \\ 
  &  &  & [3.35] &  & [2.48] &  & [3.44] & [2.49] \\ 
  & & & & & & & & \\ 
 UMD &  &  &  & 0.17$^{***}$ & 0.15$^{***}$ &  &  & 0.15$^{***}$ \\ 
  &  &  &  & [5.33] & [5.24] &  &  & [5.45] \\ 
  & & & & & & & & \\ 
 CMOM &  &  &  &  &  & 0.02 & 0.03 & -0.01 \\ 
  &  &  &  &  &  & [0.55] & [1.12] & [-0.25] \\ 
  & & & & & & & & \\ 
 Constant & 1.12$^{***}$ & 1.29$^{***}$ & 1.14$^{***}$ & 1.11$^{***}$ & 1.02$^{***}$ & 1.28$^{***}$ & 1.11$^{***}$ & 1.03$^{***}$ \\ 
  & [13.71] & [14.04] & [11.27] & [13.04] & [11.83] & [13.26] & [10.71] & [11.45] \\ 
  & & & & & & & & \\
\hline \\[-1.8ex] 
Observations & 312 & 312 & 312 & 312 & 312 & 312 & 312 & 312 \\ 
R$^{2}$ & 0.147 & 0.093 & 0.183 & 0.258 & 0.300 & 0.094 & 0.187 & 0.300 \\ 
Adjusted R$^{2}$ & 0.145 & 0.084 & 0.169 & 0.248 & 0.286 & 0.082 & 0.171 & 0.284 \\ 
Residual Std. Error & 1.720 & 1.780 & 1.695 & 1.613 & 1.571 & 1.782 & 1.694 & 1.573 \\ 
F Statistic & 53.610 & 10.520 & 13.660 & 26.620 & 21.810 & 7.969 & 11.670 & 18.650 \\ 
\hline 
\hline \\[-1.8ex] 

\end{tabular} 
\end{table}
\restoregeometry

\newgeometry{left=1.5cm, right=1cm, top=3cm, bottom=1.5cm}
\begin{table}[!htbp] \centering 
  \caption{\textbf{Factor spanning test for Earnings momentum (SUE), 2005-2018} }
  \label{} 
              \begin{flushleft}
    {\medskip\small
    This table reports the results of time-series regression of customer momentum factor SUEF on the other factors.
    CMOM, SUEF and CAR3F are factors, constructed from 2x3 sorts on ME and respective variable (past returns of customers' portfolio during the past month, SUE and CAR3). The factors use NYSE breakpoints and are constructed similarly to UMD. }
    \medskip
    \end{flushleft}
  \footnotesize
\begin{tabular}{@{\extracolsep{0pt}}lcccccccc} 
\\[-1.8ex]\hline 
\hline \\[-1.8ex] 
 & \multicolumn{8}{c}{\textit{Dependent variable:}} \\ 
\cline{2-9} 
\\[-1.8ex] & \multicolumn{8}{c}{SUEF} \\ 
\\[-1.8ex] & (1) & (2) & (3) & (4) & (5) & (6) & (7) & (8)\\ 
\hline \\[-1.8ex] 
 CAR3f & 0.32$^{**}$ &  &  &  &  &  &  &  \\ 
  & [1.99] &  &  &  &  &  &  &  \\ 
  & & & & & & & & \\ 
 Mkt.RF &  & -0.03 & -0.03 & 0.02 & 0.02 & -0.03 & -0.02 & 0.02 \\ 
  &  & [-0.59] & [-0.50] & [0.39] & [0.37] & [-0.53] & [-0.40] & [0.43] \\ 
  & & & & & & & & \\ 
 SMB &  & -0.02 & -0.02 & -0.03 & -0.02 & -0.02 & -0.01 & -0.02 \\ 
  &  & [-0.46] & [-0.33] & [-0.54] & [-0.48] & [-0.36] & [-0.16] & [-0.33] \\ 
  & & & & & & & & \\ 
 HML &  & -0.40$^{***}$ & -0.40$^{**}$ & -0.29$^{*}$ & -0.28 & -0.42$^{***}$ & -0.41$^{**}$ & -0.28 \\ 
  &  & [-3.03] & [-2.50] & [-1.86] & [-1.50] & [-3.07] & [-2.51] & [-1.50] \\ 
  & & & & & & & & \\ 
 CMA &  &  & 0.01 &  & -0.03 &  & -0.01 & -0.05 \\ 
  &  &  & [0.04] &  & [-0.22] &  & [-0.09] & [-0.32] \\ 
  & & & & & & & & \\ 
 RMW &  &  & 0.03 &  & 0.004 &  & 0.05 & 0.02 \\ 
  &  &  & [0.22] &  & [0.04] &  & [0.42] & [0.21] \\ 
  & & & & & & & & \\ 
 UMD &  &  &  & 0.19$^{***}$ & 0.19$^{***}$ &  &  & 0.19$^{***}$ \\ 
  &  &  &  & [3.21] & [3.20] &  &  & [3.19] \\ 
  & & & & & & & & \\ 
 CMOM &  &  &  &  &  & -0.03 & -0.03 & -0.003 \\ 
  &  &  &  &  &  & [-0.39] & [-0.50] & [-0.06] \\ 
  & & & & & & & & \\ 
 Constant & 1.28$^{***}$ & 1.27$^{***}$ & 1.26$^{***}$ & 1.22$^{***}$ & 1.22$^{***}$ & 1.25$^{***}$ & 1.23$^{***}$ & 1.20$^{***}$ \\ 
  & [10.18] & [9.69] & [9.97] & [9.10] & [9.18] & [9.11] & [9.50] & [8.69] \\ 
  & & & & & & & & \\ 
\hline \\[-1.8ex] 
Observations & 162 & 162 & 162 & 162 & 162 & 158 & 158 & 158 \\ 
R$^{2}$ & 0.067 & 0.269 & 0.269 & 0.401 & 0.401 & 0.273 & 0.274 & 0.403 \\ 
Adjusted R$^{2}$ & 0.061 & 0.255 & 0.246 & 0.386 & 0.378 & 0.254 & 0.245 & 0.376 \\ 
Residual Std. Error & 2.030 & 1.809 & 1.820 & 1.643 & 1.653 & 1.826 & 1.837 & 1.671 \\ 
F Statistic & 11.520 & 19.390 & 11.500 & 26.270 & 17.310 & 14.360 & 9.497 & 14.490 \\ 
\hline 
\hline \\[-1.8ex] 

\end{tabular} 
\end{table}
\restoregeometry

\newgeometry{left=1.5cm, right=1cm, top=3cm, bottom=1.5cm}
\begin{table}[!htbp] \centering 
  \caption{\textbf{Factor spanning test for Earnings momentum (CAR3), 1979-2004} }
  \label{} 
  \small
              \begin{flushleft}
    {\medskip\small
    This table reports the results of time-series regression of earning momentum factor CAR3 on the other factors.
    CMOM, SUEF and CAR3F are factors, constructed from 2x3 sorts on ME and respective variable (past returns of customers' portfolio during the past month, SUE and CAR3). The factors use NYSE breakpoints and are constructed similarly to UMD. }
    \medskip
    \end{flushleft}
\begin{tabular}{@{\extracolsep{0pt}}lcccccccc} 
\\[-1.8ex]\hline 
\hline \\[-1.8ex] 
 & \multicolumn{8}{c}{\textit{Dependent variable:}} \\ 
\cline{2-9} 
\\[-1.8ex] & \multicolumn{8}{c}{CAR3F} \\ 
\\[-1.8ex] & (1) & (2) & (3) & (4) & (5) & (6) & (7) & (8)\\ 
\hline \\[-1.8ex] 
 SUEF & 0.31$^{***}$ &  &  &  &  &  &  &  \\ 
  & [4.61] &  &  &  &  &  &  &  \\ 
  & & & & & & & & \\ 
 Mkt.RF &  & -0.03 & -0.01 & -0.01 & -0.003 & -0.03 & -0.01 & -0.004 \\ 
  &  & [-0.96] & [-0.25] & [-0.51] & [-0.12] & [-1.00] & [-0.29] & [-0.16] \\ 
  & & & & & & & & \\ 
 SMB &  & -0.01 & -0.01 & -0.03 & -0.03 & -0.02 & -0.01 & -0.03 \\ 
  &  & [-0.36] & [-0.30] & [-0.91] & [-1.21] & [-0.47] & [-0.29] & [-1.19] \\ 
  & & & & & & & & \\ 
 HML &  & 0.03 & -0.07 & 0.07 & 0.01 & 0.03 & -0.07 & 0.001 \\ 
  &  & [0.57] & [-1.21] & [1.61] & [0.13] & [0.69] & [-1.17] & [0.03] \\ 
  & & & & & & & & \\ 
 CMA &  &  & 0.19$^{**}$ &  & 0.11$^{**}$ &  & 0.19$^{**}$ & 0.12$^{***}$ \\ 
  &  &  & [2.23] &  & [2.45] &  & [2.43] & [2.63] \\ 
  & & & & & & & & \\ 
 RMW &  &  & 0.06 &  & 0.02 &  & 0.08 & 0.03 \\ 
  &  &  & [1.19] &  & [0.43] &  & [1.40] & [0.67] \\ 
  & & & & & & & & \\ 
 UMD &  &  &  & 0.14$^{***}$ & 0.13$^{***}$ &  &  & 0.12$^{***}$ \\ 
  &  &  &  & [4.41] & [4.58] &  &  & [4.65] \\ 
  & & & & & & & & \\ 
 CMOM &  &  &  &  &  & 0.06$^{***}$ & 0.07$^{***}$ & 0.04$^{**}$ \\ 
  &  &  &  &  &  & [2.81] & [2.92] & [2.05] \\ 
  & & & & & & & & \\ 
 Constant & -0.09 & 0.31$^{***}$ & 0.24$^{**}$ & 0.17$^{*}$ & 0.14 & 0.27$^{***}$ & 0.19$^{*}$ & 0.11 \\ 
  & [-0.74] & [3.24] & [2.26] & [1.82] & [1.47] & [2.86] & [1.78] & [1.22] \\ 
  & & & & & & & & \\ 
\hline \\[-1.8ex] 
Observations & 312 & 312 & 312 & 312 & 312 & 312 & 312 & 312 \\ 
R$^{2}$ & 0.147 & 0.021 & 0.058 & 0.188 & 0.200 & 0.049 & 0.092 & 0.210 \\ 
Adjusted R$^{2}$ & 0.145 & 0.011 & 0.043 & 0.178 & 0.184 & 0.037 & 0.074 & 0.192 \\ 
Residual Std. Error & 1.368 & 1.471 & 1.448 & 1.342 & 1.336 & 1.452 & 1.424 & 1.330 \\ 
F Statistic & 53.610 & 2.194 & 3.778 & 17.820 & 12.700 & 3.967 & 5.130 & 11.560 \\ 
\hline 
\hline \\[-1.8ex] 

\end{tabular} 
\end{table}
\restoregeometry

\newgeometry{left=1.5cm, right=1cm, top=3cm, bottom=1.5cm}
\begin{table}[!htbp] \centering 
  \caption{\textbf{Factor spanning test for Earnings momentum (CAR3), 2005-2018} }
  \label{} 
    \begin{flushleft}
    {\medskip\small
    This table reports the results of time-series regression of earning momentum factor CAR3 on the other factors.
    CMOM, SUEF and CAR3F are factors, constructed from 2x3 sorts on ME and respective variable (past returns of customers' portfolio during the past month, SUE and CAR3). The factors use NYSE breakpoints and are constructed similarly to UMD. }
    \medskip
    \end{flushleft}
  \footnotesize
\begin{tabular}{@{\extracolsep{0pt}}lcccccccc} 
\\[-1.8ex]\hline 
\hline \\[-1.8ex] 
 & \multicolumn{8}{c}{\textit{Dependent variable:}} \\ 
\cline{2-9} 
\\[-1.8ex] & \multicolumn{8}{c}{CAR3F} \\ 
\\[-1.8ex] & (1) & (2) & (3) & (4) & (5) & (6) & (7) & (8)\\ 
\hline \\[-1.8ex] 
 SUEF & 0.21$^{*}$ &  &  &  &  &  &  &  \\ 
  & [1.71] &  &  &  &  &  &  &  \\ 
  & & & & & & & & \\ 
 Mkt.RF &  & -0.10$^{*}$ & -0.09$^{*}$ & -0.04 & -0.04 & -0.10$^{*}$ & -0.09$^{*}$ & -0.05 \\ 
  &  & [-1.70] & [-1.67] & [-0.93] & [-0.96] & [-1.80] & [-1.82] & [-1.13] \\ 
  & & & & & & & & \\ 
 SMB &  & -0.04 & -0.04 & -0.04 & -0.05 & -0.04 & -0.05 & -0.06 \\ 
  &  & [-0.71] & [-0.69] & [-0.79] & [-0.88] & [-0.78] & [-0.80] & [-1.01] \\ 
  & & & & & & & & \\ 
 HML &  & -0.15$^{**}$ & -0.18$^{**}$ & -0.03 & -0.05 & -0.15$^{**}$ & -0.18$^{**}$ & -0.04 \\ 
  &  & [-2.20] & [-2.38] & [-0.38] & [-0.78] & [-2.03] & [-2.18] & [-0.58] \\ 
  & & & & & & & & \\ 
 CMA &  &  & 0.10 &  & 0.06 &  & 0.11 & 0.08 \\ 
  &  &  & [0.63] &  & [0.56] &  & [0.70] & [0.68] \\ 
  & & & & & & & & \\ 
 RMW &  &  & -0.005 &  & -0.03 &  & -0.01 & -0.04 \\ 
  &  &  & [-0.06] &  & [-0.49] &  & [-0.16] & [-0.72] \\ 
  & & & & & & & & \\ 
 UMD &  &  &  & 0.20$^{***}$ & 0.20$^{***}$ &  &  & 0.21$^{***}$ \\ 
  &  &  &  & [4.47] & [4.62] &  &  & [4.72] \\ 
  & & & & & & & & \\ 
 CMOM &  &  &  &  &  & 0.03 & 0.04 & 0.07 \\ 
  &  &  &  &  &  & [0.37] & [0.47] & [1.23] \\ 
  & & & & & & & & \\ 
 Constant & -0.31 & 0.02 & 0.02 & -0.03 & -0.02 & 0.02 & 0.02 & -0.02 \\ 
  & [-1.16] & [0.19] & [0.16] & [-0.28] & [-0.23] & [0.17] & [0.18] & [-0.15] \\ 
  & & & & & & & & \\ 
\hline \\[-1.8ex] 
Observations & 162 & 162 & 162 & 162 & 162 & 158 & 158 & 158 \\ 
R$^{2}$ & 0.067 & 0.149 & 0.153 & 0.360 & 0.362 & 0.155 & 0.160 & 0.379 \\ 
Adjusted R$^{2}$ & 0.061 & 0.133 & 0.126 & 0.344 & 0.338 & 0.132 & 0.127 & 0.350 \\ 
Residual Std. Error & 1.669 & 1.604 & 1.610 & 1.395 & 1.402 & 1.618 & 1.623 & 1.400 \\ 
F Statistic & 11.520 & 9.203 & 5.654 & 22.110 & 14.690 & 6.994 & 4.790 & 13.100 \\ 
\hline 
\hline \\[-1.8ex] 

\end{tabular} 
\end{table}
\restoregeometry

\begin{table}[!htbp] \centering 
  \caption{\textbf{Returns of long-short portfolios, formed on customer momentum}}
  \label{} 
  \begin{flushleft}
    {\medskip\small
    This table reports the returns of long-short portfolios, formed on the returns of customers' portfolio at a given lag. Type represents a lag, which is defined as (Beginning of return window as a number of months before the current month)-(End of return window as a number of months before the current month). This table reports the results for 2 subsamples. The first subsample (l1) is restricted to the supplier-customer pairs, where mean size of customer is less than the size of supplier. Size is defined as market equity. The second subsample (l2) is restricted to the supplier-customer pairs, where mean size of customer can not be more than twice the size of supplier. The columns l1 and l2 report returns of long-short portfolios, sorted on past customers' returns in restricted subsamples. Full sample has 143,054 observations. l1 and l2 subsamples have 7462 and 12120 observations respectively. Ew and Vw stand for equally-weighted and value-weighted portfolios.
    }
    \medskip
    \end{flushleft}
\begin{tabular}{@{\extracolsep{5pt}} cccccccccc} 
\\[-1.8ex]\hline 
\hline \\[-1.8ex] 
& & \multicolumn{2}{c}{Ew, quintiles} & \multicolumn{2}{c}{Vw, quintiles}& \multicolumn{2}{c}{Ew, deciles} & \multicolumn{2}{c}{Vw, deciles}\\
\\[-1.8ex]\hline 
\hline \\[-1.8ex] 
Type & Statistic & l1 & l2 & l1 & l2 & l1 & l2 & l1 & l2 \\ 
\hline \\[-1.8ex] 
1-1 & Mean & 1.37$^{**}$ & 0.99$^{**}$ & 1.92$^{***}$ & 1.04$^{**}$ & -0.16 & 0.45 & 0.45 & 0.93 \\ 
 & T-stat & [2.15] & [2.04] & [2.74] & [2.00] & [-0.20] & [0.66] & [0.57] & [1.24] \\ 
2-1 & Mean & 1.26$^{**}$ & 0.86$^{*}$ & 1.40$^{**}$ & 0.97$^{*}$ & 1.02 & 0.68 & 0.93 & 1.15 \\ 
 & T-stat & [2.06] & [1.68] & [2.10] & [1.72] & [1.26] & [1.01] & [1.06] & [1.53] \\ 
3-1 & Mean & 0.49 & 0.40 & 1.06 & 0.48 & 1.39$^{*}$ & 0.49 & 1.74$^{**}$ & 0.64 \\ 
 & T-stat & [0.79] & [0.78] & [1.56] & [0.91] & [1.77] & [0.72] & [1.96] & [0.84] \\ 
4-1 & Mean & 0.54 & 0.58 & 1.15 & 0.37 & 0.80 & 0.54 & 1.16 & 1.01 \\ 
 & T-stat & [0.78] & [1.05] & [1.61] & [0.64] & [1.08] & [0.70] & [1.46] & [1.23] \\ 
7-1 & Mean & -0.32 & 0.06 & -0.18 & 0.33 & 0.51 & -0.02 & 0.66 & -0.28 \\ 
 & T-stat & [-0.43] & [0.10] & [-0.22] & [0.56] & [0.70] & [-0.03] & [0.91] & [-0.41] \\ 
12-1 & Mean & -0.28 & -0.57 & -0.54 & -0.56 & 0.62 & -0.48 & 0.27 & -0.85 \\ 
 & T-stat & [-0.37] & [-0.83] & [-0.70] & [-0.81] & [0.86] & [-0.72] & [0.36] & [-1.11] \\ 
2-2 & Mean & 1.30$^{**}$ & 0.79 & 1.29$^{*}$ & 0.17 & 1.12 & 1.15$^{*}$ & 0.35 & 0.46 \\ 
 & T-stat & [1.98] & [1.55] & [1.75] & [0.31] & [1.50] & [1.78] & [0.44] & [0.70] \\ 
4-2 & Mean & 0.94 & 0.58 & 1.36$^{*}$ & 0.19 & 1.16 & 0.94 & 2.04$^{**}$ & 1.15 \\ 
 & T-stat & [1.34] & [1.05] & [1.88] & [0.32] & [1.52] & [1.31] & [2.55] & [1.52] \\ 
7-2 & Mean & -0.21 & 0.74 & -0.01 & 0.90 & 1.38$^{*}$ & 0.58 & 1.94$^{**}$ & 0.76 \\ 
 & T-stat & [-0.27] & [1.20] & [-0.01] & [1.38] & [1.74] & [0.80] & [2.46] & [1.05] \\ 
12-2 & Mean & -0.84 & -0.72 & -1.22 & -0.76 & -0.46 & -0.76 & -0.95 & -0.96 \\ 
 & T-stat & [-1.07] & [-1.03] & [-1.56] & [-1.03] & [-0.59] & [-1.09] & [-1.17] & [-1.23] \\ 
3-3 & Mean & 0.17 & 0.36 & 0.59 & 0.63 & 0.34 & 0.38 & 0.90 & 1.00 \\ 
 & T-stat & [0.27] & [0.63] & [0.85] & [1.07] & [0.46] & [0.54] & [1.21] & [1.40] \\ 
7-3 & Mean & -0.45 & -0.49 & -0.13 & -0.81 & 0.94 & -0.25 & 1.29$^{*}$ & 0.14 \\ 
 & T-stat & [-0.67] & [-0.76] & [-0.19] & [-1.15] & [1.34] & [-0.36] & [1.78] & [0.19] \\ 
12-3 & Mean & -1.02 & -0.81 & -1.43$^{*}$ & -1.04 & -0.04 & -0.30 & -0.89 & -0.48 \\ 
 & T-stat & [-1.36] & [-1.21] & [-1.83] & [-1.44] & [-0.06] & [-0.43] & [-1.24] & [-0.61] \\ 
12-7 & Mean & 0.83 & 0.57 & 0.56 & 0.47 & 1.47$^{**}$ & 0.07 & 1.01 & 0.35 \\ 
 & T-stat & [1.12] & [0.89] & [0.74] & [0.69] & [2.05] & [0.10] & [1.44] & [0.46] \\
\hline \\[-1.8ex] 
\end{tabular} 
\end{table}

\pagebreak{}

\newgeometry{left=1.25cm, right=1cm, top=3cm, bottom=1.5cm}
\begin{table}[!htbp] \centering 
  \caption{\textbf{Returns of long-short quintile portfolios, formed on customer momentum across different quintiles of relative customer size}}
  \label{} 
    \begin{flushleft}
    {\medskip\small
    This table reports the returns of long-short quintile portfolios, formed on the returns of customers' portfolio at a given lag. Type represents a lag, which is defined as (Beginning of return window as a number of months before the current month)-(End of return window as a number of months before the current month). This table reports the results separately for 5 quintiles of relative customer size, defined as mean customer size/supplier size. So for each month I divide the firms into 5 quintiles on relative customer size and then perform sort on past customers' returns within each size quintile. Reported results are the returns of long-short quintile portfolios, formed on the returns of customers' portfolio at a given lag. For example, 0.206 in the column "Q1" means that the return of long-short quintile portfolio (largest quintile by past month customers' returns minus the smallest quintile) is equal to 0.206 within the subsample of supplier-customer pairs with low (first quintile) relative customer size. 
    }
    \medskip
    \end{flushleft}
\begin{tabular}{@{\extracolsep{0pt}} cccccccccccc} 
\\[-1.8ex]\hline 
\hline \\[-1.8ex] 
Weights &  \multicolumn{7}{c}{Equal} & \multicolumn{3}{c}{Value}\\
\\[-1.8ex]\hline 
\hline \\[-1.8ex] 
Type & Statistic & Q1 & Q2 & Q3 & Q4 & Q5 & Q1 & Q2 & Q3 & Q4 & Q5 \\ 
\hline \\[-1.8ex] 
1-1 & Mean & 0.21 & 0.94$^{***}$ & 0.46 & 1.29$^{***}$ & 1.27$^{***}$ & 0.19 & 1.04$^{***}$ & 0.35 & 1.52$^{***}$ & 0.73$^{*}$ \\ 
 & T-stat & [0.78] & [2.98] & [1.40] & [4.33] & [3.71] & [0.55] & [2.64] & [0.88] & [3.92] & [1.76] \\ 
2-1 & Mean & 0.68$^{**}$ & 0.42 & 0.60$^{*}$ & 1.00$^{***}$ & 0.67$^{*}$ & 0.80$^{**}$ & 0.64$^{*}$ & 0.58 & 1.06$^{***}$ & 0.33 \\ 
 & T-stat & [2.35] & [1.31] & [1.65] & [3.15] & [1.73] & [2.34] & [1.66] & [1.50] & [2.63] & [0.70] \\ 
3-1 & Mean & 0.66$^{**}$ & 0.56$^{*}$ & 0.71$^{*}$ & 0.70$^{**}$ & 0.90$^{**}$ & 0.17 & 0.92$^{**}$ & 0.74$^{*}$ & 0.66 & 1.00$^{**}$ \\ 
 & T-stat & [2.14] & [1.72] & [1.69] & [2.12] & [2.36] & [0.47] & [2.36] & [1.78] & [1.55] & [2.28] \\ 
4-1 & Mean & 0.42 & 0.38 & 0.26 & 0.38 & 0.99$^{**}$ & 0.08 & 0.34 & 0.29 & -0.02 & 1.06$^{**}$ \\ 
 & T-stat & [1.42] & [1.19] & [0.69] & [1.08] & [2.36] & [0.21] & [0.87] & [0.65] & [-0.04] & [2.19] \\ 
7-1 & Mean & 0.35 & 0.17 & 0.44 & 0.17 & 1.26$^{***}$ & 0.15 & 0.14 & 0.05 & -0.28 & 1.21$^{**}$ \\ 
 & T-stat & [1.14] & [0.52] & [1.14] & [0.43] & [3.00] & [0.40] & [0.35] & [0.12] & [-0.61] & [2.51] \\ 
12-1 & Mean & 0.66$^{*}$ & 1.00$^{***}$ & 1.39$^{**}$ & 0.24 & 1.29$^{***}$ & 0.38 & 0.79$^{**}$ & 1.24$^{**}$ & 0.54 & 1.46$^{***}$ \\ 
 & T-stat & [1.79] & [2.61] & [2.29] & [0.43] & [2.90] & [0.88] & [1.96] & [1.97] & [0.89] & [2.72] \\ 
2-2 & Mean & 0.38 & -0.02 & 0.54 & 0.36 & 0.16 & 0.13 & 0.44 & 0.88$^{**}$ & 0.18 & 0.35 \\ 
 & T-stat & [1.37] & [-0.05] & [1.28] & [1.04] & [0.40] & [0.38] & [1.16] & [2.09] & [0.42] & [0.75] \\ 
4-2 & Mean & 0.55$^{*}$ & 0.16 & 0.43 & 0.43 & 0.79$^{**}$ & 0.01 & 0.00 & 0.38 & 0.05 & 0.97$^{**}$ \\ 
 & T-stat & [1.85] & [0.52] & [0.99] & [1.16] & [1.97] & [0.02] & [0.00] & [0.86] & [0.11] & [2.03] \\ 
7-2 & Mean & 0.63$^{*}$ & -0.18 & 0.79$^{*}$ & 0.87$^{*}$ & 1.00$^{**}$ & 0.34 & 0.18 & 0.90$^{*}$ & 0.95$^{*}$ & 1.15$^{**}$ \\ 
 & T-stat & [1.85] & [-0.48] & [1.83] & [1.91] & [2.22] & [0.89] & [0.42] & [1.94] & [1.73] & [2.21] \\ 
12-2 & Mean & 0.62$^{*}$ & 0.64 & 1.17$^{*}$ & 0.25 & 1.28$^{***}$ & 0.30 & 0.51 & 1.34$^{**}$ & 0.35 & 1.56$^{***}$ \\ 
 & T-stat & [1.73] & [1.64] & [1.88] & [0.48] & [3.04] & [0.70] & [1.18] & [2.08] & [0.59] & [2.99] \\ 
3-3 & Mean & 0.37 & 0.11 & 0.28 & 0.20 & 0.67$^{*}$ & 0.31 & 0.42 & 0.30 & 0.34 & 0.73$^{*}$ \\ 
 & T-stat & [1.20] & [0.38] & [0.81] & [0.52] & [1.85] & [0.86] & [1.11] & [0.76] & [0.74] & [1.70] \\ 
7-3 & Mean & 0.45 & 0.09 & 0.23 & 0.36 & 0.90$^{**}$ & 0.29 & 0.15 & 0.30 & 0.13 & 0.98$^{**}$ \\ 
 & T-stat & [1.30] & [0.28] & [0.56] & [0.82] & [2.34] & [0.77] & [0.37] & [0.65] & [0.25] & [2.08] \\ 
12-3 & Mean & 0.38 & 0.25 & 1.03$^{*}$ & 0.47 & 1.25$^{***}$ & 0.22 & 0.16 & 1.06$^{*}$ & 0.62 & 1.52$^{***}$ \\ 
 & T-stat & [1.05] & [0.67] & [1.69] & [0.93] & [2.97] & [0.51] & [0.39] & [1.70] & [1.08] & [3.00] \\ 
12-7 & Mean & 0.74$^{**}$ & 0.91$^{**}$ & 0.83$^{*}$ & 0.78 & 0.08 & 0.20 & 0.89$^{**}$ & 0.69 & 0.97$^{*}$ & 0.04 \\ 
 & T-stat & [2.03] & [2.38] & [1.87] & [1.54] & [0.17] & [0.50] & [2.14] & [1.47] & [1.72] & [0.08] \\  
\hline \\[-1.8ex] 
\end{tabular} 
\end{table}
\restoregeometry

\newgeometry{left=1.25cm, right=1cm, top=3cm, bottom=1.5cm}
\begin{table}[!htbp] \centering 
  \caption{\textbf{Returns of long-short decile portfolios, formed on customer momentum across different quintiles of relative customer size} }
  \label{} 
      \begin{flushleft}
    {\medskip\small
    This table reports the returns of long-short decile portfolios, formed on the returns of customers' portfolio at a given lag. Type represents a lag, which is defined as (Beginning of return window as a number of months before the current month)-(End of return window as a number of months before the current month). This table reports the results separately for 5 quintiles of relative customer size, defined as mean customer size/supplier size. So for each month I divide the firms into 5 quintiles on relative customer size and then perform sort on past customers' returns within each size quintile. Reported results are the returns of long-short decile portfolios, formed on the returns of customers' portfolio at a given lag. For example, 0.451 in the column "Q1" means that the return of long-short decile portfolio (largest decile by past month customers' returns minus the smallest decile) is equal to 0.451 within the subsample of supplier-customer pairs with low (first quintile) relative customer size. 
    }
    \medskip
    \end{flushleft}
\begin{tabular}{@{\extracolsep{-1pt}} cccccccccccc} 
\\[-1.8ex]\hline 
\hline \\[-1.8ex] 
Weights &  \multicolumn{7}{c}{Equal} & \multicolumn{3}{c}{Value}\\
\\[-1.8ex]\hline 
\hline \\[-1.8ex] 
Type & Statistic & Q1 & Q2 & Q3 & Q4 & Q5 & Q1 & Q2 & Q3 & Q4 & Q5 \\ 
\hline \\[-1.8ex] 
1-1 & Mean & 0.45 & 1.34$^{***}$ & 0.62 & 0.96$^{**}$ & 1.61$^{***}$ & 0.51 & 1.06$^{**}$ & 0.55 & 1.38$^{***}$ & 1.12$^{**}$ \\ 
 & T-stat & [1.19] & [2.93] & [1.33] & [2.08] & [3.32] & [1.11] & [2.08] & [1.00] & [2.66] & [2.12] \\ 
2-1 & Mean & 0.57 & 0.82$^{*}$ & 1.36$^{**}$ & 1.15$^{***}$ & 0.82 & 0.59 & 1.71$^{***}$ & 1.08$^{*}$ & 1.05$^{**}$ & 1.03$^{*}$ \\ 
 & T-stat & [1.52] & [1.83] & [2.31] & [2.58] & [1.61] & [1.27] & [3.38] & [1.82] & [2.14] & [1.79] \\ 
3-1 & Mean & 0.71 & 0.26 & 0.72 & 1.08$^{**}$ & 1.46$^{***}$ & 1.03$^{**}$ & 0.82$^{*}$ & 0.91 & 1.28$^{**}$ & 1.90$^{***}$ \\ 
 & T-stat & [1.59] & [0.61] & [1.17] & [2.20] & [2.75] & [2.04] & [1.68] & [1.34] & [2.22] & [3.10] \\ 
4-1 & Mean & 1.12$^{**}$ & 0.78$^{*}$ & 0.66 & 0.59 & 1.36$^{**}$ & 1.30$^{***}$ & 0.62 & 0.58 & 0.23 & 1.70$^{***}$ \\ 
 & T-stat & [2.34] & [1.76] & [1.25] & [1.13] & [2.49] & [2.60] & [1.22] & [1.08] & [0.39] & [2.66] \\ 
7-1 & Mean & 0.86$^{*}$ & 0.35 & 0.69 & 0.26 & 1.72$^{***}$ & 0.61 & 0.37 & 0.42 & 0.01 & 2.08$^{***}$ \\ 
 & T-stat & [1.74] & [0.73] & [1.26] & [0.48] & [3.12] & [1.21] & [0.69] & [0.71] & [0.01] & [3.44] \\ 
12-1 & Mean & 1.38$^{**}$ & 0.90$^{*}$ & 2.10$^{*}$ & 0.89 & 1.11$^{*}$ & 1.21$^{**}$ & 1.24$^{**}$ & 2.10$^{*}$ & 0.68 & 1.37$^{*}$ \\ 
 & T-stat & [2.56] & [1.75] & [1.93] & [1.19] & [1.73] & [1.97] & [2.42] & [1.88] & [0.88] & [1.90] \\ 
2-2 & Mean & 0.26 & -0.13 & 0.73 & 0.70 & 0.00 & -0.15 & 0.76 & 1.01 & 0.54 & 0.00 \\ 
 & T-stat & [0.68] & [-0.30] & [1.18] & [1.43] & [-0.01] & [-0.33] & [1.56] & [1.50] & [0.97] & [-0.01] \\ 
4-2 & Mean & 0.94$^{**}$ & 0.37 & -0.10 & 0.72 & 0.54 & 0.86$^{*}$ & -0.12 & 0.15 & -0.05 & 1.01$^{*}$ \\ 
 & T-stat & [2.03] & [0.80] & [-0.20] & [1.32] & [1.05] & [1.76] & [-0.23] & [0.27] & [-0.08] & [1.69] \\ 
7-2 & Mean & 0.85$^{*}$ & -0.25 & 0.95 & 0.16 & 1.68$^{***}$ & 0.68 & -0.11 & 0.89 & -0.04 & 2.08$^{***}$ \\ 
 & T-stat & [1.65] & [-0.47] & [1.34] & [0.27] & [2.70] & [1.25] & [-0.18] & [1.19] & [-0.06] & [3.02] \\ 
12-2 & Mean & 0.79 & 0.23 & 0.62 & 1.13 & 1.66$^{**}$ & 0.66 & 0.79 & 0.52 & 1.18 & 1.48$^{**}$ \\ 
 & T-stat & [1.41] & [0.41] & [1.03] & [1.54] & [2.56] & [1.04] & [1.40] & [0.86] & [1.53] & [2.08] \\ 
3-3 & Mean & 0.35 & -0.28 & -0.16 & 0.63 & 0.53 & 0.25 & 0.14 & -0.10 & 0.27 & 0.94 \\ 
 & T-stat & [0.76] & [-0.65] & [-0.33] & [1.21] & [1.00] & [0.56] & [0.29] & [-0.18] & [0.44] & [1.53] \\ 
7-3 & Mean & 0.40 & -0.57 & 0.79 & -0.29 & 0.69 & 0.70 & -0.54 & 0.66 & -0.42 & 1.09$^{*}$ \\ 
 & T-stat & [0.78] & [-1.09] & [1.14] & [-0.48] & [1.22] & [1.33] & [-0.97] & [0.91] & [-0.64] & [1.72] \\ 
12-3 & Mean & 0.60 & 0.09 & 0.01 & 0.92 & 0.72 & 0.32 & 0.29 & 0.02 & 1.00 & 0.60 \\ 
 & T-stat & [1.18] & [0.17] & [0.01] & [1.23] & [1.20] & [0.56] & [0.52] & [0.02] & [1.29] & [0.86] \\ 
12-7 & Mean & 0.73 & 0.78 & 0.50 & 1.65$^{**}$ & 0.25 & 0.60 & 0.66 & 0.41 & 1.93$^{**}$ & 0.38 \\ 
 & T-stat & [1.43] & [1.38] & [0.75] & [2.05] & [0.38] & [1.04] & [1.13] & [0.60] & [2.21] & [0.56] \\ 
\hline \\[-1.8ex] 
\end{tabular} 
\end{table}
\restoregeometry

\newgeometry{left=1.25cm, right=1cm, top=2cm, bottom=1.5cm}
\begin{table}[!htbp] \centering 
  \caption{\textbf{Fama-MacBeth regressions with interaction term}} 
  \label{} 
      \begin{flushleft}
    {\medskip\small
This table reports the results of Fama-MacBeth regression of returns on the characteristics, including the interaction term between past customers' returns and relative customers' size. Cmom1-1 is returns of customers' portfolio during the last month. Mom1-1 and Mom12-2 are past returns of stock at lags 1-1 and 12-2. OP is operating profitability. CAR3 is abnormal return during 3-day window, centered around earnings announcement. Standardized unexpected earnings  (SUE) are defined as the most recent year-to-year change in earnings per share (EPS), divided by the standard deviation of the changes in earnings during the recent eight announcements. Cmom1-1\_cme is the interaction term between returns of customers' portfolio in the last month and customer's size. Cmom1-1\_rme is the interaction term between returns of customers' portfolio in the last month and customer's relative size. Relative size of customer is defined as customer's size/supplier's size.}
    \medskip
    \end{flushleft}
  \small
\begin{tabular}{@{\extracolsep{-7pt}}lccccccccc} 
\\[-1.8ex]\hline 
\hline \\[-1.8ex] 
 & \multicolumn{9}{c}{\textit{Dependent variable:}} \\ 
\cline{2-10} 
\\[-1.8ex] & (1) & (2) & (3) & (4) & (5) & (6) & (7) & (8) & (9)\\ 
\hline \\[-1.8ex] 
 Mom12-2 &  &  &  &  &  & 0.002 & -0.004 & -0.004 & -0.004 \\ 
  &  &  &  &  &  & [0.85] & [-1.53] & [-1.45] & [-1.40] \\ 
  & & & & & & & & & \\ 
 Mom1-1 &  &  &  &  &  & -0.04$^{***}$ & -0.06$^{***}$ & -0.06$^{***}$ & -0.06$^{***}$ \\ 
  &  &  &  &  &  & [-5.13] & [-6.83] & [-6.99] & [-7.21] \\ 
  & & & & & & & & & \\ 
 Cmom & 0.05$^{***}$ & 0.05$^{***}$ & 0.04$^{***}$ & -0.07 & -0.04 & 0.05$^{***}$ & 0.05$^{***}$ & -0.01 & -0.01 \\ 
  & [3.39] & [3.33] & [3.28] & [-0.96] & [-1.15] & [3.60] & [3.78] & [-0.20] & [-0.38] \\ 
  & & & & & & & & & \\ 
 Cmom*cME &  &  &  & 0.01$^{*}$ &  &  &  & 0.01 &  \\ 
  &  &  &  & [1.71] &  &  &  & [0.96] &  \\ 
  & & & & & & & & & \\ 
 Cmom*rME &  &  &  &  & 0.05$^{**}$ &  &  &  & 0.04$^{*}$ \\ 
  &  &  &  &  & [2.39] &  &  &  & [1.82] \\ 
  & & & & & & & & & \\ 
 log(ME) &  & -0.16$^{***}$ & -0.17$^{***}$ & -0.19$^{***}$ & -0.22$^{***}$ & -0.19$^{***}$ & -0.18$^{***}$ & -0.19$^{***}$ & -0.23$^{***}$ \\ 
  &  & [-2.82] & [-3.28] & [-3.50] & [-3.76] & [-3.64] & [-3.33] & [-3.54] & [-4.03] \\ 
  & & & & & & & & & \\ 
 log(B/M) &  & 0.08 & 0.12 & 0.15 & 0.12 & 0.15 & 0.20 & 0.26$^{*}$ & 0.20 \\ 
  &  & [0.61] & [0.79] & [0.96] & [0.82] & [1.03] & [1.34] & [1.67] & [1.33] \\ 
  & & & & & & & & & \\ 
 OP &  &  & 0.48 & 0.52 & 0.42 & 0.16 & 0.58 & 0.67 & 0.58 \\ 
  &  &  & [1.03] & [1.10] & [0.89] & [0.35] & [1.22] & [1.39] & [1.20] \\ 
  & & & & & & & & & \\ 
 CAR3 &  &  & 2.76$^{**}$ & 2.90$^{**}$ & 2.84$^{**}$ &  & 3.55$^{***}$ & 3.63$^{***}$ & 3.79$^{***}$ \\ 
  &  &  & [2.39] & [2.46] & [2.46] &  & [2.91] & [2.90] & [3.08] \\ 
  & & & & & & & & & \\ 
 SUE &  &  & 0.87$^{***}$ & 0.87$^{***}$ & 0.88$^{***}$ &  & 0.91$^{***}$ & 0.91$^{***}$ & 0.92$^{***}$ \\ 
  &  &  & [12.69] & [12.78] & [12.68] &  & [12.84] & [12.89] & [12.79] \\ 
  & & & & & & & & & \\ 
 Constant & 1.40$^{***}$ & 2.30$^{***}$ & 2.17$^{***}$ & 2.24$^{***}$ & 2.41$^{***}$ & 2.30$^{***}$ & 2.23$^{***}$ & 2.28$^{***}$ & 2.51$^{***}$ \\ 
  & [5.31] & [5.48] & [5.16] & [5.30] & [5.33] & [5.58] & [5.39] & [5.49] & [5.62] \\ 
  & & & & & & & & & \\ 
\hline \\[-1.8ex] 
Observations & 47,488 & 47,488 & 47,488 & 47,488 & 47,488 & 47,488 & 47,488 & 47,488 & 47,488 \\ 
R$^{2}$ & 0.002 & 0.003 & 0.007 & 0.007 & 0.008 & 0.003 & 0.008 & 0.008 & 0.009 \\ 
Adjusted R$^{2}$ & 0.002 & 0.003 & 0.007 & 0.007 & 0.008 & 0.003 & 0.008 & 0.008 & 0.008 \\ 

\hline 
\hline \\[-1.8ex] 
 
\end{tabular} 
\end{table}

\begin{table}[!htbp] \centering 
  \caption{\textbf{Returns of portfolios, formed on customers' SUE}} 
  \label{} 
    \label{} 
      \begin{flushleft}
    {\medskip\small
This table reports the returns of quintile and decile portfolios, formed by sorts on SUE of customers' portfolio. L/S corresponds to long-short portfolio, defined as a difference in returns of tenth and first decile portfolios (D10-D1) or fifth and first quintile portfolio (Q5-Q1). Tenth portfolio (D10) is the portfolio of the firms with the largest customers' returns. Standard errors are calculated using Newey-West adjustment.}
    \medskip
    \end{flushleft}
\begin{tabularx}{\linewidth}{l*{7}{Y}}
    \toprule
    \multicolumn{7}{l}{\textbf{Panel A: Returns of equally-weighted quintile portfolios, sorted on SUE of customers}} \\
    \midrule
\\[-1.8ex]\hline 
\hline \\[-1.8ex] 
statistic & Q1 & Q2 & Q3 & Q4 & Q5 & LS \\ 
\hline \\[-1.8ex] 
Mean & 1.02$^{***}$ & 1.05$^{***}$ & 1.17$^{***}$ & 1.06$^{***}$ & 1.29$^{***}$ & 0.27 \\ 
T-stat & [3.34] & [3.55] & [4.02] & [3.58] & [4.25] & [1.43] \\ 
\hline \\[-1.8ex] 
\end{tabularx} 
\begin{tabularx}{\linewidth}{l*{7}{Y}}
    \toprule
    \multicolumn{7}{l}{\textbf{Panel B: Returns of value-weighted quintile portfolios, sorted on SUE of customers}} \\
    \midrule
\\[-1.8ex]\hline 
\hline \\[-1.8ex] 
statistic & Q1 & Q2 & Q3 & Q4 & Q5 & LS \\ 
\hline \\[-1.8ex] 
Mean & 0.47 & 0.77$^{**}$ & 0.63$^{**}$ & 0.40 & 0.82$^{**}$ & 0.36 \\ 
T-stat & [1.52] & [2.44] & [2.09] & [1.28] & [2.56] & [1.30] \\ 
\hline \\[-1.8ex] 
\end{tabularx} 
\begin{tabularx}{\linewidth}{l*{12}{Y}}
    \toprule
    \multicolumn{12}{l}{\textbf{Panel C: Returns of equally-weighted decile portfolios, sorted on SUE of customers}} \\
    \midrule
\\[-1.8ex]\hline 
\hline \\[-1.8ex] 
statistic & D1 & D2 & D3 & D4 & D5 & D6 & D7 & D8 & D9 & D10 & LS \\ 
\hline \\[-1.8ex] 
Mean & 0.89$^{**}$ & 1.18$^{***}$ & 0.92$^{***}$ & 1.15$^{***}$ & 1.17$^{***}$ & 1.15$^{***}$ & 0.84$^{***}$ & 1.26$^{***}$ & 1.22$^{***}$ & 1.42$^{***}$ & 0.54$^{**}$ \\ 
T-stat & [2.57] & [3.65] & [2.87] & [3.62] & [3.74] & [3.71] & [2.63] & [3.99] & [3.81] & [4.33] & [2.02] \\ 
\hline \\[-1.8ex] 
\end{tabularx} 

\begin{tabularx}{\linewidth}{l*{12}{Y}}
    \toprule
    \multicolumn{12}{l}{\textbf{Panel D: Returns of value-weighted decile portfolios, sorted on SUE of customers}} \\
    \midrule
\\[-1.8ex]\hline 
\hline \\[-1.8ex] 
statistic & D1 & D2 & D3 & D4 & D5 & D6 & D7 & D8 & D9 & D10 & LS \\ 
\hline \\[-1.8ex] 
Mean & 0.48 & 0.60$^{*}$ & 0.52 & 1.04$^{***}$ & 0.89$^{***}$ & 0.54$^{*}$ & 0.59$^{*}$ & 0.46 & 0.82$^{**}$ & 0.94$^{***}$ & 0.45 \\ 
T-stat & [1.37] & [1.79] & [1.41] & [2.87] & [2.63] & [1.67] & [1.72] & [1.34] & [2.42] & [2.60] & [1.31] \\ 
\hline \\[-1.8ex] 
\end{tabularx} 
\end{table}

\begin{table}[!htbp] \centering 
  \caption{\textbf{Returns of portfolios, formed on customers' CAR3}} 
  \label{} 
        \begin{flushleft}
    {\medskip\small
This table reports the returns of quintile and decile portfolios, formed by sorts on CAR3 of customers' portfolio. L/S corresponds to long-short portfolio, defined as a difference in returns of tenth and first decile portfolios (D10-D1) or fifth and first quintile portfolio (Q5-Q1). Tenth portfolio (D10) is the portfolio of the firms with the largest customers' returns. Standard errors are calculated using Newey-West adjustment.}
    \medskip
    \end{flushleft}
\begin{tabularx}{\linewidth}{l*{7}{Y}}
    \toprule
    \multicolumn{7}{l}{\textbf{Panel A: Returns of equally-weighted quintile portfolios, sorted on CAR3 of customers}} \\
    \midrule
\\[-1.8ex]\hline 
\hline \\[-1.8ex] 
statistic & Q1 & Q2 & Q3 & Q4 & Q5 & LS \\ 
\hline \\[-1.8ex] 
Mean & 1.15$^{***}$ & 0.92$^{***}$ & 1.19$^{***}$ & 1.14$^{***}$ & 1.21$^{***}$ & 0.06 \\ 
T-stat & [3.63] & [3.10] & [4.32] & [3.98] & [3.81] & [0.32] \\ 
\hline \\[-1.8ex] 
\end{tabularx} 
\begin{tabularx}{\linewidth}{l*{7}{Y}}
    \toprule
    \multicolumn{7}{l}{\textbf{Panel B: Returns of value-weighted quintile portfolios, sorted on CAR3 of customers}} \\
    \midrule
\\[-1.8ex]\hline 
\hline \\[-1.8ex] 
statistic & Q1 & Q2 & Q3 & Q4 & Q5 & LS \\ 
\hline \\[-1.8ex] 
Mean & 0.76$^{**}$ & 0.60$^{**}$ & 0.52$^{*}$ & 0.55$^{*}$ & 0.66$^{*}$ & -0.10 \\ 
T-stat & [2.31] & [2.02] & [1.88] & [1.66] & [1.94] & [-0.36] \\ 
\hline \\[-1.8ex]
\end{tabularx} 
\begin{tabularx}{\linewidth}{l*{12}{Y}}
    \toprule
    \multicolumn{12}{l}{\textbf{Panel C: Returns of equally-weighted decile portfolios, sorted on CAR3 of customers}} \\
    \midrule
\\[-1.8ex]\hline 
\hline \\[-1.8ex] 
statistic & D1 & D2 & D3 & D4 & D5 & D6 & D7 & D8 & D9 & D10 & LS \\ 
\hline \\[-1.8ex] 
Mean & 1.22$^{***}$ & 1.10$^{***}$ & 1.13$^{***}$ & 0.74$^{**}$ & 1.14$^{***}$ & 1.23$^{***}$ & 0.97$^{***}$ & 1.28$^{***}$ & 1.31$^{***}$ & 1.07$^{***}$ & -0.14 \\ 
T-stat & [3.45] & [3.31] & [3.51] & [2.36] & [3.81] & [4.16] & [3.34] & [3.98] & [3.77] & [3.18] & [-0.54] \\ 
\hline \\[-1.8ex] 
\end{tabularx} 

\begin{tabularx}{\linewidth}{l*{12}{Y}}
    \toprule
    \multicolumn{12}{l}{\textbf{Panel D: Returns of value-weighted decile portfolios, sorted on CAR3 of customers}} \\
    \midrule
\\[-1.8ex]\hline 
\hline \\[-1.8ex] 
statistic & D1 & D2 & D3 & D4 & D5 & D6 & D7 & D8 & D9 & D10 & LS \\ 
\hline \\[-1.8ex] 
Mean & 0.84$^{**}$ & 0.87$^{**}$ & 0.81$^{**}$ & 0.61$^{*}$ & 0.43 & 0.79$^{***}$ & 0.68$^{**}$ & 0.66$^{*}$ & 0.71$^{*}$ & 0.67$^{*}$ & -0.17 \\ 
T-stat & [2.17] & [2.53] & [2.27] & [1.83] & [1.32] & [2.71] & [2.00] & [1.78] & [1.89] & [1.84] & [-0.50] \\ 
\hline \\[-1.8ex] 
\end{tabularx} 
\end{table}

\clearpage

\begin{table}[!htbp] \centering 
  \caption{\textbf{Daily returns of long-short portfolios, formed on customers momentum}} 
  \label{} 
        \begin{flushleft}
    {\medskip\small
This table reports the daily returns of long-short decile portfolios, formed by sorts on customer momentum. Type corresponds to the width of the window (in trading days) over which customer momentum is calculated. Columns correspond to the length of the window over which the returns of the supplier is calculated. For example, the intersection of row 10-1 and column "15days" contains the return of long-short decile portfolio from sorts on customer momentum, where customer momentum is calculated over the past 10 days and the return of the supplier is cumulative return over the period [t, t+15]. Standard errors are calculated using Newey-West adjustment.}
    \medskip
    \end{flushleft}
\begin{tabularx}{\linewidth}{l*{7}{Y}}
    \toprule
    \multicolumn{7}{l}{\textbf{Panel A: Equal weights}} \\
    \midrule
\\[-1.8ex]\hline 
\hline \\[-1.8ex] 
type & statistic & 1day & 5days & 10days & 15days & 20days & 30days \\ 
\hline \\[-1.8ex] 
1-1 & Mean & 0.17$^{***}$ & 0.28$^{***}$ & 0.38$^{***}$ & 0.49$^{***}$ & 0.53$^{***}$ & 0.58$^{***}$ \\ 
 & T-stat & [14.25] & [11.09] & [10.50] & [10.89] & [10.20] & [8.83] \\ 
5-1 & Mean & 0.12$^{***}$ & 0.32$^{***}$ & 0.54$^{***}$ & 0.71$^{***}$ & 0.79$^{***}$ & 0.92$^{***}$ \\ 
 & T-stat & [9.85] & [12.59] & [15.06] & [16.04] & [15.04] & [14.13] \\ 
10-1 & Mean & 0.12$^{***}$ & 0.37$^{***}$ & 0.66$^{***}$ & 0.85$^{***}$ & 0.96$^{***}$ & 1.13$^{***}$ \\ 
 & T-stat & [10.45] & [15.06] & [18.47] & [18.90] & [18.15] & [17.34] \\ 
20-1 & Mean & 0.11$^{***}$ & 0.40$^{***}$ & 0.68$^{***}$ & 0.88$^{***}$ & 0.98$^{***}$ & 1.21$^{***}$ \\ 
 & T-stat & [9.49] & [15.33] & [18.36] & [19.30] & [18.82] & [18.61] \\ 
40-1 & Mean & 0.09$^{***}$ & 0.36$^{***}$ & 0.64$^{***}$ & 0.87$^{***}$ & 1.06$^{***}$ & 1.36$^{***}$ \\ 
 & T-stat & [7.54] & [13.79] & [17.29] & [18.61] & [19.49] & [20.25] \\ 
\hline \\[-1.8ex] 
\end{tabularx} 

\begin{tabularx}{\linewidth}{l*{7}{Y}}
    \toprule
    \multicolumn{7}{l}{\textbf{Panel B: Value weights}} \\
    \midrule
\\[-1.8ex]\hline 
\hline \\[-1.8ex] 
type & statistic & 1day & 5days & 10days & 15days & 20days & 30days \\ 
\hline \\[-1.8ex] 
1-1 & Mean & 0.12$^{***}$ & 0.17$^{***}$ & 0.22$^{***}$ & 0.29$^{***}$ & 0.35$^{***}$ & 0.36$^{***}$ \\ 
 & T-stat & [7.91] & [4.88] & [4.61] & [5.01] & [5.28] & [4.45] \\ 
5-1 & Mean & 0.08$^{***}$ & 0.15$^{***}$ & 0.29$^{***}$ & 0.42$^{***}$ & 0.48$^{***}$ & 0.66$^{***}$ \\ 
 & T-stat & [5.27] & [4.40] & [5.99] & [7.40] & [7.35] & [8.27] \\ 
10-1 & Mean & 0.08$^{***}$ & 0.20$^{***}$ & 0.41$^{***}$ & 0.56$^{***}$ & 0.58$^{***}$ & 0.76$^{***}$ \\ 
 & T-stat & [5.04] & [5.85] & [8.49] & [9.69] & [8.97] & [9.64] \\ 
20-1 & Mean & 0.10$^{***}$ & 0.28$^{***}$ & 0.46$^{***}$ & 0.61$^{***}$ & 0.73$^{***}$ & 0.94$^{***}$ \\ 
 & T-stat & [6.18] & [7.87] & [9.64] & [10.90] & [11.39] & [11.98] \\ 
40-1 & Mean & 0.08$^{***}$ & 0.31$^{***}$ & 0.52$^{***}$ & 0.65$^{***}$ & 0.70$^{***}$ & 0.86$^{***}$ \\ 
 & T-stat & [5.19] & [8.73] & [10.86] & [11.41] & [10.66] & [10.36] \\ 
\hline \\[-1.8ex] 
\end{tabularx} 
\end{table}

\clearpage

\begin{table}[!htbp] \centering 
  \caption{\textbf{Daily returns of long-short quintile portfolios, double-sorted on investor attention and customer momentum}} 
  \label{} 
        \begin{flushleft}
    {\medskip\small
This table reports the daily returns of long-short quintile portfolios, formed by sorts on investor attention to customer-supplier link. Attention to customer-supplier link is measured as Normalized Abnormal Trading Volume (NAV) of suppliers during the day of earnings announcement of the customer.
Type corresponds to the width of the window (in trading days) over which customer momentum is calculated. Columns correspond to the length of the window over which the returns of the supplier is calculated. For example, the intersection of row 10-1 and column "15days" contains the return of long-short decile portfolio (within the first quintile of investor`s attention) from sorts on customer momentum, where customer momentum is calculated over the past 10 days and the return of the supplier is cumulative return over the period [t, t+15], minus the the return of long-short decile portfolio (within the fifth quintile of investor`s attention) from sorts on customer momentum. In other words, this cell contains the returns of long-short portfolio, conditionally double-sorted on investor attention and customer momentum. Standard errors are calculated using Newey-West adjustment.}
    \medskip
    \end{flushleft}
\begin{tabularx}{\linewidth}{l*{7}{Y}}
    \toprule
    \multicolumn{7}{l}{\textbf{Panel A: Equal weights}} \\
    \midrule
\\[-1.8ex]\hline 
\hline \\[-1.8ex] 
type & statistic & 1day & 5days & 10days & 15days & 20days & 30days \\ 
\hline \\[-1.8ex] 
1-1 & Mean & 0.00 & -0.15 & -0.04 & -0.04 & 0.06 & 0.45 \\ 
 & T-stat & [0.85] & [-1.25] & [-0.24] & [-0.21] & [0.25] & [1.24] \\ 
5-1 & Mean & -0.00 & -0.21$^{*}$ & -0.32$^{**}$ & -0.22 & 0.03 & 0.58$^{*}$ \\ 
 & T-stat & [-0.99] & [-1.77] & [-1.97] & [-1.09] & [0.11] & [1.66] \\ 
10-1 & Mean & 0.00 & -0.00 & -0.21 & 0.17 & 0.49$^{**}$ & 0.75$^{**}$ \\ 
 & T-stat & [-0.20] & [-0.00] & [-1.31] & [0.88] & [2.09] & [2.16] \\ 
20-1 & Mean & 0.00 & -0.00 & 0.24 & 0.55$^{***}$ & 0.63$^{***}$ & 1.58$^{***}$ \\ 
 & T-stat & [-0.61] & [-0.04] & [1.61] & [2.96] & [2.90] & [4.47] \\ 
40-1 & Mean & 0.00 & 0.41$^{***}$ & 0.72$^{***}$ & 1.28$^{***}$ & 1.21$^{***}$ & 1.50$^{***}$ \\ 
 & T-stat & [0.93] & [3.40] & [4.42] & [6.68] & [5.23] & [4.07] \\ 
\hline \\[-1.8ex] 
\end{tabularx}

\begin{tabularx}{\linewidth}{l*{7}{Y}}
    \toprule
    \multicolumn{7}{l}{\textbf{Panel B: Value weights}} \\
    \midrule
\\[-1.8ex]\hline 
\hline \\[-1.8ex] 
type & statistic & 1day & 5days & 10days & 15days & 20days & 30days \\ 
\hline \\[-1.8ex] 
1-1 & Mean & -0.00 & -0.16 & -0.19 & -0.22 & -0.18 & -0.19 \\ 
 & T-stat & [-0.46] & [-1.48] & [-1.21] & [-1.13] & [-0.81] & [-0.71] \\ 
5-1 & Mean & -0.00$^{*}$ & -0.24$^{**}$ & -0.28$^{*}$ & -0.11 & 0.16 & -0.26 \\ 
 & T-stat & [-1.83] & [-2.06] & [-1.76] & [-0.59] & [0.70] & [-0.93] \\ 
10-1 & Mean & 0.00 & 0.03 & -0.04 & 0.03 & -0.09 & -0.30 \\ 
 & T-stat & [-0.79] & [0.25] & [-0.28] & [0.15] & [-0.42] & [-1.14] \\ 
20-1 & Mean & 0.00 & 0.00 & 0.18 & 0.17 & 0.20 & 0.06 \\ 
 & T-stat & [-0.13] & [0.00] & [1.20] & [0.92] & [0.90] & [0.24] \\ 
40-1 & Mean & 0.00 & 0.36$^{***}$ & 1.03$^{***}$ & 1.17$^{***}$ & 1.60$^{***}$ & 1.48$^{***}$ \\ 
 & T-stat & [0.86] & [3.21] & [6.77] & [6.39] & [7.42] & [5.61] \\ 
\hline \\[-1.8ex] 
\end{tabularx} 
\end{table}

\end{document}